\documentclass[a4paper,11pt]{article}
\pdfoutput=1 % if your are submitting a pdflatex (i.e. if you have
\usepackage{jheppub} % for details on the use of the package, please
\usepackage{amssymb}
\usepackage{mathtools}
\usepackage{slashed}
\usepackage{cleveref}
\usepackage{dsfont}
\usepackage{amssymb}
\usepackage{mathtools}
\usepackage{slashed}
\usepackage{cleveref}
\usepackage[mathscr]{eucal}
\usepackage[shortlabels]{enumitem}
\usepackage[table]{xcolor}
\usepackage[makeroom]{cancel}
\usepackage{bbding}
\usepackage{subfig}
\usepackage{booktabs}
\usepackage[]{graphicx}
\usepackage{xltabular}
\usepackage{floatrow}
\usepackage{subfig}  
\usepackage{xcolor,colortbl}
\definecolor{Gray}{gray}{0.9}
\usepackage{caption}

\allowdisplaybreaks

\newcommand{\epm}{e^+e^-}

\newcommand{\sia}{SIA$^{\text{\tiny thr}}$}

\newcommand*\rot{\rotatebox{90}}
\newfloatcommand{capbtabbox}{table}[][\FBwidth]

\preprint{JLAB-THY-23-3836}

\title{\boldmath 
Full treatment of the thrust distribution in single~inclusive $\epm \to h \, X$ processes}

\author{M. Boglione$^{1,2}$}
\affiliation{${}^1$ Dipartimento di Fisica, Universit\`a di Torino, Via P.~Giuria 1, I-10125 Torino, Italy}
\affiliation{${}^2$ INFN, Sezione di Torino, Via P.~Giuria 1, I-10125 Torino, Italy}                

\author{and A. Simonelli$^{3,4}$}

\affiliation{${}^3$ Department of Physics, Old Dominion University, Norfolk, VA 23529, USA}
\affiliation{${}^4$
Theory Center, Jefferson Lab, Newport News, Virginia 23606, USA}

\emailAdd{boglione@to.infn.it}
\emailAdd{andsim@jlab.org}

 \abstract{Extending the transverse momentum dependent  factorization to thrust dependent observables entails a series of difficulties, ultimately associated to the behavior of soft radiation. As a consequence, the definition of the transverse momentum dependent functions has to be revised, while preserving (and possibly extending) their universality properties. Moreover, the regularization of the rapidity divergences generates non trivial correlations between rapidity and thrust. In this paper, we show how to deal with these correlations in a consistent treatment of the thrust dependence of $\epm \to h\,X$ cross section, where the hadron transverse momentum is measured with respect to the thrust axis. In this framework all results obtained in the past few years properly fit together, leading to a remarkable phenomenological description of the experimental measurements. 
}

\begin{document} 

\maketitle

\section{Introduction\label{sec:intro}}

The thrust distribution associated to single-inclusive hadron production from $\epm$ annihilations (\sia), sensitive to the transverse momentum of the detected hadron with respect to the thrust axis, is one of the most challenging processes where transverse momentum dependent (TMD) factorization needs to be extended beyond the standard scheme in which it was originally formulated~\cite{Collins:1984kg,Collins:1989bt,Aybat:2011zv}. 
Interest has grown since the beginning of 2019, when the BELLE collaboration released the measurements~\cite{Seidl:2019jei} of the $\epm \to hX$ cross section. 
The factorization properties of this process have been studied using two main schemes, one based on Soft Collinear Effective Field Theory (SCET)~\cite{Kang:2020yqw,Makris:2020ltr}, and one based on the Collins Soper Sterman (CSS) formalism~\cite{Boglione:2020cwn,Boglione:2020auc,Boglione:2021wov}.
Within the nearly $2$-jet configuration, all the above studies agree in the classification of the underlying kinematic structure in terms of three different kinematic regions, each corresponding to a different factorization theorem. 
Adopting the nomenclature from Ref.~\cite{Makris:2020ltr} and according to the most recent work of Ref.~\cite{Boglione:2021wov}, this  classification can be traced back to the distance of the outgoing hadron from its jet axis.
\begin{figure}
\centering
\includegraphics[width=10cm]{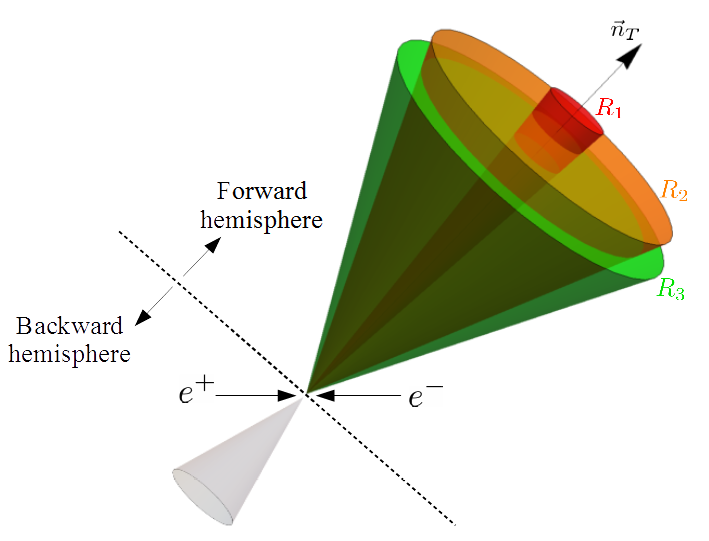}
\caption{Pictorial representation of the \sia. The three independent kinematic regions are explicitly represented in the jet cone where the hadron is detected. The central region ($R_2$) covers the widest portion of the phase space and it is located in between $R_1$ and $R_3$, neither very close to the thrust axis $\vec{n_T}$ ($R_1$), nor very close to the jet boundaries  ($R_3$).} 
\label{fig:cone}
\end{figure}
Region 1 ($R_1$) and Region 3 ($R_3$) are located at the boundaries of the available phase space, either extremely close to the thrust axis ($R_1$) or in the outer region of the jet cone, far from the jet axis ($R_3$).  
Region 2 ($R_2$), located in the intermediate range between $R_1$ and $R_3$, covers the widest portion of the phase space and 
is of primary importance to understand the experimental data correctly. 
Due to its location inside the jet cone, we will also refer to Region 2 as the central region.

While SCET and CSS schemes find the same results when examining the factorization properties of $R_1$ and $R_3$, some tension arises in the case of $R_2$. We will show that these discrepancies originate from genuine non-perturbative effects, while  perturbative QCD alone would lead to the same result. We will conclude that  
the two factorization theorems refer to two different kinematic regions, even if they were originally devised to describe the same configuration. In particular, adopting the same classification used in Ref.~\cite{Boglione:2021wov}, the SCET result is associated to a kinematic region located ``halfway" between $R_2$ and $R_3$, where one would expect to have to apply some matching prescription. This is totally unexpected as, usually,  matching regions do not allow for properly defined factorization theorems.

According to Refs.~\cite{Boglione:2020auc, Boglione:2020cwn,Boglione:2021wov} the extended factorization theorem devised in the CSS approach for the study of $R_2$ shows some peculiarities, as it  \emph{requires} the rapidity regulator (introduced to prevent the rapidity divergences associated to TMD contributions) to be a function of thrust and transverse momentum, thereby introducing non trivial correlations among rapidity, transverse momentum and thrust\footnote{Note that thrust and  hadronic  transverse momentum are measured quantities.}.
Such relation is by far the most controversial and debated feature of the factorization theorem valid in Region 2, as obtained in the CSS-based framework. Incidentally, one could note that  the SCET treatment  does not involve any issue regarding the rapidity regulators. 
A first attempt to explore and clarify the correlation among  rapidity regulators and thrust, although rather naive, was already presented in Ref.~\cite{Boglione:2020auc} and investigated phenomenologically in Ref.~\cite{Boglione:2022nzq}. In this paper, the equation encoding such a correlation will be formally devised without any approximation, and it will appear as a natural consequence of the kinematics underlying Region 2. 

There is another unique aspect regarding the factorization theorem valid in Region 2, addressed in both SCET and CSS formalisms and widely discussed in Ref.~\cite{Boglione:2020cwn}, concerning the definition of the TMD Fragmentation Function (FF) appearing in the final cross section, which does not coincide with the usual definition adopted in standard TMD factorization~\cite{Collins:2011zzd,Aybat:2011zv}. As a matter of fact, the usual TMDs include, by definition, the non-perturbative effects introduced by soft gluon radiation. In Region 2, the same definition cannot be used, simply  because such effects are not there. 
Notably, this does not undermine the universality of TMDs, since the two definitions can be unambiguously related~\cite{Boglione:2020auc}. In fact, these effects can properly be described by a novel  non-perturbative function, $M_S(b_T)$, called the "soft model"~\cite{Boglione:2020cwn}, which can be considered as the soft counterpart of the non-perturbative core of the TMDs. Similarly, $M_S$ cannot be computed, but should be extracted from experimental data. Then, the standard definition is recovered by multiplying the TMD FF associated with $R_2$ by the square root of the soft model, see Appendix~\ref{app:tmdff}.
Troubles in TMD universality like those described above are not specific of \sia, they are rather side effects of extending TMD factorization beyond the so-called benchmark processes (semi-inclusive deep inelastic scattering (SIDIS), Drell Yan (DY) processes), in which only two hadrons appear, either in the initial or final states.

As discussed in Ref.~\cite{Boglione:2020auc}, it might appear that any extension to processes including a different number of hadrons requires the introduction of a new non-perturbative function, in order to perform global phenomenological analyses. 
Clearly, as long as the total number of non-perturbative unknowns is reasonably low, the phenomenological treatment is feasible and the idea of universality is preserved, even if slightly weakened. 
On the other hand, introducing new non-perturbative functions provides novel  perspectives on the inner non-perturbative core of QCD. The case of \sia \, is the first remarkable example of this enrichment: the extraction of the unpolarized TMD FF from $R_2$ data can be used as an input for phenomenological analyses on double-inclusive $\epm$ annihilation (DIA) data (which belongs to the benchmark processes) in order to access the soft model. In turn, achieving a more detailed knowledge on the soft model provides a useful insight about the non-perturbative properties of the QCD vacuum, which would not be accessible within the sole standard TMDs.

\bigskip

Given its many peculiarities and the new perspectives on TMD phenomenology that it offers, the \sia \, process is definitely worth investigating. This paper is devoted to address and solve those theoretical issues that have inspired long debates in the last couple of years.  
More formally, we consider the cross section of $\epm \to h\,X$, as a function of the  fractional energy $z$, the transverse momentum $P_T$ and the thrust $T$, where the reference jet axis is given by the thrust axis~\cite{Seidl:2019jei}. For simplicity, we only consider the production of a spinless hadron, but the generalization to the polarized case is straightforward, as shown in Ref.~\cite{Boglione:2020cwn}. 
The cross section is conveniently written by decoupling the initial state, encoded into the leptonic tensor $L_{\mu\,\nu}$, from the final state, described by the hadronic tensor $W^{\mu\,\nu}$. Projecting the latter onto its relevant Lorentz structures leads to the definition of two Lorentz-invariant structure functions, $F_1$ and $F_2$. In the $2$-jet limit, these functions are not independent, as they are related through $F_2 = -\frac{2}{z} \, F_1$ apart from power suppressed terms. Therefore, in this limit the cross section is simply written as:
\begin{align}
    \label{eq:xs_F1}
    \frac{d \sigma}{d z \, d^2 \vec{P}_T \, d T} = \sigma_B \, z \, F_1 +\;
    \parbox{1cm}{\centering p.s. terms},
\end{align}
where $\sigma_B = {4\pi \alpha^2}/{3 Q^2}$ is the Born cross section. A much more detailed presentation of the above results can be found in Ref.~\cite{Boglione:2021wov}.

\bigskip

This paper is organized as follows. In Section~\ref{sec:id_R2} we show how to properly define and identify the central kinematic region. In Section~\ref{sec:T_rap} we present the relation between rapidity regulator, thrust and transverse momentum that characterizes the factorization theorem of Region 2, which is the subject of Section~\ref{sec:ft_R2}. The relation with the SCET result, together with the pecularity of the matching with Region 3 is addressed in Section~\ref{sec:match}. Finally, we present a phenomenological study, based on this new formalism, in Section~\ref{sec:pheno}. 
The Appendices regard the notation used~\eqref{app:notation}, a review on the definition of the unpolarized TMD FF~\eqref{app:tmdff} and a detailed discussion about the functions appearing in the factorized cross section in the central kinematic region~\eqref{app:gst},~\eqref{app:scthr-fact},~\eqref{app:thr-fun}.

\bigskip

\section{Identifying the central region \label{sec:id_R2}}

The derivation of the TMD factorization theorem for the $\epm \to h\,X$ thrust distribution is highly non-trivial, especially in the kinematics of Region 2, where the detected hadron is neither very close to the thrust axis nor to the outer boundaries of the jet.
At the same time, this thrust distribution  is one of the most interesting observables, as it shows properties that make it particularly relevant for TMD physics.
In this paper, we will not focus on the details of factorization itself and on  how the various contributions are separated out, as these aspects have been thoroughly  investigated in Ref.~\cite{Boglione:2021wov}, where a formal proof was obtained in the Collins factorization formalism of Ref.~\cite{Collins:2011zzd}.  
Rather, in this Section we will show that, once all terms contributing to the final cross section have been identified and properly separated, the theorem devised in Ref.~\cite{Boglione:2021wov} for Region 2 is actually the only viable solution. 
Let us therefore inspect the various types  of radiation contributing to the final state of the process. 
\begin{itemize}
 \item
{\textbf{Collinear forward emissions}}.
 These include gluons radiated collinearly along the thrust axis, in the same hemisphere of the detected hadron. Their contribution plays a pivotal role in producing TMD effects. Consequently, the dependence on transverse momentum in the corresponding term becomes crucial and is conveniently accounted for by considering its Fourier conjugate, $b_T$, and working in the impact parameter space. 
 Collinear radiation in the forward hemisphere is described by a function depending on both transverse momentum and thrust, which one might be tempted to identify  with a (Fourier transformed) generalized fragmenting jet function (FJF), $\mathcal{G}_{h/j}$. This, however, would not be  completely correct, as the hadron is detected in Region 2, sufficiently far from the boundaries of the jet, and its  contribution to the jet spreading is negligible. In other words, its transverse momentum is not large enough to significantly affect the thrust dependence of the cross section, which is instead a characteristic of generalized FJFs. The correct way to include forward collinear radiation in Region 2 is then to consider only the low transverse momentum approximation of a generalized FJF or, equivalently, its large-$b_T$ asymptotic  counterpart in the impact parameter space. This function will be denoted by $\mathcal{G}_{h/j}^{\text{asy}}$.
Notice that the whole $z$-dependence is encoded in this term.
\item
{\textbf{Soft gluon emissions}}.
 The classification of soft radiation is less straightforward. Indeed, the soft gluons that flows in the hemisphere opposite to $h$ must be treated likewise backward collinear radiation. On the contrary, soft gluons that flows in the same hemisphere of $h$ could, in principle, produce some TMD effects. However, here we are considering the hadron detected in the central region of the jet, $R_2$, far enough from the thrust axis so that its transverse momentum is sufficiently large not to be significantly affected by soft gluon  radiation\footnote{This implies that in Region 2 the thrust axis is not affected by soft recoiling: this is  relevant when developing a jet algorithm suitable for this kinematic configuration.}. As a consequence, in Region 2, the contribution of soft gluon radiation is integrated over the transverse momentum components. The corresponding function is the generalized soft thrust function $\mathscr{S}$,  reviewed in Appendix~\ref{app:gst}. 
\item 
 {\textbf{Collinear backward emissions}}. They include the contributions of the gluons emitted collinearly, but opposite, to the thrust axis. These gluons flow in the hemisphere opposite to the detected hadron, hence they cannot affect its transverse momentum. Their contribution is integrated over the transverse momentum components and it is embedded in the jet thrust function $J$.
    \item 
{\textbf{Virtual gluon emissions}}.  These are short-distance contributions that dress the vertex $\gamma\,q\,\overline{q}$, and are encoded into the hard function $H$.
\end{itemize}
The various types of radiation described above do not resolve completely the kinematics of Region 2. In fact, the hallmark of a kinematic region is fixed only once the overlapping among the various momentum regions is properly taken into account and any eventual double counting is canceled. 
Subtle issues are in general hidden in the overlapping between soft and collinear momentum regions. The corresponding double counting is dealt with  by subtracting out the soft-collinear radiation contributions (i.e. gluons with large rapidity but low energy) from the collinear part of the cross section. In the case of \sia \, there are two overlappings of this kind, one in each of the hemispheres. 
Clearly, soft-collinear backward gluons cannot affect the transverse momentum of the detected hadron, hence their contribution depends only on thrust. It corresponds to the (left) soft-collinear thrust function $\mathscr{Y}_L$, reviewed in Appendix~\ref{subsec:scthr-fun}. 
On the other hand, %at first sight 
the role of forward emitted soft-collinear gluons might seem 
ambiguous, being them hybrids between soft and collinear radiations: they may either produce relevant TMD effects, like collinear gluons, or be totally irrelevant for the transverse momentum of the detected hadron, like soft gluons. 
Crucially, different ways of dealing with this ambiguity lead to different factorization theorems. 

This implies that the underlying kinematics of Region 2 must be associated to only one of the two possible interpretations of  soft-collinear radiation in the forward hemisphere and, consequently, that the other one refers to a \emph{different} kinematic region.
In particular, we anticipate that opting for the TMD-relevance of forward soft-collinear radiation leads to the factorization theorem devised in Ref.~\cite{Boglione:2021wov}, while the result presented in Ref.~\cite{Makris:2020ltr} implies the TMD-irrelevance of such contributions.
It is important to stress that these are not different results referring to the same physics, but they really describe different kinematic configurations. 
At this stage, the question is determining which choice can be regarded as truly describing Region 2.

Clearly the identification of Region 2 essentially depends on the correct identification of the other two kinematic regions, the factorization theorems of which are well-established at present. Framing the kinematic of the central region of the phase space is indeed crucial to assign the correct role to forward soft-collinear radiation. 
In Region 1, all the contributions to the forward hemisphere are relevant for TMD effects, while in Region 3 this only applies for collinear radiation. This suggests that the only way to define a truly independent Region 2 is to consider as TMD-relevant the corresponding forward soft-collinear contribution. The other possibility would make the kinematic structure of Region 2 almost indistinguishable from that of Region 3, undermining its fundamental role of ``central'' region of the phase space.
This is represented in the following scheme of forward radiation:
\newline
\begin{center}
\begin{tabular}{c || c || c || c ||}
  {}   &  soft  & soft-collinear & collinear \\
  \hline\hline
  $R_1$   &   
  \cellcolor{blue!15}TMD-relevant &  \cellcolor{blue!15}TMD-relevant & \cellcolor{blue!15}TMD-relevant	\\
  \hline\hline
  $R_2$   &   TMD-irrelevant   & \cellcolor{blue!15}TMD-relevant & \cellcolor{blue!15}TMD-relevant	\\
  \hline\hline
  $R_3$   &   TMD-irrelevant   &  TMD-irrelevant & \cellcolor{blue!15}TMD-relevant	\\
  \hline
\end{tabular}
\end{center}

\vspace{0.3cm}

Each of the three kinematic regions shows its distinctive kinematic structure. This is their finger-print, which makes them unique and independent. Notice how the scheme above presents a  totally ``symmetric" structure among the three regions.
The soft-collinear radiation in $R_2$ is described by the large-$b_T$ asymptotic behavior of the (right) soft-collinear thrust factor, introduced in Ref.~\cite{Boglione:2021wov}. Such function can be regarded as the soft approximation of the collinear radiation term, embodied into the low-$P_T$ (or equivalently large-$b_T$) generalized FJF. Remarkably, it also coincides with the Thrust-TMD collinear soft function of Ref.~\cite{Makris:2020ltr}. In this paper, we will simply refer to it as the soft-collinear thrust factor and denote it as $\mathcal{C}_R$.  It is reviewed in Appendix~\ref{app:scthr-fact}.

We are now able to write a first (schematic) version of the factorization theorem of Region 2. Here, the cross section factorizes as:
\begin{align}
\label{eq:FactT_R2_schematic}
& d\sigma_{R_2} \sim |H|^2 \, J \, \frac{\mathcal{S}}{\mathcal{Y}_L} \, \frac{\mathcal{G}^{\text{\tiny asy}}_{h/j}}{\mathcal{C}_R},
\end{align}
where the formula holds in the $b_T$ space (Fourier conjugate of the transverse momentum $k_T$) and in the $u$ space (Laplace conjugate of the thrust, $T$). In Eq.~\eqref{eq:FactT_R2_schematic} the explicit dependence on these variables has been omitted for clarity.
\begin{figure}
\centering
\includegraphics[width=10cm]{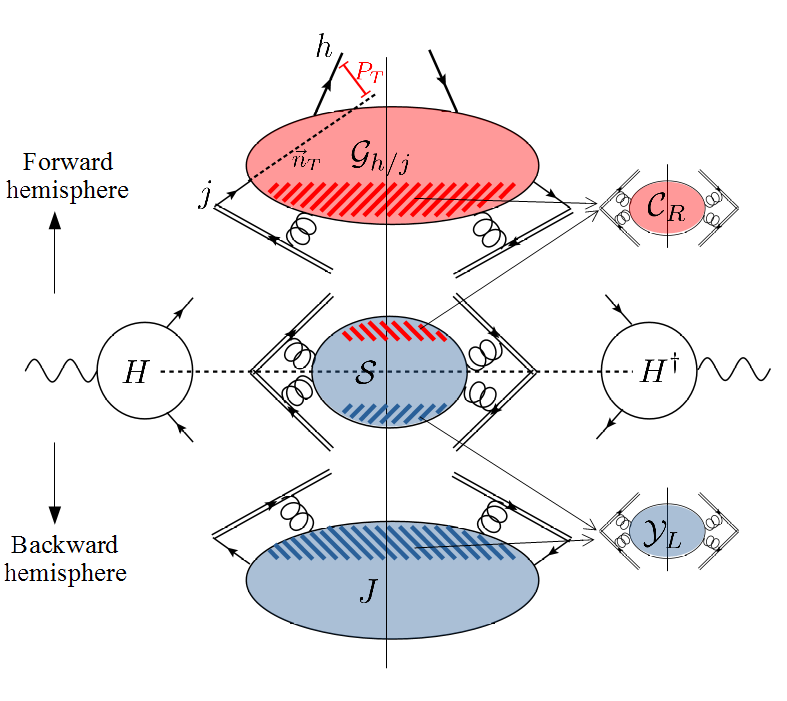}
\caption{Factorized momentum regions associated with the central kinematic region $R_2$. Each blob corresponds to one of the terms appearing in the cross section in Eq.~\eqref{eq:FactT_R2_schematic}. Red blobs correspond to TMD-relevant contributions, while blue blobs are TMD-irrelevant. The overlapping between soft and collinear momentum regions is represented by the hatched areas. In the pictorial representation of the contribution of the $\mathcal{G}_{h/j}$ function we also indicate the thrust axis, $\vec{n}_T$, and the momentum $P_T$ of the detected hadron transverse to $\vec{n}_T$.
Double fermionic lines represent Wilson lines.} 
\label{fig:reg_dec}
\end{figure}
The momentum regions associated to each of the terms appearing in Eq.~\eqref{eq:FactT_R2_schematic} are represented in Fig.~\ref{fig:reg_dec}.

\bigskip

\section{Thrust and rapidity regulators \label{sec:T_rap}}

Despite its appealing appearance, the forward-radiation scheme devised for defining Region 2 hides a potentially lethal pitfall regarding the factorization within this region. More specifically, troubles arise when dealing with the regularization of rapidity divergences. 
In \sia, part of the rapidity divergences are naturally prevented by the thrust dependence. This is a general feature of thrust observables. 
The $2$-jet limit $\tau \to 0$ (where as usual $\tau = 1-T$) corresponds to removing the regulator and exposing the rapidity divergences in fixed order calculations. Nevertheless,  this limit corresponds to a physically meaningful result: resummation shows that thrust-dependent cross sections vanish at $\tau=0$, ensuring the ``independence" of the rapidity regulator. 
Thrust dependence is rarely considered in these terms. In fact, being a \emph{measurable} quantity,  it is usually treated as an external variable and not just a mere mathematical tool like a regulator. 
On the other hand, each term contributing to TMD observables is affected by the presence of rapidity divergences, which cancel out only when the full cross section is assembled. In order to make each term separately finite, many different regularization procedures have been proposed in the past  years~\cite{Collins:2011zzd,Chiu:2011qc,Chiu:2012ir,Echevarria:2011epo,Li:2016axz,Li:2016ctv,Echevarria:2016scs,Vladimirov:2020umg}.
Any of these procedures introduces rapidity regulators that artificially prevent the generation of rapidity divergences.
Unlike thrust, these regulators really are artificial mathematical tools, hence the final result should not depend on them. More specifically, the final result should coincide with the limit in which such regulators are removed, a condition which is clearly satisfied when they disappear in the final cross-section.
This is indeed the most common situation and one of the characterizing features of standard TMD factorization theorems.
In \sia, these regulators coexist with the thrust dependence. 
Therefore, it should not be surprising that the two mechanisms are intertwined generating, in turn, a correlation between thrust and rapidity regulators.

\subsection{Intertwining of rapidity regularizations \label{subsec:intw_rap}}

The effects of this interconnection permeate the whole kinematics of \sia, but they become evident in Region 2. This must be traced back to the different role played by soft and soft-collinear radiation in the forward hemisphere. In most cases, this two contributions are considered on an  equal footing as far as TMD effects are concerned:  they are either integrated over transverse momentum, as in Region 3, or Fourier transformed and considered in $b_T$-space, as in Region 1 and standard TMD factorization.
Since the regularization of rapidity divergences is particularly relevant in soft and soft-collinear sectors, and because it is heavily affected by how TMD effects are taken into account, whenever these two types of radiation are considered differently, the question about the cancellation of rapidity divergences rises urgently. For this reason, Region 2 is very peculiar and, at present, unique.

In the following, we will show some examples of the intertwining between thrust and rapidity regulators in \sia. 
For this reason, we have to explicitly select one of the various methods proposed to regularize TMD rapidity divergences.
Among them,  one of the cleanest is the Collins' regularization prescription.  It consists in tilting the Wilson lines associated with soft approximations off the light-cone. This  regularization is controlled by two parameters $y_1$ and $y_2$, one for each hemisphere (or relevant directions in standard TMD factorization). They act like cut-offs in the integration over  rapidity, preventing the corresponding terms from being divergent.  
The Wilson lines are restored to their original light-cone directions in the limits $y_1 \to +\infty$ and $y_2 \to -\infty$.
We will adopt this regularization scheme, in the footsteps of our previous  work~\cite{Boglione:2020cwn,Boglione:2020auc,Boglione:2021wov}. 

Let's first consider the soft and soft-collinear sectors of Region 3.
These consist in the generalized soft thrust function $\mathscr{S}$, properly subtracted, where the subtraction terms are the  left and right soft-collinear thrust functions $\mathscr{Y}_{L,R}$. Each of these functions is integrated over transverse momentum and all of them are rapidity regulated according to the Collins' prescription. Naively, one may expect that the final combination still depends on thrust as well as on the rapidity regulators. Instead, the whole dependence on the rapidity cut-offs is washed away in the subtraction mechanism and the sole thrust is enough to take into account all the rapidity divergences. This is elegantly expressed by the following theorem~\cite{Boglione:2021wov}:
\begin{align}
    \label{eq:soft_mix}
    &S\left(\tau\right) = \frac{\mathscr{S}\left( \tau ,y_1, y_2\right)}{\mathscr{Y}_L\left( \tau , y_2\right)\,\mathscr{Y}_R\left( \tau , y_1\right)},
\end{align}
where the l.h.s of the previous equation is the usual soft thrust function appearing in the thrust-distributions of $\epm$ annihilation. The previous equation just reflects the fact that any information regarding TMD effects vanishes  in integrated quantities, provided that they are correctly defined (i.e. properly subtracted).
Notice that the intertwining between thrust and $y_2$ is always carried out as in Eq.~\eqref{eq:soft_mix}, as all the regions share the same structure in the backward hemisphere.

There are also cases where the intertwining between thrust and rapidity regulators leads exactly to the opposite conclusion: the whole thrust-dependence disappears and only rapidity regulators survive. This clearly applies only for combinations in the forward hemisphere, hence regarding the intertwining between thrust and $y_1$.
Remarkably, this feature allows to define a TMD Fragmentation Function in Region 1 and Region 2. In fact, consider the (forward) soft-collinear and collinear sectors of these two regions. Collinear radiation is captured by the low-$P_T$ generalized FJF $\mathcal{G}_{h/j}^{\text{asy}}$. This must be properly subtracted with the soft-collinear term, embodied by the (right) soft-collinear thrust factor $\mathcal{C}_R$ which brings in the dependence on $y_1$. Both these functions depend on thrust, but such dependence does not survive the subtraction mechanism and the final result depends only on $y_1$. Most importantly, it coincides with a TMD FF, as expressed by the following theorem~\cite{Boglione:2021wov}:
\begin{align}
    \label{eq:coll_mix}
    &\widetilde{D}_{h/j}\left(z,  b_T, y_1\right) = \cfrac{\widetilde{\mathcal{G}}_{h/j}^{\text{asy}}\left(z,  b_T, \tau\right)}{\widetilde{\mathcal{C}}_R\left(  b_T, \tau, y_1 \right)},
\end{align}
where the tilde denotes Fourier transformed quantities. Remarkably, this TMD FF is not defined as usual, as it only involves collinear and soft-collinear sectors, leaving out any soft physics effect. Hence, it can be re-casted as in Eq.~\eqref{eq:fact_def}.
This does not undermines its universality properties, as reviewed in Ref.~\cite{Boglione:2020cwn} and discussed in Appendix~\ref{app:tmdff}.
Collins-Soper (CS) evolution defines how the TMD FF above depends on the rapidity cut-off $y_1$. The other term of the cross-section carrying a dependence on $y_1$ is the function accounting for soft radiation. Depending on whether this is relevant for TMD effects, the overall $y_1$-dependence in the final cross section has a difference fate. 

If, as in Region 1, soft gluons are TMD-relevant, then the soft function evolves, with respect to $y_1$, exactly as the TMD FF in Eq.~\eqref{eq:coll_mix} but in the opposite direction. Therefore, all the terms depending on $y_1$ cancel among themselves in the final result and the limit in which the rapidity cut-off is removed is trivial, just as in standard TMD factorization. 

On the other hand, in Region 2 the soft gluons are not relevant for TMD effects. The soft function cannot evolve as a TMD FF, if only because it is integrated over transverse momentum. 
Therefore, in this case the combination of soft and collinear (subtracted) radiation leaves out a dependence on $y_1$, which must be considered as being large and positive, ultimately infinite. 
But $y_1$ is not the only rapidity regulator that survives in the final combination, as the generalized soft thrust function keeps track of the thrust $\tau$ also in the forward hemisphere. Hence, $\tau$ and $y_1$ are both present in the final cross section and both regularize the rapidity divergences. This suggests that there is a \emph{redundancy} of regulators, signaling that one can be expressed in terms of the other. In particular, the rapidity cut-off $y_1$ should be a function of thrust, such that when it is removed, also $\tau$ is removed. In other words, the limit $y_1 \to \infty$ should coincide with the limit $\tau \to 0$, or, more specifically,  $u \to \infty$ in the Laplace conjugate space of thrust\footnote{\label{ftnot:kin}
Another clue regarding the mutual dependence between thrust and rapidity cut-off comes from a simple kinematic argument~\cite{Boglione:2020auc}. In fact, since~\cite{Makris:2020ltr} $P_T \leq z \, \sqrt{\tau} \, Q$, then the rapidity of the detected hadron is naturally bounded from below by $-\log{\sqrt{\tau}}$, suggesting the relation $y_1 \sim -\log{\sqrt{\tau}}$, which satisfies the limits of thrust and rapidity cut-off discussed above. This relation has been tested in a recent phenomenological analysis~\cite{Boglione:2022nzq} providing good agreement with BELLE data, but it lacks of a proper formal justification.}.

\subsection{Rapidity regulator as a function of thrust \label{subsec:cond_y1}}

The intertwining between thrust and rapidity regulator is ultimately traced back to how soft and soft-collinear sectors overlap each other: requesting a perfect cancellation of double countings  inevitably forces $y_1$ to be tuned to the value where the range of soft transverse momentum do actually coincide with the range of soft-collinear transverse momentum. 
In fact, the generalized soft thrust function roughly constraint the total transverse momentum as\footnote{At 1-loop the rapidity of the real gluon is $y = \log{(k_T/{\tau Q})} \leq y_1$, which then constraints the transverse momentum. At higher orders, the total transverse momentum is constrained similarly, as any correction to its maximum size is suppressed in soft-collinear limit (large rapidities, low transverse momenta).} $k_T \lesssim \tau \, Q e^{y_1}$, which translates in $k_T \lesssim {Q\,e^{y_1}}/{u_E}$ in the Laplace conjugate space. Here $u_E = u e^{\gamma_E}$, with $\gamma_E$ being the Euler-Mascheroni constant. On the other hand, the soft-collinear factor at the denominator of Eq.~\eqref{eq:fact_def} roughly fixes its maximum size as $k_T \lesssim {c_1}/{b_T}$, with $c_1 = 2 e^{-\gamma_E}$. 
Equating this two upper limits makes the two ranges of transverse momentum to coincide and the subtraction mechanism to work. The equation implies:
\begin{equation}
\label{eq:y1SOL_pert}
\overline{y}_1 = L_u - L_b,
\end{equation}
where we have introduced $L_u = \log{u_E}$ and $L_b = \log{({b_T \,Q}/{c_1})}$. Thus, $\overline{y}_1$ correctly diverges as $u$ becomes large.
Notice that this estimate regards the limits of integration over transverse momentum encountered in perturbative calculations and hence this argument is blind to non-perturbative effects. This is manifest in the $b_T$ dependence of the rapidity cut-off, which becomes negative, and even divergent, at large distances where non-perturbative effects come into play. 
Later on, we will extend this solution also to the non-perturbative regime. 

This value of the rapidity cut-off is very special. Not only it allows for a proper separation of the terms in the final result, but it is also a critical point of the factorized cross section when regarded as a function of $y_1$. In particular, it is its only \emph{minimum}, being the factorized cross section a convex function of the rapidity cut-off. 
We first prove the property of convexity. Given that the first derivative of the (logarithm of the) cross section with respect to $y_1$ is the difference of the evolution kernels of the  generalized soft thrust function and the TMD FF, i.e. $\widehat{G}_R \left(u,y_1\right) - \widetilde{K}(b_T)$, the sign of the second derivative corresponds to the slope of the right G-kernel $\widehat{G}_R$ with respect to $y_1$. Following Appendix~\ref{subsec:gst_evo}, the right G-kernel and the Collins-Soper kernel $\widetilde{K}$ are basically the same function, although evaluated at different argument. In particular, the former is obtained by replacing $b_T$ with some (unknown) positive-definite monotonically increasing function of $c_1 \, {u_E}/{Q} \, e^{-y_1}$, which coincides with identity at perturbative level. 
The slope of $\widetilde{G}_R$ with respect to $y_1$ is thus opposite to the slope of $\widetilde{K}$ with respect to $b_T$, and this is known to be negative. Confirmations are given both at perturbative level (the coefficients of perturbative expansion are known up to four loops~\cite{Collins:2017oxh,Moch:2018wjh,Moch:2017uml,Henn:2019swt}, and also in the non-perturbative regime, where both phenomenological extractions ~\cite{Konychev:2005iy,Scimemi:2019cmh,Bacchetta:2019sam,Cerutti:2022lmb,Bacchetta:2022awv,Bury:2022czx,Barry:2023qqh,Boglione:2022nzq,DAlesio:2020wjq,DAlesio:2022brl} and lattice computations ~\cite{Ebert:2018gzl,Ebert:2019tvc,Schlemmer:2021aij,LatticeParton:2020uhz,Shanahan:2021tst} agree on considering $g_K$ as a positive-definite monotonically increasing function of $b_T$.
Therefore, $\widehat{G}_R$ must have a positive slope with respect to $y_1$. This concludes the proof that the factorized cross section is a convex function of the rapidity cut-off. The minimum is found by imposing the vanishing of the first derivative of the factorized cross section with respect to $y_1$, which is equivalent to find the value of the rapidity cut-off that makes the cross section CS-invariant.
Neglecting non-perturbative effects, this corresponds to the following condition:
\begin{align}
\label{eq:CS_cond_pert}
\widehat{G}_R^{(\text{pert.})} \left(a_S(\mu), L_R \right) = \widetilde{K}^{(\text{pert.})}(a_S(\mu), L_b) 
%\quad \leftrightarrow \quad  
%L_R = L_b
%\begin{aligned}
%&L_R = L_b \\
%\text{i.e. } &\mu_R = \mu_b
%\end{aligned}
%\stackrel{L_R = L_b}{\text{i.e. } \mu_R = \mu_b} 
\quad \leftrightarrow \quad  
\overline{y}^{(\text{pert.})}_1 = L_u - L_b,
\end{align}
where we used the fact that, at perturbative level, the right G-kernel is obtained from the CS-kernel by replacing $\mu_b$ with $\mu_R = {Q e^{y_1}}/{u_E}$. 

\bigskip

As mentioned above, the solution $\overline{y}^{(\text{pert.})}_1$ coincides exactly with the condition of Eq.~\eqref{eq:y1SOL_pert}, which was obtained by requiring that the subtraction of soft-collinear radiation perfectly cancels the overlapping terms. The limitation of an approach based solely on perturbative QCD is noticeable by the negative divergent behavior of the rapidity cut-off at large distances, indicating a break down of the factorization itself. Consider however that any factorization theorem obtained for Region 2 could never be extended at very large values of $b_T$, as this operation would be roughly equivalent to consider it at extremely low values of $P_T$, where another kinematics (Region 1) is supposed to describe the underlying physics of \sia. 
Nevertheless, we anticipate that the inclusion of non-perturbative effects will have a positive impact on the estimate of the rapidity cut-off, healing its unwelcomed negative divergent behavior and replacing it with a constant. This will set a \emph{finite} lower limit for $u$ (and hence a minimum value of thrust) below which the factorization cannot be trust anymore. The larger $b_T$ is, the more this limit will constrain the validity of the factorization theorem at low $P_T$. Ultimately, the very low transverse momentum behavior will be described in terms of Region 2 only if the thrust is large enough. This agrees with physical intuition: close to the exact $2$-jet limit the whole phase space shrinks and its boundaries become narrower and narrower.

In the following, we will show how to get an estimate of the rapidity cut-off that takes into account the most possible amount of information regarding non-perturbative effects and, in particular, that properly considers the impact of large distances. 
This is achieved by solving the condition:
\begin{align}
\label{eq:CS_cond}
\widehat{G}_R\left(a_S(\mu), L_R; \,u_E \, e^{-y_1}
% f\left(c_1 \, \frac{u_E}{Q} e^{-y_1}\right)
\right) = \widetilde{K}(a_S(\mu), L_b;\,b_T) 
\end{align}
outside its strict perturbative domain. This is emphasized by the arguments following the semicolon in the functions appearing in the equation above, which are intended to represent any dependence outside the logarithms get by perturbative computations. Extending the conclusions obtained in perturbation theory, the solution of Eq.~\eqref{eq:CS_cond} will be that value $\overline{y}_1$ of the rapidity cut-off that on one side allows for a perfect cancellation of the double counting due to the soft-collinear overlap, and on the other coincides with the minimum of the factorized cross section when regarded as a function of the rapidity regulator.
The first step consists in identify and isolate the non-perturbative effects associated to large distances from the perturbative-dominated contributions in $\widetilde{K}$. Following one of the most common procedures of standard TMD factorization, we adopt the $b^\star$-prescription reviewed in Appendix~\ref{app:tmdff}. Next, we exploit the RG-invariance of Eq.~\eqref{eq:CS_cond} to evaluate both sides to a convenient scale. Choosing $\mu = \mu_b^\star$ we get a double advantage. In fact, not only $\mu_b^\star$ can always be considered safely large enough to avoid the Landau pole (by construction) but, since it is the reference scale for the evolution of the CS-kernel, it also allows to simplify the r.h.s of Eq.~\eqref{eq:CS_cond} as much as possible. In particular, it becomes $ \widetilde{K}|_{\mu_b^\star} = \widetilde{K}_\star(a_S(\mu_b^\star)) - g_K(b_T)$, where $\widetilde{K}_\star$ and $g_K$ account for the perturbative and the non-perturbative effects of the CS-kernel, respectively. 
The whole evolution mechanism has been relegated to the l.h.s., where the scale changes from $\mu_R$ to $\mu_b^\star$ in a perturbative-controlled way, being totally ruled by the anomalous dimension $\gamma_K$. But the l.h.s also includes non-perturbative contributions, which should be organized in a $g_K$ function as in $\widetilde{K}$, in this case dependent on some combination of $c_1{u_E}/{Q} e^{-y_1}$ that replaces $b_T$. Unfortunately, the exact functional form of the argument of $g_K$, and even of $g_K$ itself, is still unknown at present days. This prevents Eq.~\eqref{eq:CS_cond} from being inverted exactly. Therefore, we have to introduce some sensible approximation.  First, notice that since $g_K$ behaves\footnote{This behavior is strongly suggested by theory~\cite{Collins:2014jpa} and largely confirmed by most recent phenomenological extractions~\cite{Cerutti:2022lmb,Bacchetta:2022awv,Bury:2022czx,Barry:2023qqh,Boglione:2022nzq,DAlesio:2022brl}.}  as $\sim b_T^2$ at low-$b_T$ , then the first non-perturbative correction to $\widehat{G}_R$ must be at least of order $\mathcal{O}( e^{-2\,y_1})$. This is the same size of the terms neglected in the factorization theorem because of the largeness of the rapidity cut-off. Therefore, we could neglect all non-perturbative contributions to $\widehat{G}_R$ by simply advocating the founding requirement $y_1 \gtrsim 0$, which must be true also for the solution $\overline{y}_1$. However, we prefer to impose a more stringent constraining assumption, requiring that the scale $\mu_R$ can be considered large enough to be not affected by the Landau pole and hence suitable for perturbative expansions. 
At perturbative level, this is consistent with the smallness of $b_T$, as it can be easily checked from the solution obtained in Eq.~\eqref{eq:CS_cond_pert}.
When non-perturbative effects are considered, this request would correspond to have the lowest possible value of $\mu_R$ of the same order of magnitude of $c_1/b_{\text{\tiny MAX}}$. 
Given these assumptions, we can expand the l.h.s of Eq.~\eqref{eq:CS_cond} to the leading power of $y_1$, as this is the same approximation holding for rapidity cut-offs dependent terms in factorization theorems. This step requires to express the strong coupling at scale $\mu_R$ in terms of $a_S(\mu_b^\star)$ and the approximation is valid as long as $2\beta_0 \, a_S(\mu_b^\star) \log{{\mu_b^\star}/{\mu_R}} = 2\beta_0 \,a_S(\mu_b^\star) \, (L_u - L_{b^\star} - y_1) \lesssim 1$. Eq.~\eqref{eq:CS_cond} becomes:
\begin{align}
\label{eq:y1SOL_eq}
&\frac{\gamma_K^{[1]}}{2 \beta_0} \, \log{\left( 
1 - 2 \, \beta_0 \, a_S(\mu_b^\star) \, (L_u - L_{b^\star} - y_1) 
\right)}
+ \mathcal{O}\left( \frac{1}{y_1}\right) +
\parbox{6em}{suppressed\\ non-pert.\\ terms of $\widehat{G}_R$}
=
\widetilde{K}|_{\mu_b^\star},
%\substack{\text{suppressed} \\ \text{non-pert. terms}}
\end{align}
which can now be inverted to find the solution $\overline{y}_1$, valid under the assumptions discussed above. We obtain: 
\begin{align}
\label{eq:y1SOL}
&\overline{y}_1 = L_u - L_{b^\star} \, \left( 1 + 
\frac{1 - e^{\frac{2 \beta_0}{\gamma_K^{[1]}}\, (\widetilde{K}_\star(a_S(\mu_b^\star)) - g_K(b_T)) }}{2 \beta_0 \, a_S(\mu_b^\star) L_{b^\star}}
\right),
\end{align}
that generalizes the result of Eq.~\eqref{eq:y1SOL_pert}. In this regard, notice that the low-$b_T$ expansion of the solution above matches with the perturbative estimate $\overline{y}^{(\text{pert.})}_1$ of the rapidity cut-off. In fact, letting $g_K$ to behave as $\sim g_2 b_T^2 + \mathcal{O}(b_T^4)$ at small $b_T$, $\overline{y}_1$ can be expanded as:
\begin{align}
\label{eq:y1SOL_lowbT}
&\overline{y}_1 
%= L_u - L_{b^\star}
%\frac{e^{-\frac{2 \beta_0}{\gamma_K^{[1]}} g_K(b_T) } - 1}{2 \beta_0 a_S(\mu_b^\star)} 
%+
%\frac{e^{-\frac{2 \beta_0}{\gamma_K^{[1]}} g_K(b_T) }}{\gamma_K^{[1]}} \left( \widetilde{K}^{[2]} \, a_S(\mu_b^\star) + \mathcal{O}\left( 
%a_S^2(\mu_b^\star)
%\right)
%\right)
%%%%%%%%%%%
%%%%%%%%%%%
%+ \mathcal{O}\left( 
%a_S(\mu_b^\star); \, \frac{b_T^2}{a_S(\mu_b^\star)}
%\right) 
=
\overline{y}^{(\text{pert.})}_1 + 
 \mathcal{O}\left( 
a_S(\mu_b); \, \frac{b_T^2}{a_S(\mu_b)};\,\frac{b_T^2}{b^2_{\text{\tiny MAX}}}
\right).
\end{align}
On the other hand, notice how $\overline{y }_1$ may become negative at large-$b_T$, but \emph{never} divergent. This shows that the non-perturbative effects associated to large distances have been properly taken into account in solving Eq.~\eqref{eq:CS_cond}.  

The assumptions on which the solution $\overline{y}_1$ relies induce constraints both on the domain of validity of the factorization theorem of Region 2 and also on the large-$b_T$ behavior of the Collins-Soper kernel. In particular, how far the physics described by Region 2 can be extended at low $P_T$ and large values of thrust depends significantly on if $g_K$ has an asymptotic divergent behavior or if instead it saturates to a certain constant $g_0>0$ in the limit $b_T \to \infty$. 
For simplicity, let's define the term into round brackets in Eq.~\eqref{eq:y1SOL} as $\alpha(b_T, b_{\text{\tiny MAX}})$. Then, requiring the rapidity cut-off to be large and positive sets a condition relating $u$ and $b_T$, consequently extended also to their conjugate variables $\tau$ and $q_T = P_T/z$. We have:
\begin{align}
\label{eq:y1SOL_hp1}
\overline{y}_1 > 0 
\quad \Rightarrow \quad 
u_E > \left( \frac{b_T^\star(b_{\text{\tiny MAX}}) \, Q}{c_1} \right)^{\alpha(b_T, b_{\text{\tiny MAX}})} \equiv u_E^{\text{\tiny MIN}}(b_T, b_{\text{\tiny MAX}}).
\end{align}
As long this condition is satisfied, the errors associated with the factorization of Region 2 are under control. 
More specifically, the lower $u_E^{\text{\tiny MIN}}$ is, the more the constraint on the positivity of the rapidity cut-off does not prevent the factorization to be pushed to lower values of the thrust.
At small $b_T$, the inequality above becomes $u_E \gtrsim {b_T\,Q}/{c_1}$, as expected perturbatively. Thus, $u_E^{\text{\tiny MIN}}$ vanishes at short distances, implying that the factorization is not constrained by Eq.~\eqref{eq:y1SOL_hp1} at large transverse momentum, regardless of the value of the thrust. Nevertheless, the applicability of $R_2$ factorization at large transverse momentum must still take into account that the larger $q_T$ becomes, the more the kinematics typical of Region 3 becomes dominant, as well as the more the thrust decreases, the more the contributions associated to higher order topologies have to be considered. Besides these considerations, the positivity of the rapidity cut-off does not provide any hints about where a description of the data in terms of Region 2 cease to be valid at large $q_T$.

The behavior of  $u_E^{\text{\tiny MIN}}$ is more constraining at large distances . Here, the positivity of the rapidity cut-off is not ineffective and it indeed provides valuable information about the description of data at low $q_T$, in terms of the kinematics of the central kinematic region.
This limit strongly depends on the large-$b_T$ behavior of $g_K$:
\begin{align}
\label{eq:uEmin_asy}
u_E^{\text{\tiny MIN}}(b_T, b_{\text{\tiny MAX}}) 
\overset{b_T \to \infty}{\to}
\left(\frac{b_{\text{\tiny MAX}} \, Q}{c_1}\right)^{\alpha^{\text{\tiny asy}}(g_0)},
\quad \text{with }
\alpha^{\text{\tiny asy}}(g_0) = 
 1 + 
\frac{1 - A_1 \, e^{-\frac{2 \beta_0}{\gamma_K^{[1]}}\,g_0}}{A_2},
\end{align}
where $\log{A_1} = {2 \beta_0}/{\gamma_K^{[1]}}\, \left.\widetilde{K}_{\star}\right|_{b_{\text{\tiny MAX}}}$ and $A_2 = 2 \beta_0 \, a_S(c_1/b_{\text{\tiny MAX}}) \log{\left({b_{\text{\tiny MAX}} \, Q}/c_1 \right)}$ are constants depending on $Q$, $b_{\text{\tiny MAX}}$ and the perturbative accuracy.
At $Q \approx 10$ GeV and $b_{\text{\tiny MAX}} \approx 1$ GeV$^{-1}$, their values are $A_1 \approx 1$ and $A_2 \approx 0.8$. 
As a general feature, the value of $u_E^{\text{\tiny MIN}}$ increases as $b_T$ gets larger and larger. Therefore, the size of the errors of the factorization of Region 2 increases as well at low transverse momentum. This is totally consistent with the kinematics of \sia, as the lower is $q_T$, the more the kinematics of Region 1 becomes dominant. In other words, we should not expect to use the factorized cross section obtained in the central kinematic region at very small transverse momentum. Nevertheless, the size of such errors explicitly depends on the behavior of the Collins-Soper kernel at large $b_T$. In fact, since $\alpha^{\text{\tiny asy}}(g_0) < \alpha^{\text{\tiny asy}}(\infty)$, $u_E^{\text{\tiny MIN}}$ tends to a lower limit if $g_K$ is assumed to saturate at large distances instead of diverging. 
If the thrust is sufficiently large, such different limit might be not relevant. However, for moderate-high values of $T$ the description of the data at low $q_T$ might benefit of a model in which $g_K$ does saturate at large $b_T$, as this case would be associated with smaller errors of factorization.
The same suggestion is enforced by the other (two) assumptions that have been imposed to obtain the solution in Eq.~\eqref{eq:y1SOL}. In particular, we have assumed that the scale $\mu_R$ can be considered large enough to avoid the influence of the Landau pole. This condition is satisfied if $\overline{\mu}_R \approx c_1/b_{\text{\tiny MAX}}$ at large $b_T$, which means:
\begin{align}
\label{eq:y1SOL_hp2}
\overline{\mu}_R 
\overset{b_T \to \infty}{\to}
Q^{1-\alpha^{\text{\tiny asy}}(g_0)} \, \left(\frac{c_1}{b_{\text{\tiny MAX}}}\right)^{\alpha^{\text{\tiny asy}}(g_0)}
\approx \frac{c_1}{b_{\text{\tiny MAX}}}
\quad\Rightarrow\quad
\alpha^{\text{\tiny asy}}(g_0) \approx 1
\end{align}
By inspecting the explicit expression for $\alpha^{\text{\tiny asy}}$ in Eq.~\eqref{eq:uEmin_asy}, a finite $g_0$ pushes the constraint above in right direction. 
Moreover, the approximation of $\widehat{G}_R$ at leading power of $y_1$ is based on the requirement:
\begin{align}
\label{eq:y1SOL_hp3}
2\beta_0 \,a_S(\mu_b^\star) \, (L_u - L_{b^\star} - y_1) \lesssim 1
\quad 
\overset{b_T \to \infty}{\Rightarrow} \quad 
e^{-\frac{2 \beta_0}{\gamma_K^{[1]}}\,g_0} 
\gtrsim 0,
\end{align}
which is automatically satisfied for any value of $b_T$ if $g_0$ is finite.

All these considerations suggest that $g_0 < \infty$ improves the accuracy of the factorization of Region 2 at low transverse momentum. 
More generally, they confirm that the behavior of the Collins-Soper kernel at large distances crucially affects the physics of \sia \, at low transverse momentum. 
This is extremely relevant for present days studies on TMD physics. In fact, phenomenological analyses based on standard TMD factorization and in particular on extractions from SIDIS and (especially) Drell-Yan data  seem to be very sensitive to the low/moderate $b_T$ behavior of $g_K$, but almost blind to its large distance asymptotic. The process of \sia, specifically in its central kinematic region, would then play a pivotal role in global phenomenological analyses, offering a totally new tool to investigate the non-perturbative properties of the CS-kernel.
Furthermore, despite there have been theoretical arguments in the past in favor of a constant limit\cite{Collins:2014jpa} of $g_K$, most modern TMD phenomenological analyses opt for a divergent behavior, ranging from sublinear~\cite{Boglione:2022nzq,DAlesio:2020wjq,DAlesio:2022brl} to quartic power~\cite{Bacchetta:2019sam}. In this regard, the considerations above regarding the relation between the asymptote of the CS-kernel and the errors associated with the factorization of \sia \, in the central kinematic region can be considered as a novel insight from the theoretical perspective, in favor of saturation models. We will present in Section~\ref{sec:pheno} a phenomenological analysis that exploits a non-diverging $g_K$, showing how this choice leads to an extremely good agreement with experimental data. 

\bigskip

We conclude this Section by showing how the solution $\overline{y}_1$ obtained for the rapidity cut-off agrees with the kinematic boundaries of \sia \, (see footnote on Pag.~\pageref{ftnot:kin}). This follows by expressing $\overline{y}_1$ in thrust-space. However, a direct inverse Laplace transform would lead to the plus distribution $\left({1}/{\tau}\right)_+$ instead of a function. 
Thus, we express the rapidity cut-off in the Laplace conjugate space in terms of $\overline{\zeta} = Q^2 \, 
e^{-2 \, \overline{y}_1}$. This is the common variable used to express Collins-Soper evolution in standard TMD factorization. The inverse Laplace transform of this new variable gives:
\begin{align}
\label{eq:y1SOL_thrust_1}
\overline{\zeta} = Q^2 \, e^{-2 \, \overline{y}_1}
\quad \overset{\text{LT}^{-1}}{\Rightarrow} \quad 
\overline{\xi} = 
\tau \, Q^2  \, e^{-2 \gamma_E} \, 
\left( \frac{b_T^\star(b_{\text{\tiny MAX}}) \, Q}{c_1} \right)^{-\alpha(b_T, b_{\text{\tiny MAX}})}.
\end{align}
We might then define the following rapidity cut-off in thrust space as the logarithm of the quantity above:
\begin{align}
\label{eq:y1SOL_thrust_2}
\overline{Y}_1 = - \frac{1}{2} \log{\frac{\overline{\xi}}{Q^2}} 
= -\log{\sqrt{\tau}} + \; \parbox{3em}{log-$b_T$\\ terms } .
\end{align}
Therefore, the sole kinematic argument misses all the $b_T$-dependent corrections to the rapidity cut-off. This is important for the comparison with previous works on this subject~\cite{Boglione:2020auc,Boglione:2022nzq}, where the kinematic argument is assumed to be valid and the whole correlation between thrust and transverse momentum induced by the rapidity cut-off is neglected.
Notice, however, that all the considerations on the size of the rapidity cut-off only hold for $y_1$ in the Laplace conjugate space and they should not extended straightforwardly to $\overline{Y}_1$ as defined in Eq.~\eqref{eq:y1SOL_thrust_2}.

\bigskip

\section{Factorization theorem in the central region \label{sec:ft_R2}}

In this Section we present our final results for the factorization theorem in Region 2 of \sia. In the previous Section, all kinds of radiation contributing to this kinematics have been identified and organized, as schematically presented in Eq.~\eqref{eq:FactT_R2_schematic}. As widely discussed above, the factorized cross section must be evaluated at the very special value of the rapidity cut-off obtained in Eq.~\eqref{eq:y1SOL} as a solution of the condition of Eq.~\eqref{eq:CS_cond}. This corresponds to the minimum of the cross section regarded as a function of the rapidity regulator and it is the only value of $y_1$ that truly factorizes all terms in Region 2, as it avoids the double counting induced by the soft-collinear overlapping. Exploiting the evolution equations associated to the various contributing terms of the factorization theorem, the cross section is re-casted in a form suitable for counting the powers of logarithms, both on thrust (logs of $u$) and on transverse momentum (logs of $b_T$), at the desired accuracy. This will be the final result, ready for phenomenological applications.

\bigskip

The computation starts from Eq.~\eqref{eq:FactT_R2_schematic} and it is conveniently carried out in the Laplace-Fourier conjugate space of  thrust and transverse momentum. Without specifying any specific value for the rapidity cut-off and using Eq.~\eqref{eq:coll_mix} to express the subtracted collinear radiation contribution in terms of a TMD FF, the factorization theorem reads:
\begin{align}
\label{eq:FactT_R2_prelim1}
&\frac{ d\sigma_{R_2}(y_1)}{d z \, d u \, d^2 \vec{b}_T} =
\sigma_B N_C \, \sum_j e_j^2 \,
|H|^2\left( a_S(\mu), \log{{\mu}/{Q}}\right) \, 
\widehat{J}\left( a_S(\mu), L_J \right) 
\notag \\
&\qquad \times
\frac{\widehat{\mathcal{S}}\left( a_S(\mu), L_S, L_L, L_R;\,u \right)}{\widehat{\mathcal{Y}}_L\left( a_S(\mu), L_L \right)} \,
\widetilde{D}_{h/j}\left( a_S(\mu),z,b_T,\mathcal{L}_b,\log{\sqrt{\zeta}/{\mu}};\,b_T\right),
\end{align}
where we have explicitly shown the overall multiplying constants and all the details of the functions' arguments. Next, we use the evolution equations of each of the terms appearing in the equation above. The RG-invariance is automatically satisfied, as the sum $\gamma_H + \gamma_J - 1/2 \, \gamma_K \, y_1 + 1/2 \, \gamma_S + \gamma_D \equiv 0$ vanishes for \emph{any} value of the rapidity cut-off. After some manipulation, the factorization theorem can be re-casted as a product of three factors:
\begin{align}
\label{eq:FactT_R2_prelim2}
&\frac{ d\sigma_{R_2}(y_1)}{d z \, d u \, d^2 \vec{b}_T} =
\sigma_B \, N_C \, \sum_j e_j^2 \,
\left[
 F_u \left( u\right) \times 
F_{\text{corr.}} \left( u, b_T, y_1 \right)
\times
\left. \widetilde{D}_{h/j}(z,b_T) \right|_{\substack{\mu = Q \\ y_1 = 0}}.
\right]
\end{align}
The last factor is the (unpolarized) TMD FF evaluated at the \emph{same} scales at which it is usually evaluated in standard TMD factorization. However, crucially, here it is  defined differently, following the factorization definition instead of the commonly used square root definition~\cite{Boglione:2020cwn}. In particular, the TMD FF appearing in the equation above is totally free from non-perturbative soft contamination. 
A comprehensive review on how to compute the TMD FF by merging its non-perturbative core, associated with the non-perturbative effects genuinely produced by the transverse momentum of the detected hadron, with its perturbative behavior at small distances can be found in Appendix~\ref{app:tmdff}.

The first factor $F_u$ depends solely on $u$ and it is strictly related to the thrust distribution of full inclusive $\epm$ annihilation (fIA$^{\text{\tiny thr}}$), widely studied in the past~\cite{Catani:1991kz,Catani:1992ua,Dokshitzer:1997ew,Schwartz:2007ib,Becher:2008cf,Monni:2011gb}. Specifically, it is given by:
\begin{align}
\label{eq:Fu_def}
& F_u \left( u\right) = 
\left( \;
\parbox{5.7em}{$H$, $\widehat{J}$, $\widehat{\mathcal{S}}/\widehat{\mathcal{Y}_L}$ \\
at ref. scales}
\right) \,
\text{exp}\left\{
\int_{\mu_J}^Q \frac{d \mu'}{\mu'} \, \gamma_J \left( a_S(\mu'), L'_J\right)
+
\frac{1}{2} \int_{\mu_S}^Q \frac{d \mu'}{\mu'} \, \gamma_S \left( a_S(\mu'), L'_S\right)
\right\},
\end{align}
where the exponent in the equation above matches \emph{exactly} with half of the analogue evolution factor in fIA$^{\text{\tiny thr}}$. On the other hand, in the first terms all the logarithmic dependence has been killed by setting the scales at their reference values, i.e. $\mu=Q$ in the hard function, $\mu=\mu_J$ in the jet thrust function, $\mu = \mu_S$ and $y_{1,2} = 0$ in the left-subtracted generalized soft thrust function. The result is:
\begin{align}
\label{eq:u_refscale}
&\left( \,
\parbox{5.7em}{$H$, $\widehat{J}$, $\widehat{\mathcal{S}}/\widehat{\mathcal{Y}_L}$ \\
at ref. scale}
\right)
=
|H|^2\left( a_S(Q), 0\right) \, 
\widehat{J}\left( a_S(\mu_J), 0 \right) 
\,
\frac{\widehat{\mathcal{S}}\left( a_S(\mu_S), 0, 0, 0;\,u \right)}{\widehat{\mathcal{Y}}_L\left( a_S(\mu_S),0 \right)}.
\end{align}
Notice that the evaluation at the reference scales does not affect any eventual extra-logarithimic $u$-dependence. Such dependence signals the presence of non-perturbative effects genuinely associated with thrust and becoming manifest close to the $2$-jet limit $\tau \to 0$. As long as these effects can be neglected, Eq.~\eqref{eq:u_refscale} can be expanded in powers of $a_S$ and the issue associated to $\mu_J$ and $\mu_S$ approaching the Landau pole can be handled by the sole resummation. However, it is reasonable to expect that describing data at very high thrust requires the inclusion of a non-perturbative function $f_{\text{\tiny NP}}(\tau)$ that takes over the resummation when $T \lesssim 1$. Assuming that the generalized soft thrust function has the same non-perturbative asymptotic behavior of the usual soft thrust function at large values of $u$, a phenomenological analysis could exploit the non-perturbative models investigated and used in past studies of fIA$^{\text{\tiny thr}}$. Some examples of how to include such non-perturbative effects can be found in Refs.~\cite{Korchemsky:1999kt,Korchemsky:2000kp,Gardi:2001ny,Gardi:2002bg,Hoang:2007vb,Ligeti:2008ac,Abbate:2010xh,Davison:2009wzs}

These strong similarities with the very well-known case of full inclusive $\epm$ annihilation make the treatment of the factor $F_u$ particularly simple. In fact, borrowing the results obtained in fIA$^{\text{\tiny thr}}$ and combining them with the considerations above, at next-to-next-to-next leading log (NNLL) we have:
\begin{align}
\label{eq:Fu_resum}
&F_u \left( u\right) = 
\left( 
1 + a_S(Q) H^{[1]}
\right)
\text{exp}\left\{
L_u \, g_1(\lambda_u) + g_2(\lambda_u) + \frac{1}{L_u} g_3(\lambda_u)
\right\},
\end{align}
where $\lambda_u = \beta_0 \, a_S(Q) L_u$ and the functions $g_i$ are:
\begin{subequations}
\label{eq:gi_u}
\begin{align}
&g_1(\lambda) = 
- \frac{\gamma_K^{[1]}}{8 \beta_0 \, \lambda} \,
\left(
(1-2\lambda)\,\log{(1-2\lambda)}
-
2 (1-\lambda)\,\log{(1-\lambda)}
\right),
\label{eq:g1}\\
&g_2(\lambda) = 
- \frac{\gamma_K^{[1]}}{16 \beta_0^2}
\, \frac{\beta_1}{\beta_0} \,
\left(
2 \log{(1-2\lambda)} + \log^2{(1-2\lambda)}
- 4 \log{(1-\lambda)} - 2 \log^2{(1-\lambda)}
\right)
\notag \\
&\quad
+ \frac{\gamma_K^{[2]}}{8 \beta_0^2}\,
\left( 
\log{(1-2\lambda)} - 2 \log{(1-\lambda)}
\right)
- \frac{\gamma_J^{[1]}}{2 \beta_0}\,
\log{(1-\lambda)},
\label{eq:g2}\\
&g_3(\lambda) = 
- \frac{\gamma_K^{[1]} \, \lambda}{8 \beta_0^3} \, \frac{\beta_1^2}{\beta_0^2}
\, \Bigg(
\frac{\lambda^2}{(1-2\lambda)\,(1-\lambda)}  \left( 1 + \frac{\beta_0 \, \beta_2}{\beta_1^2} \right) 
\notag \\
&\quad
+ \frac{\log{(1-2\lambda)}}{1-2\lambda}
\left( 2\lambda + (1-2\lambda)\,\frac{\beta_0 \, \beta_2}{\beta_1^2} \right) 
+ \frac{\log^2{(1-2\lambda)}}{2 (1-2\lambda)}
- 2 \left[ 2\lambda \leftrightarrow \lambda \right]
\Bigg)
\notag \\
&\quad
+ \frac{\gamma_K^{[2]}}{8 \beta_0^3}
\frac{\beta_1}{\beta_0}
\frac{\lambda}{(1-2\lambda)\,(1-\lambda)}
\, \left( 
3\lambda^2 + 
(1-\lambda)\,\log{(1-2\lambda)}
-2(1-2\lambda)\,\log{(1-\lambda)}
\right)
\notag \\
&\quad
- \frac{\gamma_K^{[3]}}{8 \beta_0^3}
\frac{\lambda^3}{(1-2\lambda)\,(1-\lambda)}
- \frac{\gamma_J^{[1]}}{2 \beta_0^2}
\frac{\beta_1}{\beta_0}
\frac{\lambda}{1-\lambda}
\, \left( 
\lambda + \log{(1-\lambda)}
\right)
+ \frac{\gamma_J^{[2]}}{2 \beta_0^2}
\frac{\lambda^2}{1-\lambda}
\notag \\
&\quad
+ \frac{\gamma_S^{[2]}}{2 \beta_0^2}
\frac{\lambda^2}{1-2\lambda}
+ \frac{J^{[1]}}{\beta_0}
\frac{\lambda}{1-\lambda}
+ \frac{S^{[1]}}{\beta_0}
\frac{\lambda}{1-2\lambda},
\label{eq:g3}
\end{align}
\end{subequations}
where all anomalous dimensions and constant terms can be found in Appendices.

Finally, the factor $F_{\text{corr.}}$ encodes the correlation between $u$ and $b_T$, as well as the whole dependence on the rapidity cut-off. It encodes the authentic spirit of factorization in Region 2, as this highly non-trivial correlation is primarily induced by the special mechanism that intertwines the rapidity regulator with thrust and transverse momentum, as described in Section~\ref{subsec:cond_y1}. It is given by:
\begin{align}
\label{eq:Fcorr_def}
&F_{\text{corr.}}\left( u, b_T, y_1 \right) = 
\text{exp}\left\{
\frac{1}{2} \Phi(\mu_S, \mu_R)
\right\},
\end{align}
where the function $\Phi(\mu_1, \mu_2)$ is defined as:
\begin{align}
\label{eq:Phi_def}
&\Phi(\mu_1, \mu_2) = 
\int_{\mu_1}^{\mu_2}
\frac{d\mu'}{\mu'}
\left(
\widehat{g}(a_S(\mu'))
- \gamma_K(a_S(\mu')) \, \log{\frac{\mu'}{\mu_1}}
\right)
-
\log{\frac{\mu_2}{\mu_1}} \,
\left.\widetilde{K}(b_T)\right|_{\mu = \mu_2}
\end{align}
and it has the properties:
\begin{subequations}
\label{eq:Phi_prop}
\begin{align}
&\Phi(\mu_1, \mu_2) = 
- \Phi(\mu_2, \mu_1), 
\label{eq:Phi_prop_1}\\
&\Phi(\mu_1, \mu_2) = 
\Phi(\mu_1, \mu_0) + \Phi(\mu_0, \mu_1), 
\label{eq:Phi_prop_2}\\
&\frac{\partial \Phi(\mu_0, \mu_R)}{\partial y_1} = 
\widehat{G}_R(u,y_1) + \widetilde{K}(b_T)
\quad \Rightarrow \quad
\left. \frac{\partial \Phi(\mu_0, \mu_R)}{\partial y_1}
\right|_{\overline{y}_1} \equiv 0.
\label{eq:Phi_prop_3}
\end{align}
\end{subequations}
In the last equation we have applied the condition for the rapidity cut-off found in Eq.~\eqref{eq:CS_cond}, which solution is $\overline{y}_1$ obtained in Eq.~\eqref{eq:y1SOL}. Using these properties, the correlation factor can be re-casted as:
\begin{align}
\label{eq:Fcorr_dec_1}
&F_{\text{corr.}}\left( u, b_T, y_1 \right) = 
\text{exp}\left\{
\frac{1}{2} \Phi(\mu_S, Q)
\right\}
\text{exp}\left\{
-\frac{1}{2} \Phi(\mu_b^\star, Q)
\right\}
\text{exp}\left\{
-\frac{1}{2} \Phi(\mu_R, \mu_b^\star)
\right\}.
\end{align}
The integrations involved in the first two exponentials depend now solely on $u$ and $b_T$, respectively, making the treatment of thrust and transverse momentum simpler and more transparent. 
Before dealing with it, consider first the last exponential, which deserves a different treatment compared to the other two.
In fact, this is where the whole dependence on $y_1$ is encoded. It must be treated following the same procedure that we have applied when we solved the condition for the rapidity cut-off of Eq.~\eqref{eq:CS_cond}, by isolating and considering the leading power of $y_1$ according to the hypotheses that the rapidity cut-off is large and positive $y_1 \gtrsim 0$, enough to have the scale $\mu_R$ far from the Landau pole and such that the condition $2\beta_0 \, a_S(\mu_b^\star) \log{{\mu_b^\star}/{\mu_R}} \lesssim 1$ is satisfied. The resummation of thrust and transverse momentum has to be pursued only \emph{after} that the rapidity cut-off dependence has been properly organized. This is consistent with the priority that has the rapidity cut-off dependence at the very core of the factorization procedure, where the asymptotic limit of large $y_1$ has precedence even with respect to the limit on the spacetime dimension $D \to 4$ in perturbative calculations. 
Given all such considerations, the last exponent of Eq.~\eqref{eq:CS_cond} is written as:
\begin{align}
\label{eq:Phi3_y1}
& \Phi(\mu_R,\mu_b^\star) = 
\left(
L_u - L_{b^\star} -  y_1
\right) \left(
-\frac{\gamma_K^{[1]}}{2 \beta_0} \left(
1 + \frac{1-x}{x}\log{(1-x)}\right)
-
\left.\widetilde{K}(b_T)\right|_{\mu = \mu_b^\star}
\right),
\end{align}
where $x = 2\beta_0 \, a_S(\mu_b^\star) \, (L_u - L_{b^\star} - y_1)$. 
It can be readily checked that first derivative of the above result vanishes when it is evaluated at $y_1 = \overline{y}_1$. 
We have neglected any correction suppressed by $\mathcal{O}(1/y_1)$ and the non-perturbative contributions to $\widehat{G}_R$, as in Eq.~\eqref{eq:y1SOL_eq}. More precisely, the first neglected term is suppressed by $\mathcal{O}({1}/{(L_u - L_{b^\star} - y_1)})$. On the solution $\overline{y}_1$, this corresponds to a suppressed correction of order $\mathcal{O}(L_{b^\star}^{-1})$. Therefore, the $b_T$-dependence is automatically cut out at NLL accuracy by the leading power approximation applied to the rapidity cut-off dependence. Keeping higher order corrections of $1/y_1$ would allow to consider the logarithms $L_{b^\star}$ beyond NLL. However, in that case Eq.~\eqref{eq:CS_cond} could not be solved analytically. 
For this reason, we will not include terms higher than NLL for the $b_T$-dependence in the overall exponent of the factorized cross section. 
Such considerations are particularly relevant if the $b_T$-dependence is treated on an equal footing with the  $u$-dependence. In fact, the $b^\star$-prescription prevents the logarithms $L_{b^\star}$ to be larger than $\sim \log{({Q}/\text{GeV})}$. This mechanism is absent in the $u$-dependent terms and resumming the logarithms $L_u$ is necessary to describe as far as possible the behavior of the cross section when they become large.
Therefore, perturbative calculations of $b_T$-dependent contributions cannot be spoiled by large \emph{divergent} logarithms of $b_T$, ultimately making the fixed order sensible and numerically close to the resummed expression. Nevertheless, we will adopt a very common strategy, expressing  $b_T$-dependent quantities by re-organizing the perturbative expansion in powers of $L_{b^\star}$. Clearly, the TMD FF appearing in the factorized cross section in Eq.~\eqref{eq:FactT_R2_prelim2} should be consider at the same NLL-accuracy for consistency.
We can now proceed to evaluate Eq.~\eqref{eq:Phi3_y1} on the solution  $y_1 = \overline{y}_1$:
\begin{align}
\label{eq:Phi3_y1SOL}
&\left.\Phi(\mu_R,\mu_b^\star)\right|_{y_1 = \overline{y}_1} = 
B\left(g_K(b_T)\right)
\left(
L_{b^\star} \, r_1(\lambda_b^\star) +  r_2(\lambda_b^\star)
\right),
\end{align}
where $\lambda_b^\star = 2 \beta_0 \, a_S(Q) L_{b^\star}$ and the functions $r_i$ and the factor $B$ are defined as:
\begin{subequations}
\label{eq:ri}
\begin{align}
&r_1(\lambda) = 
\frac{1-\lambda}{\lambda},
\label{eq:r1}\\
&r_2(\lambda) = 
\frac{1}{2\beta_0} \log{(1-\lambda)} \frac{\beta_1}{\beta_0};
\label{eq:r2}\\
&B\left(g_K(b_T)\right) = 
g_K(b_T) - \frac{\gamma_K^{[1]}}{2\beta_0}
\left(
1 - e^{-\frac{2 \beta_0}{\gamma_K^{[1]}} g_K(b_T)}
\right).
\label{eq:B}
\end{align}
\end{subequations}
Notice that keeping only the leading power of the rapidity cut-off makes the $u$-dependence of the solution $\overline{y}_1$ to be totally irrelevant for the exponent in Eq.~\eqref{eq:Phi3_y1}. Moreover, notice that the $B$-factor vanishes for $g_K \to 0$, i.e. at small $b_T$. This is indeed a non-perturbative correction to the transverse momentum dependence not brought by the TMD FF.

The other source of $b_T$-dependence in the correlation factor in Eq.~\eqref{eq:Fcorr_dec_1} comes from the second exponent. Following the same procedure adopted for the previous contribution, it can be written as:
\begin{align}
\label{eq:Phi2_nnll}
&\Phi(\mu_b^\star, Q) = L_{b^\star} \, 
\left(h_1({\lambda_b^\star}) - \kappa_1({\lambda_b^\star}) +  g_K(b_T) \right) + 
h_2({\lambda_b^\star})  -  \kappa_2({\lambda_b^\star}).
\end{align}
The functions $\kappa_i$ define the Collins-Soper kernel at NLL:
\begin{align}
\label{eq:CSkernel_NNLL}
&\left.\widetilde{K}(b_T)\right|_{\mu = Q} = 
\kappa_1 (\lambda_b^\star) + \frac{1}{L_{b^\star}}\kappa_2(\lambda_b^\star) - g_K(b_T),
\end{align}
with:
\begin{subequations}
\label{eq:kappai}
\begin{align}
&\kappa_1(\lambda) = \frac{\gamma_K^{[1]}}{2\beta_0} \log{(1-\lambda)},
\label{eq:kappa1}\\
&\kappa_2(\lambda) = \frac{\gamma_K^{[1]}}{4 \beta_0^2}
\frac{\beta_1}{\beta_0} \frac{\lambda^2}{1-\lambda} \left(
1+\frac{\log{(1-\lambda)}}{\lambda}
\right)
- \frac{\gamma_K^{[2]}}{4 \beta_0^2} \frac{\lambda^2}{1-\lambda};
\label{eq:kappa2}
\end{align}
\end{subequations}
On the other hand, the functions $h_i$ are given by:
\begin{subequations}
\label{eq:hi}
\begin{align}
&h_1(\lambda) = 
- \frac{\gamma_K^{[1]}}{2 \beta_0} \,
\left(
1 + \frac{1-\lambda}{\lambda}
\log{(1-\lambda)}
\right),
\label{eq:h1}\\
&h_2(\lambda) = 
- \frac{\gamma_K^{[1]}}{4 \beta_0^2}
\frac{\beta_1}{\beta_0}
\left( 
\lambda + \log{(1-\lambda)}
+ \frac{1}{2} \log^2{(1-\lambda)}
\right)
+\frac{\gamma_K^{[2]}}{4 \beta_0^2}
\left( 
\lambda + \log{(1-\lambda)}
\right).
\label{eq:h2}
\end{align}
\end{subequations}
Notice that the function $g_K$ appearing into Eq.~\eqref{eq:Phi2_nnll} represents another non-perturbative contribution to the transverse momentum distribution not brought by the TMD FF.

Finally, we have to organize the power of $L_u$ and $L_{b^\star}$ in the first exponent of  Eq.~\eqref{eq:Fcorr_dec_1}. Unlike the $b_T$-dependence, there are no constraints for the minimum allowed power of $L_u$. Therefore, the thrust dependence will be resummed up to NNLL. This leads to:
\begin{align}
\label{eq:Phi1_nnll}
& \Phi(\mu_S, Q) = L_u \,
\left( h_1(2\lambda_u) - \kappa_1(\lambda_b^\star) + g_K(b_T) \right) + 
h_2(2\lambda_u) - \frac{2\lambda_u}{\lambda_b^\star}\kappa_2(\lambda_b^\star) +
\frac{1}{L_u} h_3(2\lambda_u),
\end{align}
where the only function not encountered yet is $h_3$, defined as:
\begin{align}
\label{eq:h3}
&h_3(\lambda) = 
-\frac{\gamma_K^{[1]}}{8 \beta_0^3}
\frac{\beta_1^2}{\beta_0^2}
\frac{\lambda^3}{2(1-\lambda)} \,
\Bigg(
1 + \frac{1 + (1-\lambda)}{\lambda} \frac{\beta_0 \beta_2}{\beta_1^2}
\notag \\
&\quad
+ 2\frac{\log{(1-\lambda)}}{\lambda}
\left(
1 + \frac{1-\lambda}{\lambda} \frac{\beta_0 \beta_2}{\beta_1^2}
+ \frac{1}{2} \frac{\log{(1-\lambda)}}{\lambda}
\right)
\Bigg)
\notag \\
&\quad
+ \frac{\gamma_K^{[2]}}{8 \beta_0^3}
\frac{\beta_1}{\beta_0}
\frac{\lambda^2}{1-\lambda}
\, \left(
\frac{1 + (1-\lambda)}{2}+\lambda+\frac{\log{(1-\lambda)}}{\lambda}
\right) 
- \frac{\gamma_K^{[3]}}{16 \beta_0^3}
\frac{\lambda^3}{1-\lambda}
+ \frac{\widehat{g}^{[2]}}{4 \beta_0^2}
\frac{\lambda^2}{1-\lambda}.
\end{align}
Again, the function $g_K$ in  Eq.~\eqref{eq:Phi1_nnll} encodes (the last) non-perturbative $b_T$-dependent contribution not brought by the TMD FF.

Combining all these results together, the correlation factor $F_{\text{corr.}}$ evaluated on the solution $\overline{y}_1$ is written as:
\begin{align}
\label{eq:Fcorr_res}
&\left.F_{\text{corr.}}(u,b_T,y_1)\right|_{y_1 =\overline{y}_1}  = 
\text{exp}\left\{
L_u \, m_1(\lambda_u, \lambda_b^\star; b_T)
+
m_2(\lambda_u, \lambda_b^\star)
+
\frac{1}{L_u} \, m_3(\lambda_u)
\right\} 
\notag \\
&\quad
\text{exp}\left\{
L_{b^\star}  \, n_1(\lambda_b^\star; b_T)
+
n_2(\lambda_b^\star; b_T)
\right\}.
\end{align}
In this way, the NNLL resummation of $L_u$ and the NLL resummation of $L_{b^\star}$ is transparent. The functions $m_i$ and $n_i$ are the combinations:
\begin{subequations}
\label{eq:mi}
\begin{align}
&m_1(\lambda_u, \lambda_b^\star; b_T) = 
\frac{1}{2} \left( h_1(2\lambda_u) - \kappa_1(\lambda_b^\star) + g_K(b_T) \right),
\label{eq:m1}\\
&m_2(\lambda_u, \lambda_b^\star) =
\frac{1}{2}h_2(2\lambda_u) - \frac{\lambda_u}{\lambda_b^\star}\kappa_2(\lambda_b^\star),
\label{eq:m2}\\
&m_3(\lambda_u) =\frac{1}{2} h_3(2\lambda_u);
\label{eq:m3}
\end{align}
\end{subequations}
and:
\begin{subequations}
\label{eq:ni}
\begin{align}
&n_1(\lambda_b^\star; b_T) = 
-\frac{1}{2} \left( 
h_1({\lambda_b^\star}) - \kappa_1({\lambda_b^\star}) +  g_K(b_T)
+ B\left(g_K(b_T)\right) \, r_1(\lambda_b^\star)
\right),
\label{eq:n1}\\
&n_2(\lambda_b^\star; b_T) =
-\frac{1}{2} \left( 
h_2({\lambda_b^\star})  -  \kappa_2({\lambda_b^\star})
+ B\left(g_K(b_T)\right)\,  r_2(\lambda_b^\star)
\right).
\label{eq:n2}
\end{align}
\end{subequations}
where the functions $h_i$ have been collected in Eqs.~\eqref{eq:hi} and~\eqref{eq:h3}, the functions $\kappa_i$ in Eqs.~\eqref{eq:kappai}, while the functions $r_i$ and the coefficient $B$ can be found in Eqs.~\eqref{eq:ri}.
Now we possess all the necessary ingredients to write the factorized cross section in the Laplace-Fourier conjugate space at NNLL-accuracy in $u$ and NLL-accuracy in $b_T$. This reads:
\begin{align}
\label{eq:FactT_R2_LapFour}
&\frac{ d\sigma_{R_2}}{d z \, d u \, d^2 \vec{b}_T} =
\sigma_B \, N_C \, \sum_j e_j^2
\left( 
1 + a_S(Q) H^{[1]}
\right)
\notag \\
&\quad
\times
\text{exp}\left\{
L_{b^\star}  \, n_1(\lambda_b^\star; b_T)
+
n_2(\lambda_b^\star; b_T)
\right\} 
\left. \widetilde{D}^{\text{\tiny NLL}}_{h/j}(z,b_T) \right|_{\substack{\mu = Q \\ y_1 = 0}}
\notag \\
&\quad
\times
\text{exp}\left\{
L_u \, f_1(\lambda_u, \lambda_b^\star; b_T) + 
f_2(\lambda_u, \lambda_b^\star) +
\frac{1}{L_u} f_3(\lambda_u)
\right\},
\end{align}
where $ d\sigma_{R_2} \equiv  d\sigma_{R_2}(y_1)|_{y_1 = \overline{y}_1}$.
Notice that only the last line is relevant for the thrust resummation and hence for the inverse Laplace transform. 
The functions $f_i$ simply collect all the $u$-dependent terms:
\begin{subequations}
\label{eq:fi}
\begin{align}
&f_1(\lambda_u, \lambda_b^\star; b_T) = 
g_1(\lambda_u) + m_1(\lambda_u, \lambda_b^\star; b_T),
\label{eq:f1}\\
&f_2(\lambda_u, \lambda_b^\star) =
g_2(\lambda_u) + m_2(\lambda_u, \lambda_b^\star),
\label{eq:f2}\\
&f_3(\lambda_u) = g_3(\lambda_u) + m_3(\lambda_u).
\label{eq:f3}
\end{align}
\end{subequations}
where the functions $g_i$ have been collected in Eqs.~\eqref{eq:gi_u} and the functions $m_i$ in Eqs.~\eqref{eq:mi}.
The inverse Laplace transform can be carried out analytically up to the desired logarithm-accuracy~\cite{Catani:1992ua,Monni:2011gb}. This is an undeniable advantage for estimating the cross-section in thrust space, as the numerical computation is extremely hard because of the highly oscillating nature of the integrand. 
The analytic inversion is conveniently performed on  the cumulative thrust distribution $R$, which is obtained by integrating the cross section in thrust space up to a certain $\tau_{\text{\tiny MAX}} \equiv \omega$. At NNLL the result reads:
\begin{align}
\label{eq:cumulativeR_NNLL}
&R(z,\omega,b_T) = 
\sigma_B \, N_C \, \sum_j e_j^2 
\left( 
1 + a_S(Q) H^{[1]}
\right)
\notag \\
&\quad
\times 
\text{exp}\left\{
L_{b^\star}  \, n_1(\lambda_b^\star; b_T)
+
n_2(\lambda_b^\star; b_T)
\right\} 
\left. \widetilde{D}^{\text{\tiny NLL}}_{h/j}(z,b_T) \right|_{\substack{\mu = Q \\ y_1 = 0}}
\notag \\
&\quad
\times
\left( 
1 + a_S(Q) C_1
\right)
\text{exp}\left\{
L f_1(\bullet) + f_2^{\text{\tiny eff.}}(\bullet) + \frac{1}{L} f_3^{\text{\tiny eff.}}(\circ)
\right\}
\frac{1}{\Gamma\left( 1 - \gamma_1 (\bullet)\right)}
\Bigg\{
1 + 
\notag \\
&\quad+
\frac{1}{L} \left[
\left( \gamma_2 (\circ) + \gamma_E \, \rho_1(\lambda) \right) 
\psi_0\left(1-\gamma_1 (\bullet)\right)
+\frac{1}{2}\rho_1(\lambda) \left( 
\psi^2_0\left(1-\gamma_1 (\bullet)\right)
- \psi_1\left(1-\gamma_1 (\bullet)\right)
\right)
\right]
\Bigg\},
\end{align}
where $L = -\log{\omega}$ and $\lambda = \beta_0 a_S(Q) L$. The symbol $\bullet$ in the functions's arguments stay for the dependence on $\lambda,\lambda_b^\star; b_T$, while the symbol $\circ$ represents the argument $\lambda,\lambda_b^\star$.
We have introduced the functions $\gamma_i$:
\begin{subequations}
\label{eq:gammai}
\begin{align}
&\gamma_1(\lambda, \lambda_b^\star; b_T) = 
f_1(\lambda, \lambda_b^\star; b_T) + \lambda \,\frac{\partial}{\partial \lambda} f_1(\lambda, \lambda_b^\star; b_T) 
\notag \\
&\quad=
\frac{1}{2} g_K(b_T) + \frac{\gamma_K^{[1]}}{4\beta_0}
\left( 
2 \log{(1-2\lambda)} -  \log{(1-\lambda)} -  \log{(1-\lambda_b^\star)}
\right),
\label{eq:gamma1}\\
&\gamma_2(\lambda, \lambda_b^\star) =
\lambda \,\frac{\partial}{\partial \lambda} f_2(\lambda, \lambda_b^\star; b_T) 
\notag \\
&\quad=
\frac{\gamma_K^{[1]}}{4\beta_0^2}\frac{\beta_1}{\beta_0} 
\Bigg(
\lambda \left(
\frac{1}{(1-\lambda)\,(1-2\lambda)} - 
\frac{1+\log{(1-\lambda_b^\star)}}{(1-\lambda_b^\star)}
\right)
\notag \\
&\quad
+ \frac{2\lambda}{1-2\lambda} \log{(1-2\lambda)}
- \frac{\lambda}{1-\lambda} \log{(1-\lambda)}
\Bigg) 
\notag \\
&\quad
+\frac{\gamma_K^{[2]}}{4\beta_0^2} \lambda
\left( 
\frac{1}{1-\lambda_b^\star} + \frac{1}{1-\lambda} - \frac{2}{1-2\lambda}
\right)
+ \frac{\gamma_J^{[1]}}{2\beta_0} \frac{\lambda}{1-\lambda};
\label{eq:gamma2}
\end{align}
\end{subequations}
the function $\rho_1$:
\begin{align}
&\rho_1(\lambda) = 
\lambda \,\frac{\partial}{\partial \lambda} \gamma_1(\lambda, \lambda_b^\star; b_T) 
=
 \frac{\gamma_K^{[1]}}{4\beta_0} \lambda
\left(
\frac{1}{1-\lambda} -\frac{4}{1-2\lambda}
\right);
\label{eq:rho1}
\end{align}
and the functions $f_2^{\text{\tiny eff.}}$ and $f_3^{\text{\tiny eff.}}$, which play the role of effective NLL and NNLL term in the exponent, respectively:
\begin{subequations}
\label{eq:feffi}
\begin{align}
&f_2^{\text{\tiny eff.}}(\lambda,\lambda_b^\star; b_T) = 
f_2(\lambda,\lambda_b^\star) + \gamma_E \, \gamma_1(\lambda_u, \lambda_b^\star; b_T),
\label{eq:f2eff} \\
&f_3^{\text{\tiny eff.}}(\lambda,\lambda_b^\star) = 
f_3(\lambda) + \gamma_E \, \gamma_2(\lambda_u, \lambda_b^\star) + \frac{1}{2} \gamma_E^2 \, \rho_1(\lambda)
- \frac{\lambda }{\beta_0} \, C_1.
\end{align}
\end{subequations}
The subtraction of the term involving the constant $C_1$ ensures that $f_3^{\text{\tiny eff.}}$ starts its perturbative expansion from $\mathcal{O}(a_S^2 \times \text{logs})$. It is given by:
\begin{align}
&C_1 = \frac{1}{2}\gamma_E \, \left( \gamma_J^{[1]} - \frac{3}{4} \gamma_E \, \gamma_K^{[1]} \right).
\label{eq:C1}
\end{align}
The cross section in thrust space is then simply obtained as the derivative of $R$ with respect to $\omega$ and evaluating the result on $\omega = 1-T$. 
The Fourier transform on $b_T$ finally gives the factorized cross section resummed at NNLL in thrust and NLL in tranverse momentum:
\begin{align}
\label{eq:FactT_R2}
&\frac{ d\sigma_{R_2}}{d z \, d T \, d^2 \vec{P}_T} =
- \frac{\sigma_B \, N_C}{1-T} \, \sum_j e_j^2 
\left( 
1 + a_S(Q) H^{[1]}
\right)
\notag \\
&\quad
\times 
\int \frac{d^2\vec{b}_T}{(2\pi)^2} e^{i \vec{b}_T \cdot {\vec{P}_T}/{z}} \,
\text{exp}\left\{
L_{b^\star}  \, n_1(\lambda_b^\star; b_T)
+
n_2(\lambda_b^\star; b_T)
\right\} 
\left. \widetilde{D}^{\text{\tiny NLL}}_{h/j}(z,b_T) \right|_{\substack{\mu = Q \\ y_1 = 0}}
\notag \\
&\quad
\times
\left( 
1 + a_S(Q) C_1
\right)
\text{exp}\left\{
L f_1(\bullet) + f_2^{\text{\tiny eff.}}(\bullet) + \frac{1}{L} f_3^{\text{\tiny eff.}}(\circ)
\right\}
\notag \\
&\quad
\times
\frac{1}{\Gamma\left( 1 - \gamma_1 (\bullet)\right)}
\Bigg\{
\gamma_1 (\bullet) + \frac{1}{L} \Bigg[
\rho_1(\lambda) 
\psi_0\left(1-\gamma_1 (\bullet)\right) + 
\gamma_2(\circ) \left(
1 + \gamma_1 (\bullet)\psi_0\left(1-\gamma_1 (\bullet)\right)
\right)
\notag \\
&\quad+
\frac{1}{2}\gamma_1 (\bullet) \rho_1(\lambda) \left(
\psi_0^2\left(1-\gamma_1 (\bullet)\right) + 2\gamma_E \, \psi_0\left(1-\gamma_1 (\bullet)\right)
- \psi_1\left(1-\gamma_1 (\bullet)\right) 
\right)
\Big)
\Bigg]
\Bigg\}.
\end{align}
For reference, Eq.~\eqref{eq:hard_1loop} contains the hard coefficient $H^{[1]}$, the functions $n_i$ have been presented in Eqs.~\eqref{eq:ni}, while all ingredients required to build the TMD FF at NLL have been collected in Appendix~\ref{app:tmdff}, the functions $f_i$ can be found in Eqs.~\eqref{eq:fi} and~\eqref{eq:feffi}, the functions $\gamma_i$ are shown in Eqs.~\eqref{eq:gammai} and finally the function $\rho_1$ is obtained from Eq.~\eqref{eq:rho1}.
Unlike the inverse Laplace tranform, the analytic inversion of the Fourier transform is extremely difficult to be performed analytically, because the TMD FF involves  collinear FFs (see Eq.~\eqref{eq:tmd_evosol_star}) which are implemented as grids of numbers rather than closed analytical expressions. 
In conclusion, we will rely on \eqref{eq:FactT_R2} for our phenomenological analysis. 

\bigskip

\section{Matching regions \label{sec:match}}

In this section, we explore the transition from Region 2 (the ``bulk" of the phase space) to the Region 3 (one of the two boundaries of the phase space). Given the kinematic structure of Region 2 as described in Section~\ref{sec:id_R2}, the only sector that may allow to devise a smooth matching would necessarily correspond to forward soft-collinear radiation. Keeping the definition of soft and collinear contributions fixed, we now interpret soft-collinear gluons as TMD-irrelevant. Therefore, within this assumption, they do not produce any significant transverse deviation to the detected hadron, playing the same role that they play in Region 3. 
The following forward radiation scheme represents the current situation:
\newline
\begin{center}
\begin{tabular}{c || c || c || c ||}
  {}   &  soft  & soft-collinear & collinear \\
  \hline\hline
  $R_1$   &   
  \cellcolor{blue!15}TMD-relevant &  \cellcolor{blue!15}TMD-relevant & \cellcolor{blue!15}TMD-relevant	\\
  \hline\hline
   $R_2$   &   
  TMD-irrelevant &  \cellcolor{blue!15}TMD-relevant & \cellcolor{blue!15}TMD-relevant	\\
  \hline\hline
  $M_{2,3}$   &   TMD-irrelevant   &   TMD-irrelevant & \cellcolor{blue!15}TMD-relevant	\\
  \hline\hline
  $R_3$   &   TMD-irrelevant   &  TMD-irrelevant & \cellcolor{blue!15}TMD-relevant	\\
  \hline
\end{tabular}
\end{center}
where we labeled as $M_{2,3}$ the matching region allowing the transition from Region 2 to Region 3. Notice that its kinematics  is almost indistinguishable from that of Region 3. The (right) soft-collinear thrust function $\mathscr{Y}_R$, reviewed in Appendix~\ref{subsec:scthr-fun}, describes the forward soft-collinear gluons in both regions. The only difference is in the contribution associated with collinear radiation, which is a generalized FJF in both cases but approximated at low-$P_T$ in $M_{2,3}$.
In this regard, let's consider again how we have defined Region 2 in Section~\ref{sec:id_R2}.
Solving the ambiguity of the role of soft-collinear radiation in Region 2 by associating to it the kinematic structure of $M_{2,3}$ would have lead to a scheme representing just two kinematic regions, $R_1$ and $R_3$, with $R_2$ being relevant only for matching purposes. However, relying on this interpretation leads to an incomplete picture of the kinematic structure of the whole process. In fact,  in this case  Region 2 and Region 3 would have shared strong similarities,  in striking contrast with the sharp differences with Region 1. Therefore, it would be rather hard to determine what the matched regions actualy are: while $R_3$ is surely involved, $R_1$ seems to be too far to be affected. In other words, while the limit $R_3 \to R_2$ would follow naturally from the low-$P_T$ approximation of the generalized FJF, the limit $R_1 \to R_2$ would definitely be more obscure\footnote{In  Ref.~\cite{Makris:2020ltr} the authors propose to match $R_1$ to $R_2$ (there defined as $M_{2,3}$) through the fixed order of perturbative QCD, naively as $R_1 \to \text{F.O.} \to R_2$. Besides the formal correctness of such procedure (all the regions must be recovered from full QCD within some approximation), it automatically cuts off all the non-perturbative contributions of both regions, which are most likely relevant for matching.}. 
Such issues would reflect an unpleasant asymmetry of the physics of \sia\, strongly enforcing  our definition of Region 2 as the proper description of the ``bulk" of the phase space of the process.

\bigskip

Remarkably, factorization does hold in the matching region $M_{2,3}$, which is a totally new and fascinating feature: matching prescriptions are usually required to join different factorization theorems associated to different kinematic regions, but they never rely on a sound and solid factorization theorem. 
In contrast, in $M_{2,3}$ the cross section factorizes as follows:
\begin{align}
\label{eq:FactT_M_schematic}
& d\sigma_{M_{2,3}} \sim |H|^2 \, J \, \frac{\mathcal{S}}{\mathcal{Y}_L \, \mathcal{Y}_R} \, \mathcal{G}_{h/j}^{\text{\tiny asy}},
\end{align}
where, in the sake of clarity, we omitted any finer detail as in Eq.~\eqref{eq:FactT_R2_schematic}.
The combination of soft and soft-collinear terms results in the usual soft thrust function, as follows from the theorem in Eq.~\eqref{eq:soft_mix}.
In contrast to Region 2, the hadronization process in the matching region is not described by a TMD FF. However, further manipulations in  the Laplace-Fourier conjugate space allow to rearrange all terms \emph{as if} a proper TMD FF contributed to the final result. This is simply achieved by multiplying and dividing the factorized cross section by the soft-collinear thrust factor $\mathcal{C}_R$ and exploiting the theorem presented in Eq.~\eqref{eq:coll_mix}. The factorized cross section in Laplace-Fourier space reads:
\begin{align}
\label{eq:FactT_M23_prelim1}
&\frac{ d\sigma_{R_{M_{2,3}}}}{d z \, d u \, d^2 \vec{b}_T} =
\sigma_B N_C \, \sum_j e_j^2 \,
|H|^2\left( a_S(\mu), \log{{\mu}/{Q}}\right) \, 
\widehat{J}\left( a_S(\mu), L_J \right) 
\notag \\
& \times
\widehat{S}\left(a_S(\mu),L_S;\,u\right) \, 
\widetilde{\widehat{\mathcal{C}}}_R\left( a_S(\mu), \mathcal{L}_b, L_R; u, b_T\right) 
\widetilde{D}_{h/j}\left( a_S(\mu),z,b_T,\mathcal{L}_b,\log{\sqrt{\zeta}/{\mu}};\,b_T\right).
\end{align}
Notice that the combination $\mathcal{C}_R \, D_{h/j}$ is naturally CS-invariant and there is no left over dependence on the rapidity regulator (see also Appendix~\ref{app:scthr-fact}). This reflects the fact that in the matching region the sole thrust is sufficient for dealing with rapidity divergences, or, in other words, that in this case rapidity divergences are naturally regulated by the thrust.
 
Crucially, the theorem above coincides with the result obtained in the SCET formalism for the kinematic region there identified as Region 2. This can be readily checked by comparing our result in Eq.~\eqref{eq:FactT_M23_prelim1} with the factorization theorem presented in Eq.~(2.21) of Ref.~\cite{Makris:2020ltr}: the two formulae coincide. This represents a very important step forward in the understanding of \sia. Our formalism, based on the Collins factorization procedure, is not only able to recover the result obtained in the effective theory, but it also properly frames the result  into the whole kinematic structure of the process. In particular, we do not consider it as the description of the core of the ``bulk" of the phase space, but instead as the bridge that links it to one of its external edges, namely  Region 3. While Eq.~\eqref{eq:FactT_M23_prelim1} cannot be used for a direct extraction of TMD FFs, as they are not really contributing to it, this factorization theorem would be incredibly helpful for constraining the non-perturbative content of the generalized FJFs. In fact, the matching condition with Region 3 requires:
\begin{align}
\label{eq:matchR3_1}
&\widetilde{\widehat{\mathcal{G}}}_{h/j}\left(z,  b_T, u\right) 
\xrightarrow[b_T]{\text{\tiny large}}
\widetilde{\widehat{\mathcal{C}}}_R\left( b_T, u; y_1 \right) \,
\widetilde{D}_{h/j}\left( z,  b_T; y_1 \right),
\end{align}
or equivalently:
\begin{align}
\label{eq:matchR3_2}
&\widetilde{\widehat{\mathcal{G}}}_{h/j}^{NP}\left(z,  b_T, u\right) 
\sim
\Delta_{\mathcal{C}}(b_T,u) \,
M_D(z,b_T;\,j,h) \,
\text{exp} \left\{
-\frac{1}{2}\,g_K(b_T)\,\log{\frac{Q}{M_h}}
\right\},
\end{align}
where we have used the expressions for the non-perturbative content of the TMD FF and the soft-collinear thrust factor obtained from the last lines of Eqs.~\eqref{eq:tmd_evosol_star},~\eqref{eq:CR_evosol}, respectively. This kind of constraint might be extremely important for future phenomenological studies on generalized FJFs, which will likely play a crucial role in some of the processes  investigated by the future EIC~\cite{AbdulKhalek:2021gbh}.

The matching condition with the factorization theorem of Region 2 is conveniently explored by noticing that the two theorems in Eq.~\eqref{eq:FactT_R2_prelim1} and Eq.~\eqref{eq:FactT_M23_prelim1} are proportional to each other in the Laplace-Fourier conjugate space. Isolating the soft-collinear subtraction term $\widehat{\mathcal{Y}}_R$, implicitly encoded into the soft-thrust function, we can express the factorized cross section for the matching region $M_{2,3}$ in terms of the factorized cross section obtained for Region 2:
\begin{align}
\label{eq:matchR2_1}
&\frac{ d\sigma_{R_{M_{2,3}}}}{d z \, d u \, d^2 \vec{b}_T} =  
\mathcal{R}\left( a_S(\mu), \mathcal{L}_b, L_R; u, b_T\right)
\,
\frac{ d\sigma_{R_2}(y_1)}{d z \, d u \, d^2 \vec{b}_T}.
\end{align}
Both the coefficient $\mathcal{R}$ and the cross section of Region 2 are temporarily considered as functions of a generic rapidity cut-off $y_1$, yet however such dependence is canceled out in their product. The ratio coefficient $\mathcal{R}$ is given by:
\begin{align}
\label{eq:matchR2_R}
&\mathcal{R}\left( a_S(\mu), \mathcal{L}_b, L_R; u, b_T\right)
=
\frac{\widetilde{\widehat{\mathcal{C}}}_R\left( a_S(\mu), \mathcal{L}_b, L_R; u, b_T\right)}{\widehat{\mathcal{Y}}_R\left( a_S(\mu), L_R \right)} 
\notag \\
&\quad=
\text{exp} \left\{
-\frac{1}{2} \Phi(\mu_R^\star, \mu_R)
\right\}
\Delta_{\mathcal{C}}(b_T,u) \, \text{exp} \left\{
-\frac{1}{2}g_K(b_T) \, \overline{y}_1^\star
\right\},
\end{align}
where the perturbative contribution at the reference scales $\mu^{(0)} = \mu_b^\star$ and $y_1^{(0)} =  L_u - L_b^\star \equiv   \overline{y}_1^\star$, given by ${\widetilde{\widehat{\mathcal{C}}}_{R \,\star}\left( a_S(\mu_b^\star)\right) }/{
\widehat{\mathcal{Y}}_{R \,\star}\left( a_S(\mu_b^\star)\right)
}$, is trivially one. This follows from the fact that the constant terms in the perturbative expansions of $\widetilde{\widehat{\mathcal{C}}}_R$ and $\widehat{\mathcal{Y}}_R$ are the same. Such property can be checked by comparing the 1-loop expressions of Eqs.~\eqref{eq:YgenR_1loop} and~\eqref{eq:CR_1loop}, respectively. The function $\Phi$ in the equation above has been defined in Eq.~\eqref{eq:Phi_def}. Crucially, the ratio $\mathcal{R}$ gets closer and closer to one for $b_T \to 0$, provided that is is evaluated on the solution of the rapidity cut-off condition $\overline{y}_1$ defined in Eq.~\eqref{eq:y1SOL}. In fact, in this case the first exponent vanishes because the rapidity regulator collapses onto its reference value $y_1 \to L_u - L_b^\star \to L_u - L_b$, while both the non-perturbative terms converge to $1$ at low-$b_T$ by definition\footnote{The exponent involving the $g_K$ functions behaves as $\sim b_T^2 \, \log{b_T} \to 0$ for small $b_T$s. }. This implies that the two factorization theorems coincide at small distances, where the perturbative regime is dominant:
\begin{align}
\label{eq:matchR2_smallbT}
&\frac{ d\sigma_{R_{M_{2,3}}}}{d z \, d u \, d^2 \vec{b}_T} 
\xleftarrow[b_T]{\text{\tiny small}}
\left. \frac{ d\sigma_{R_2}}{d z \, d u \, d^2 \vec{b}_T}\right|_{y_1 = \overline{y}_1}.
\end{align}
As anticipated, perturbative QCD gives blindly the same answer. Used alone, it cannot properly predict the correct region decomposition of \sia. Moreover, this implies that the result obtained in the SCET formalism does coincide with our result for Region 2 at any order in perturbation theory, but they differ when the non-perturbative contributions are included.
In conclusion, we have proved that the matching of Region 2 and Region 3 is mediated by the factorization theorem holding in the matching region $M_{2,3}$. This is encoded into the relation $d \sigma_{R_3} \xrightarrow{\text{\tiny large }b_T} d \sigma_{M_{2,3}}  \xleftarrow{\text{\tiny small }b_T} d \sigma_{R_2} $, based on the matching condition of generalized FJFs of Eq.~\eqref{eq:matchR3_1} and on the property of the ratio coefficient in Eq.~\eqref{eq:matchR2_R} to converge to one in the perturbative regime, $\mathcal{R}(\overline{y}_1) \to 1$ for $b_T \to 0$.

It is important to stress that the two regions $R_2$ and $R_3$ match anyway, independently from the fact that their matching region $M_{2,3}$ hosts a factorization theorem. Clearly, having a factorization theorem provides much more powerful constraints that undeniably help the phenomenological analyses.
Intuition suggests that the matching region between Region 1 and Region 2, $M_{1,2}$, may host a factorization theorem as well. 
This would restore a complete symmetry among the kinematic regions, including the matching regions.
At present days however, there are no evidences about the existence of such a factorization theorem.

\bigskip

\section{Phenomenology \label{sec:pheno}}

In this Section the formalism developed so far on experimental data will be applied to the phenomenological analysis of relevant experimental data. In particular, we consider the \sia cross section measured by the BELLE collaboration~\cite{Seidl:2019jei} at $Q = 10.58$ GeV.
For simplicity, we restrict our analysis to the case where the detected hadron is a charged pion. 

In the past few years, very few attempts of describing these data have been performed, none of them really capable of fully capturing the rich information encoded in these experimental measurements, especially as far as the thrust dependence was concerned. This dependence turned out to be extremely difficult to be properly addressed, signaling that the underlying  theories were not fully consistent and complete.
In this regards, we mention Ref.~\cite{Kang:2020yqw} based on the SCET formalism and Ref.~\cite{Modarres:2021ffg} based on the $k_T$-factorization formalism~\cite{Kimber:2001sc,Martin:2009ii}.  
Some improvements in the description of the thrust dependence have been achieved in Ref.~\cite{Boglione:2022nzq}, based on a strongly simplified version of the formalism presented here. However, as discussed in Section~\ref{sub:approx_formalism}, the simplifications entailed in that analysis 
reduced the complexity of the correlation between  thrust and the rapidity cut-off to a naive kinematic constraint.

In this paper we present a phenomenological analysis of the whole $2$-jet region $0.8 \leq T \leq 1$ of these data. We show how only taking into account all the non-trivial issues entailed in \sia\, discussed in the previous sessions, allows us to provide a  satisfactory description of the measured cross section with respect to \emph{all} the binned kinematic variables, $z$, $P_T$ and $T$, obtaining a remarkable and unprecedented agreement with the experimental data.

In order to be safely far from higher order topologies of the final state, here we only consider data with $T \geq 0.8$. This allows us not to be concerned about matching with fixed order calculations associated with $3$ or more independent final state jets. The selection of data points strictly belonging to the central kinematic region, $R_2$, is based on the selection algorithm presented in Ref.~\cite{Boglione:2021wov}. 
This leads to 230 data points, corresponding to the bins represented in Table~\ref{tab:data_sel}.
\begin{table}[b]
\begin{center}
\begin{tabular}{ |c|c|c|c|c|c|c|c|c|c|c|c|c|c|c| } 
 \hline
 $T$ &  \multicolumn{12}{|c|}{$z$} & $P_T/z$ max & N \\  
 \toprule
 \hfill & \rot{$0.20-0.25$} & \rot{$0.25-0.30$} & \rot{$0.30-0.35$} & \rot{$0.35-0.40$} & \rot{$0.40-0.45$} & \rot{$0.45-0.50$} & \rot{$0.50-0.55$} & \rot{$0.55-0.60$} & \rot{$0.60-0.65$} & \rot{$0.65-0.70$} & \rot{$0.70-0.75$} & \rot{$0.75-0.80$} & \hfill & \hfill \\
 \hline
 $0.80 - 0.85$ & \cellcolor{blue!25}\hfill & \cellcolor{blue!25}\hfill & \cellcolor{blue!25}\hfill &
 \cellcolor{blue!25}\hfill &
 \cellcolor{blue!25}\hfill &
 \cellcolor{blue!25}\hfill &
 \cellcolor{blue!25}\hfill &
 \cellcolor{blue!25}\hfill &
 \cellcolor{blue!25}\hfill &
 \cellcolor{blue!25}\hfill &
 \hfill & \hfill &
 $0.16 \, Q$ & 57 \\
 \hline
 $0.85 - 0.90$ &  \hfill & \cellcolor{blue!25}\hfill &
 \cellcolor{blue!25}\hfill &
 \cellcolor{blue!25}\hfill &
 \cellcolor{blue!25}\hfill &
 \cellcolor{blue!25}\hfill &
 \cellcolor{blue!25}\hfill &
 \cellcolor{blue!25}\hfill &
 \cellcolor{blue!25}\hfill &
 \cellcolor{blue!25}\hfill &
 \cellcolor{blue!25}\hfill & 
 \hfill &
 $0.15 \, Q$ & 60 \\
 \hline
 $0.90 - 0.95$ &  \hfill & \hfill &
 \cellcolor{blue!25}\hfill &
 \cellcolor{blue!25}\hfill &
 \cellcolor{blue!25}\hfill &
 \cellcolor{blue!25}\hfill &
 \cellcolor{blue!25}\hfill &
 \cellcolor{blue!25}\hfill &
 \cellcolor{blue!25}\hfill &
 \cellcolor{blue!25}\hfill &
 \cellcolor{blue!25}\hfill & 
 \cellcolor{blue!25}\hfill &
 $0.14 \, Q$ & 61 \\
 \hline
 $0.95 - 1.00$ &  \hfill & \hfill &
 \hfill &
 \cellcolor{blue!25}\hfill &
 \cellcolor{blue!25}\hfill &
 \cellcolor{blue!25}\hfill &
 \cellcolor{blue!25}\hfill &
 \cellcolor{blue!25}\hfill &
 \cellcolor{blue!25}\hfill &
 \cellcolor{blue!25}\hfill &
 \cellcolor{blue!25}\hfill & 
 \cellcolor{blue!25}\hfill &
 $0.13 \, Q$ & 52 \\
 \hline
\end{tabular}
\end{center}
\caption{$T$, $z$ and $P_T/z$ bins corresponding to the BELLE measurements \cite{Seidl:2019jei}. Bins selected for our phenomenological analysis are highlighted by a blue shading. Detailed explanations on the selecting criteria are given in the text.}
\label{tab:data_sel}
\end{table}
Notice how the selected $z$ bins are progressively shifted to higher $z$ values as thrust increases. In fact, Region $2$ is expected to be centered around  relatively large values of $z$, and this feature would be more and more evident as the $2$-jet limit is approached. This originates from the fact that the transverse motion of the detected hadrons with low values of fractional energy is more likely affected by the deflection caused by soft radiation, revealing the typical physics associated with Region $1$. 
At the same time, the cut in $P_T$ is progressively reduced in our selection, as  larger values of transverse momentum are dangerously close to the phase space boundaries of Region $3$. 
Such cut is more and more stringent, consistently with the progressive shrinking of the available phase space as the $2$-jet limit is approached~\cite{Boglione:2021wov}. Table~\ref{tab:data_sel} summarizes our choice of $T$, $z$ and $P_T$ bins: those selected to be used in our analysis are highlighted in blue. 
We perform our analysis by approximating each bin with its central value.

\bigskip

As already addressed in Section~\ref{sec:ft_R2}, where we discuss the contribution of the term $F_u$ to the final cross section defined in Eq.~\eqref{eq:Fu_def}, the process we are considering does not only include non-perturbative effects associated with low transverse momentum, but it also entails non-perturbative contributions  genuinely associated with thrust. These start to be significant around the peak region~\cite{Korchemsky:1999kt,Korchemsky:2000kp,Gardi:2001ny,Gardi:2002bg,Hoang:2007vb,Ligeti:2008ac,Abbate:2010xh,Davison:2009wzs} and they become more and more dominant as the $T = 1$ limit is approached. 
Because of the intertwining of thrust with the Collins-Soper kernel, the TMD FF should be ideally extracted in a region where only the non-perturbative effects associated to transverse momentum are dominant, while those associated with thrust can safely be neglected. This would prevent to deal with the correlation between $g_K$ and $f_{\text{\tiny NP}}$, which are the functions that are primarily involved in the mechanism relating thrust and rapidity at the non-perturbative level. 
The BELLE data under consideration show a peak in thrust roughly around $T\approx0.90$, for any value of $z$ and $P_T$. This implies that only the very first bin in thrust considered in our selection can be considered far enough from the peak to be marginally affected by non-perturbative thrust effects, reducing the ideal kinematics where to extract the TMD FF to a single bin, i.e. $T=0.825$. 

For this reason, we organize our analysis in two steps.
\begin{enumerate}
    \item We perform a preliminary fit on the first bin $0.80 \leq T \leq 0.85$ (first highlighted line in Table~\ref{tab:data_sel}. Being safely far from the region where non-perturbative effects related to  thrust become significant, we will exploit the fit on this $T$-bin to determine the functional form of the TMD FF, i.e. the parametrization of the functions $M_D$ and $g_K$, see Eq.~\eqref{eq:tmd_evosol_star}.
    \item We extend the fit to all the $T$-bins within our selection by relying on the functional form of the TMD FF obtained in the extraction of fit 1., and by including the non-perturbative effects associated with thrust in an appropriate way (this will be  described in Section \ref{subsec:pheno_step2}).
\end{enumerate}
In the following, we will present the details of these two subsequent steps.

\subsection{Determination of the functional form of the TMD FF \label{subsec:step0}}

In this preliminary step, we restrict our analysis to the bin corresponding to $T = 0.825$, with $z$ and $P_T$ chosen according to our selection, illustrated  Table~\ref{tab:data_sel}. In total, this brings us to  57 data points. Such a preliminary step is necessary to determine the functional form of the non-perturbative content of the TMD, as this specific kinematics allows us to neglect all non-perturbative effects associated with thrust.
In this regard, we stress that we perform our phenomenological analysis within the standard prescriptions of the CSS approach, as thoroughly  reviewed in Appendix~\ref{app:tmdff}.
In particular, we separate the perturbative from the non-perturbative regime in $b_T$-space by introducing a scale $b_{\text{\tiny MAX}}$ and a function $b^\star$ that smoothly interpolates from short to large distances. Moreover, we adopt the $b_{\text{\tiny MIN}} = c_1/Q$ prescription to regularize the UV divergences associated with the integral of the TMD FF at large transverse momentum. 
The collinear FFs used in the OPE, Eq.~\eqref{eq:wilson_coeffs}, are obtained from the NNFF set of Ref.~\cite{Bertone:2017tyb}. 

Choosing a functional form for the TMD FF corresponds to select the functions $M_D$ and $g_K$, defined in Eqs.~\eqref{eq:model_def} and~\eqref{eq:gK_def}, respectively.
We assign 2 free parameters to each non-perturbative function, corresponding to a total of 4 free parameters associated with the TMD FF.  
$M_D$ and $g_K$ are further constrained by requiring them to assume some specific behavior at small and large $b_T$, respectively. In the following, we will specify our choices.

\paragraph{TMD model ${\boldsymbol{M_D(z,b_T)}}$}
We require $M_D$ to have a Gaussian small $b_T$ behavior, i.e. $M_D \sim e^{- c_2 b_T^2} \times \dots$ for $b_T \to 0$. This choice is mainly driven by previous successful phenomenological analyses, sensitive to the details associated with moderate-short distances. Furthermore, we require $M_D$ to decay exponentially at large $b_T$, i.e. $M_D \sim e^{- c_1 b_T}$ for $b_T \to \infty$, as suggested in Ref.~\cite{Collins:2014jpa}. These constraints are well satisfied by a functional form proportional to a modified Bessel function of the second kind, corresponding to a power law in
momentum space first used in Ref.~\cite{Boglione:2017jlh} and vaguely reminiscent of the diquark spectator models presented in Refs.~\cite{Bacchetta:2008af,Bacchetta:2007wc}.
In the impact parameter space it has the following general expression:
\begin{align}
    \label{eq:MD_def}
    M_D(z,b_T) = \frac{2}{\Gamma(p(z)-1)} \left(\frac{b_T \, m(z)}{2}\right)^{p(z)-1} \, K_{p(z)-1}\left( b_T \, m(z)\right)
\end{align}
where the free parameters $p > 1$ and $m > 0$ have been allowed a $z$ depedence as in Ref.~\cite{Boglione:2022nzq}. This parametrization of the $M_D$ function had already been used in Refs.~\cite{Boglione:2017jlh,Boglione:2022nzq}].
Since experimental data suggest a peak in the width of the measured cross sections around $z \approx 0.6$, we conveniently express $p(z)$ and $m(z)$ in terms of the width $W(z)$ of $M_D$ in momentum space and the ratio $0 \leq R(z) \leq 1$ between its peak and its maximum value. This change of variables is defined as: 
\begin{align}
    \label{eq:MD_pars}
    &p(z) = \frac{1}{2}\left( \frac{3}{1-R(z)} - 1 \right),\qquad
    m(z) = \frac{W(z)}{z} \sqrt{\frac{3}{1-R(z)}}.
\end{align}
We then parametrize the two functions $W(z)$ and $R(z)$ as follows:
\begin{align}
    \label{eq:MD_pars_new}
    &R(z) = 1 - \alpha \, \frac{f(z)}{f(z_0)} \text{ with }
    f(z) = z \, (1-z)^{\frac{1-z_0}{z_0}},
    \qquad
    W(z) = \frac{m_{\pi}}{R(z)^2}.
\end{align}
with $z_0$ and $\alpha$ being the two free parameters to be determined fitting the data and ${m_\pi = 139.6}$ MeV being the mass of the charged pions. Notice that the expression of $W(z)$ is the same used for model II of Ref.~\cite{Boglione:2022nzq}. The parametrization above is designed in such a way that the width of $M_D$ in momentum space has a maximum in $z=z_0$, roughly corresponding to the maximum of the width of the whole cross section. 

\paragraph{$\boldsymbol{g_K(b_T)}$ function.}
We require the non-perturbative model of the Collins-Soper kernel to have a quadratic behavior at low $b_T$, i.e. $g_K \sim g_2 b_T^2 + \dots$ for $b_T \to 0$. This choice agrees with lowest order theoretical arguments~\cite{Collins:2014jpa} as well as with most modern phenomenological analyses~\cite{Cerutti:2022lmb,Bacchetta:2022awv,Barry:2023qqh}. Moreover, we require $g_K$ to saturate to a constant value at large $b_T$, i.e. $g_K \to g_0$ for $b_T \to \infty$. 
This specific behavior has never been tested before in any phenomenological analysis, despite some encouraging theoretical arguments~\cite{Collins:2014jpa} preceding this work.
As we have already stressed in Section~\ref{subsec:cond_y1}, a constant behavior of the CS-kernel at large distances is not only suggested, but it is desirable to decrease the errors associated to the factorization of Region 2 to moderate values of thrust at low transverse momentum. In fact, if $g_K \to g_0 < \infty$, the constraint on the positivity of the rapidity regulator has a lower impact at low $p_T$ compared to a divergent $g_K$. 
Finally, we only consider even functions of $b_T$, as suggested by other phenomenological work~\cite{Bacchetta:2019sam,Cerutti:2022lmb,Bacchetta:2022awv,Scimemi:2019cmh,Moos:2023yfa,Barry:2023qqh,DAlesio:2022brl} and theoretical arguments~\cite{Collins:2014jpa,Vladimirov:2020umg}.
Applying all these constraints, we restrict our analysis to the following two functions:
\begin{subequations}
    \label{eq:gK_models}
\begin{align}
    &g_K^{A}(b_T) = g_0 \, \tanh{\left( \beta^2 \, \frac{b_T^2}{b_{MAX}^2}\right)} 
    \label{eq:gKA},
    \\
    &g_K^{B}(b_T) = g_0 \, \tanh{\left( \beta^2 \, b_T^\star \, b_T \right)}. 
    \label{eq:gKB}
\end{align}
\end{subequations}
where we fix $b_{MAX} = 0.7 \text{ GeV}^{-1}$. 
Notice that while $g_K^B$ contains all even powers of $b_T$, $g_K^A$ only includes terms of order $b_T^{2+4\,k}$ with $k \geq 0$. The function $g_K^B$ is indeed inspired by the functional form used in Ref.~\cite{Scimemi:2019cmh} and modified in order to saturate to a constant value ($g_0$) at large $b_T$. 

\bigskip

\begin{figure}
\begin{floatrow}
\hspace{-1.7cm}
\capbtabbox{%
\rowcolors{2}{white}{Gray}
\begin{tabular}{|c|c|}
  \hline
  $\chi^2/{\text{d.o.f.}}$ & $0.6183$\\
  \toprule
  $z_0$ & $0.5521^{+0.0415}_{-0.0398}$ \\ 
  $\alpha$ & $0.3644^{+0.0250}_{-0.0282}$ \\ 
  %\hline
  $g_0$ & $0.2943^{+0.0329}_{-0.0261}$ \\ 
  $\beta$ & $4.7100^{+1.9856}_{-1.9856}$ \\ 
  \hline
\end{tabular}
}{%
\vspace*{6.5mm}
\captionsetup{labelformat=empty}
\caption{Minimal $\chi^2_{\text{\tiny d.o.f.}}$ and parameters obtained by fitting the bin corresponding to $T = 0.825$, using the $g_K^B$ parametrization, Eq.~\eqref{eq:gKB}.}%
\label{tab:step0}
}
\ffigbox{
\includegraphics[width=9cm]{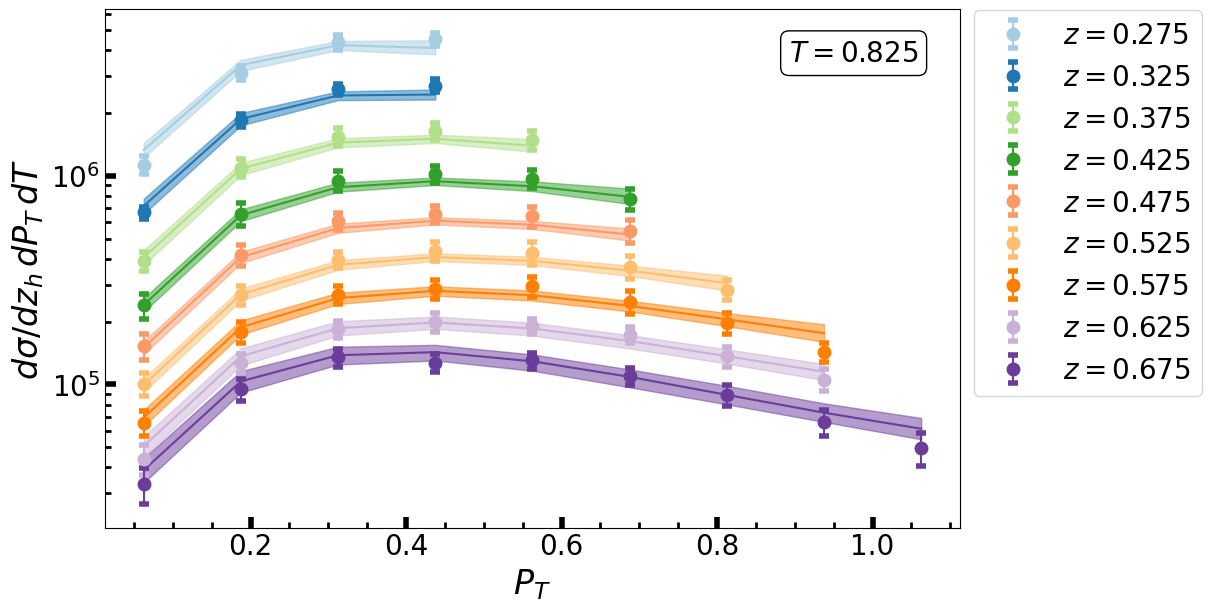}
}
{%
\caption{The cross section obtained from the preliminary fit of the $T=0.825$ bin is compared to the BELLE measurements of Ref.~\cite{Seidl:2019jei}. Error bands represent the statistical uncertainty of the fit at 2$\sigma$ confidence level.}
\label{fig:fit_prelim}}
\end{floatrow}
\end{figure}

The preliminary fit gives equivalent results in both cases, A and B. In fact, not only the value of the corresponding $\chi^2$ but also that of the parameters of the functions are well consistent with each other within statistical errors. For this reason, here we explicitely show only the results associated to case B, Eq.~\eqref{eq:gKB}. In Table~\ref{tab:step0} we show the fitted values of the 4 free parameters of the fit, which determine the shape of the TMD FF, together with the value of the $\chi^2$ per degrees of freedom. 
In Fig.~\ref{fig:fit_prelim} we show the comparison of our computation of the cross section with experimental data. 
As underlined in Ref.~\cite{Boglione:2022nzq}, the selected data allow to determine the large $b_T$ behavior of the TMD with high precision, but they loosely constrain the behavior at short distances.
This feature is particularly evident for the $g_K$ function, where the estimate of the $g_2$  coefficient renders $g_2 = 6.53^{+13.08}_{-2.39}$, corresponding to a fast and early transition from quadratic to constant behavior, already at small values of $b_T$ where perturbative contributions are expected to be dominant. Such imbalance towards non-perturbative contributions is due to the relatively small center of mass energy at which the thrust distribution is measured. Previous work on the phenomenology of thrust distributions of fully inclusive $\epm$ annihilations~\cite{Davison:2009wzs}  highlighted the significant role played by non-perturbative effects already for $Q \lesssim 35$ GeV.  
Therefore, although we selected a bin that might be considered sufficiently far from the peak of the thrust distribution, a certain amount of correlation between thrust and transverse momentum, in the correspoding non-perturbative sectors, might be sizeable. Clearly, the correlation between the thrust dependence and the Collins-Soper kernel enhances this effect.
In the next Section, we will introduce a specific function that takes care of the non-perturbative effects genuinely associated with thrust, which will partially relieve the function $g_K$ from the burden of describing alone the whole non-perturbative behaviour of the cross section, constraining its shape at short distances much more precisely.
We stress that the aim of this preliminary fit is to test the validity of the functional forms chosen for the parametrizations of the $M_D$ and $g_K$ functions. Consequently, the specific values of the free parameters obtained in this fit should not be taken at face value, as they might (and will) change in the final fit, especially those associated to $g_K$.

\subsection{Phenomenological analysis of the \texorpdfstring{\sia}{SIA thr}~cross section in the central region \label{subsec:pheno_step2}}

\begin{figure}
\centering
\subfloat[The \sia ~cross section obtained from our final fit for parametrization A is compared to the BELLE measurements~\cite{Seidl:2019jei}. The error bands represent the statistical uncertainty of the fit at 2$\sigma$ confidence level. The  uncertainties generated by the error on the collinear FFs are not shown,  not to compromise the readability of the plots. In general they would amount to a factor of 4-5 times the size of the corresponding statistical errors. They are shown in Fig. 8 of Ref.~\cite{Boglione:2022nzq}.
]{\includegraphics[width=15cm]{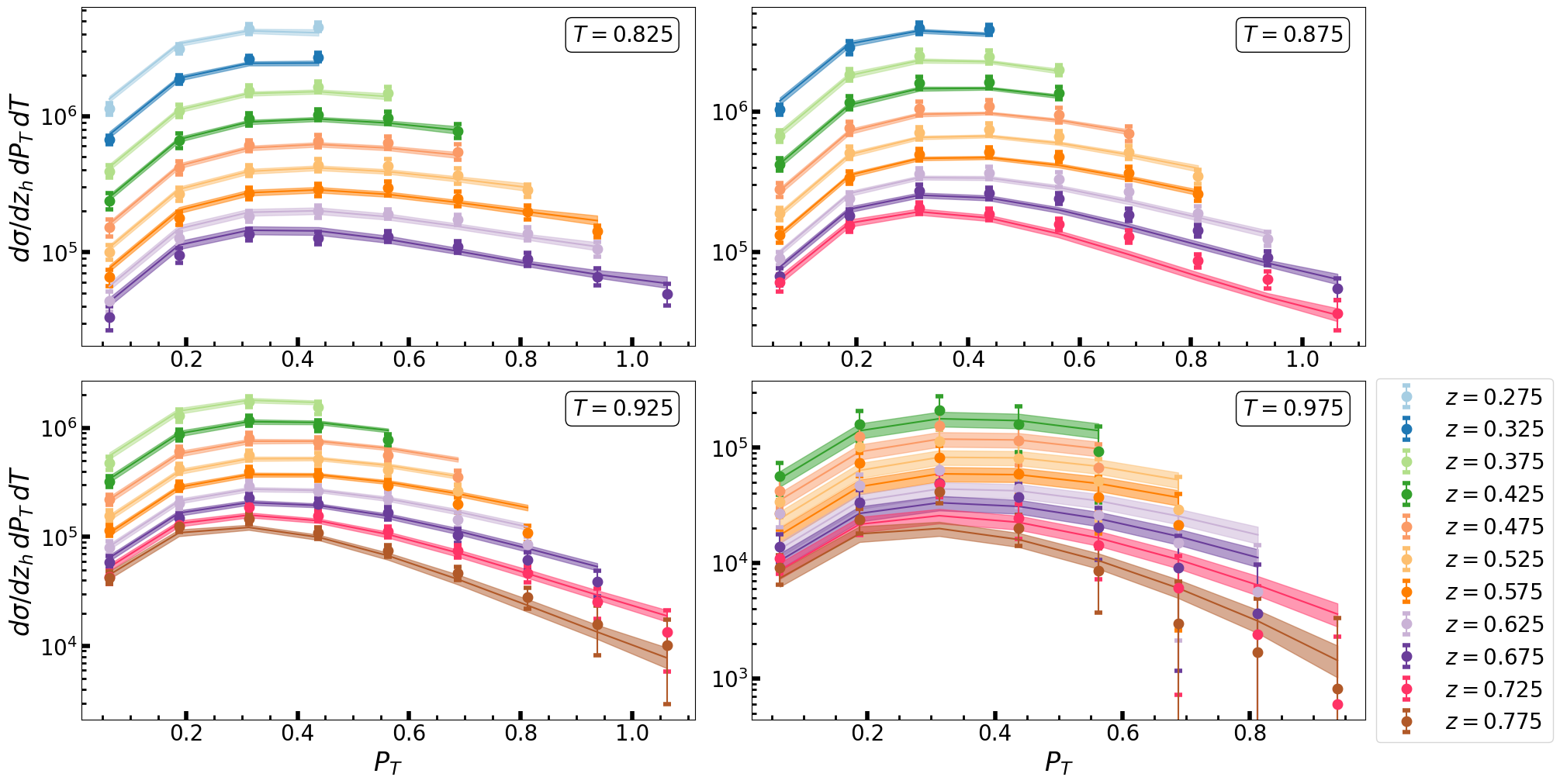}
\label{fig:step2_A}}
\\
\subfloat[The \sia ~cross section obtained from our final fit for parametrization B is compared to the BELLE measurements~\cite{Seidl:2019jei}. The error bands represent the statistical uncertainty of the fit at 2$\sigma$ confidence level.  Uncertainties generated by the error on the collinear FFs are not shown, not to compromise the readability of the plots. In general they would amount to a factor of 2-3 times the size of the corresponding statistical errors. They are shown in Fig. 8 of Ref.~\cite{Boglione:2022nzq}.]{
\includegraphics[width=15cm]{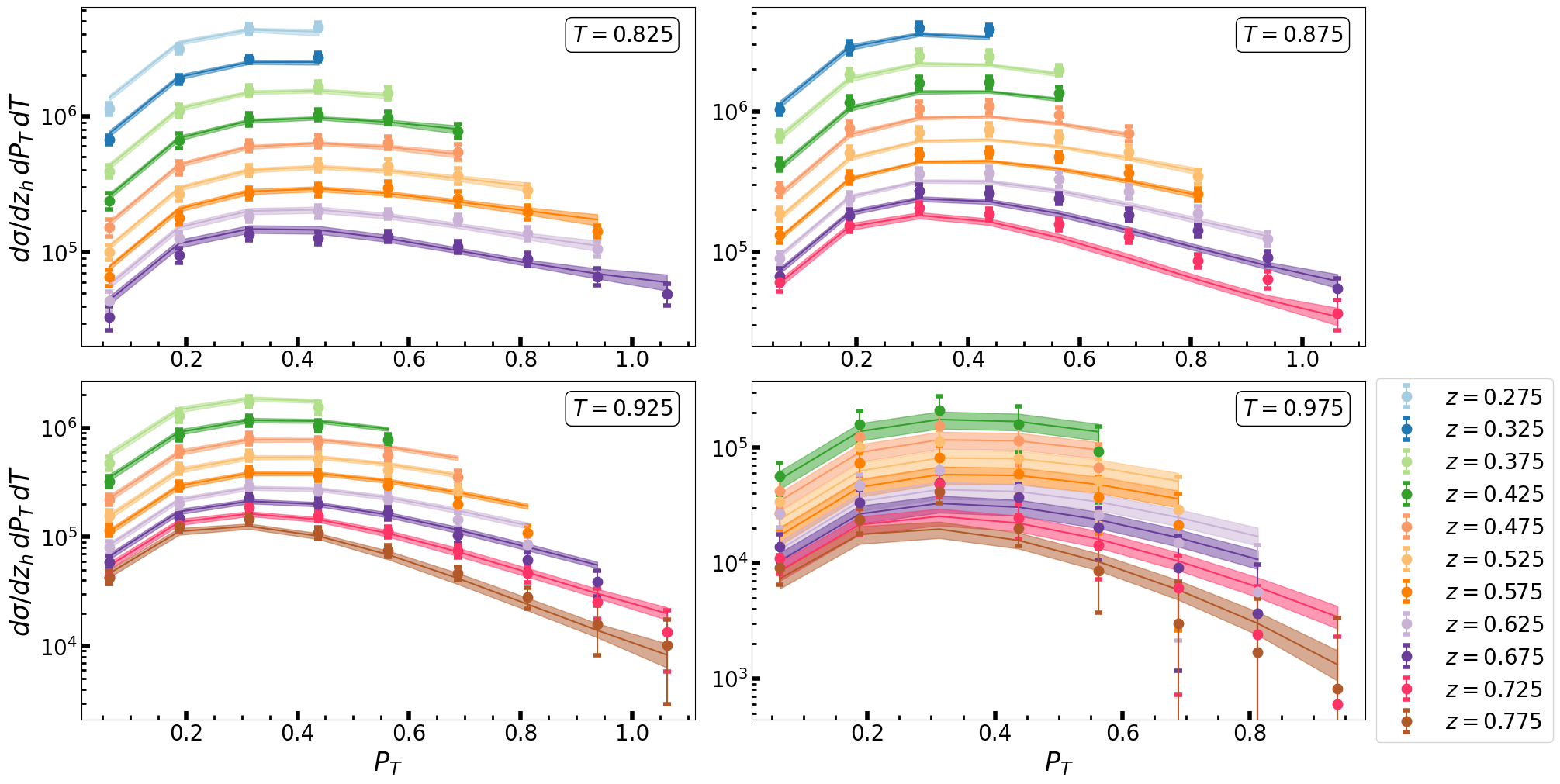}
\label{fig:step2_B}}
\end{figure}

\begin{figure}
\centering
\includegraphics[width=15cm]{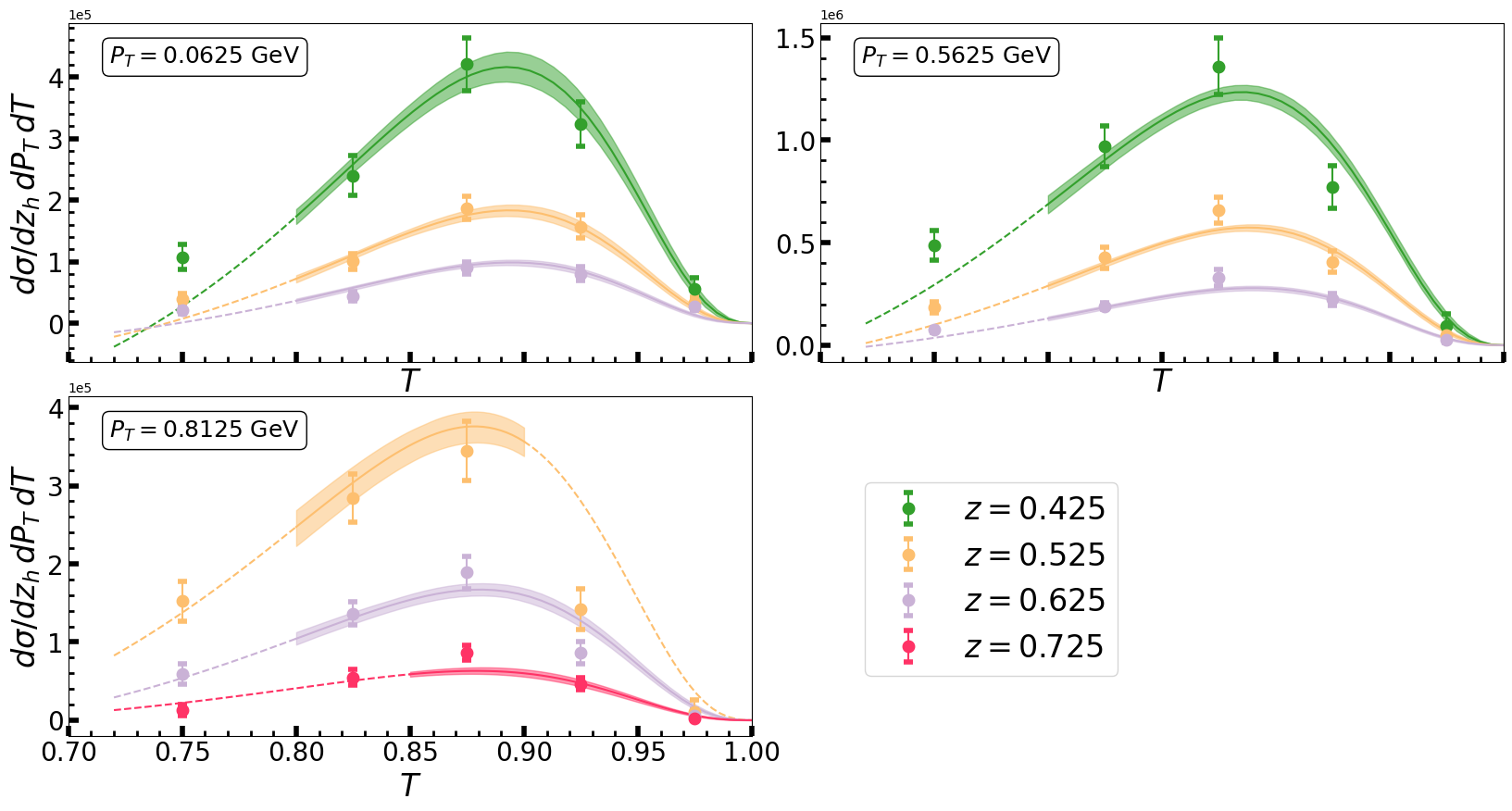}
\caption{The thrust distribution of the \sia ~cross section, obtained from our final fit, compared to the BELLE experimental data, for selected values of $P_T$ and $z$.  Here we show only model B, as model A would result in analogous curves. The lines representing the results of our fit are extended to lower values of thrust than those actually included in the fit (dotted lines) only to show that our prediction are in reasonable agreement with data also in the T-range where they are not fitted, especially for larger values of $z$.  
The uncertainty bands represent the statistical uncertainty of the fit at 2$\sigma$ confidence level.} 
\label{fig:Tdep}
\end{figure}

We are now ready for our final fit, including all 230 data points from the selection presented in Table~\ref{tab:data_sel}). In order to achieve a successful description of the thrust dependence of the SIA$^{\text{thr}}$ cross section, a specific prescription to include the non-perturbative effects associated with thrust has to be applied. In the past years, several different methods have been proposed for the treatment of  event shape observables in the non-perturbative region. Most of them apply the necessary modifications to the resummed expressions at the level of the cumulative cross section of Eq.~\eqref{eq:cumulativeR_NNLL},~\cite{Korchemsky:1999kt,Korchemsky:2000kp,Gardi:2001ny,Gardi:2002bg,Hoang:2007vb,Ligeti:2008ac,Abbate:2010xh,Davison:2009wzs}, 
Here, we opt for a simpler implementation. Our strategy  consists in shifting the resummed factorized cross section of Eq.~\eqref{eq:FactT_R2} by computing it in $T-T_0$ and multiplying it by a shaping function $f_{\text{\tiny NP}}(\tau)$. This function is required to satisfy $f_{\text{\tiny NP}}(\tau) \to 1$ as $\tau$ increases and $f_{\text{\tiny NP}}(\tau) = 0$ in the $2$-jet limit $\tau \to 0$. 
According to this prescription, the cross section used to fit the experimental data becomes:
\begin{align}
    \label{eq:xs_thrustNP}
    \frac{d \sigma}{dz\,dT\,d^2 \vec{P}_T} = 
    \left. \frac{d \sigma^{\text{res.}}}{dz\,dT\,d^2 \vec{P}_T}
    \right|_{T-T_0} \, f_{\text{\tiny NP}}(1-T)
\end{align}
where the first factor refers to the cross section of Eq.~\eqref{eq:FactT_R2}, shifted by $T_0$.
Given the correlation between the thrust dependence and the CS-kernel, the choice of $f_{\text{\tiny NP}}$ is not independent from the choice of $g_K$. We employ the following functions for the two models presented in Eqs. ~\eqref{eq:gK_models}:
\begin{subequations}
    \label{eq:fNP_models}
\begin{align}
    &f_{\text{\tiny NP}}^{A} = \tanh{\left( \rho^2 \tau^2 \right)},
    \label{eq:fNPA}
    \\
    &f_{\text{\tiny NP}}^{B} = \tanh^2{(\rho \tau)}.
    \label{eq:fNPB}
\end{align}
\end{subequations}
With this prescription, we are able to extend our phenomenological analyses to all the bins of our selection by introducing 2 extra free parameters, $T_0$ and $\rho$. We then have a total of 6 free parameters. Their values, in the two cases A and B, are reported in Tables~\ref{tab:step2_A} and~\ref{tab:step2_B}, respectively.
\begin{table}[b]%
\centering
\subfloat[][Minimal $\chi^2_{\text{\tiny d.o.f.}}$ and parameters obtained by fitting our selection of BELLE data of Ref.~\cite{Seidl:2019jei}, using the parametrizations for case A, Eq.~\eqref{eq:gKA} for $g_k$ and Eq.~\eqref{eq:fNPA} for $f_{NP}$, Eq.~\eqref{eq:MD_def} for $M_D$.]{
\rowcolors{2}{white}{Gray}
\begin{tabular}{|c|c|}
  \hline
  $\chi^2/{\text{d.o.f.}}$ & $1.0749$\\
  \toprule
  $z_0$ & $0.5335^{+0.0194}_{-0.0180}$ \\ 
  $\alpha$ & $0.3403^{+0.0114}_{-0.0122}$ \\ 
  %\hline
  $g_0$ & $0.1044^{+0.0446}_{-0.0742}$ \\ 
  $\beta$ & $1.6765^{+0.8150}_{-0.8150}$ \\ 
  %\hline
  $T_0$ & $0.0617^{+0.0295}_{-0.0134}$ \\ 
  $\rho$ & $7.7205^{+0.2834}_{-0.2099}$ \\
  \hline
\end{tabular}%
\label{tab:step2_A}
}
\qquad \qquad
\subfloat[][Minimal $\chi^2_{\text{\tiny d.o.f.}}$ and parameters obtained by fitting our selection of BELLE data of Ref.~\cite{Seidl:2019jei}, using the parametrizations for case B, Eq.~\eqref{eq:gKB} for $g_k$ and Eq.~\eqref{eq:fNPB} for $f_{NP}$, Eq.~\ref{eq:MD_def} for $M_D$.]{
\rowcolors{2}{white}{Gray}
\begin{tabular}{|c|c|}
  \hline
  $\chi^2/{\text{d.o.f.}}$ & $1.3421$\\
  \toprule
  $z_0$ & $0.5334^{+0.0192}_{-0.0189}$ \\ 
  $\alpha$ & $0.3394^{+0.0127}_{-0.0134}$ \\ 
  %\hline
  $g_0$ & $0.1205^{+0.0305}_{-0.0367}$ \\ 
  $\beta$ & $2.0610^{+2.1042}_{-0.5193}$ \\ 
  %\hline
  $T_0$ & $0.0467^{+0.0117}_{-0.0077}$ \\ 
  $\rho$ & $8.1643^{+0.3053}_{-0.3011}$ \\
  \hline
\end{tabular}%
\label{tab:step2_B}
}
\label{tab:step2}
\end{table}
As anticipated, the parameters of $M_D$, $z_0$ and $\alpha$, vary only very mildly with respect to the preliminary fit. On the other hand, the free parameters associated to $g_K$ are now significantly different compared to those in Table~\ref{tab:step0} and the behavior at short distances is considerably more constrained. 
In fact, the coefficient weighing the quadratic behavior  at small $b_T$ of the $g_K$ function is now estimated to be $g_2 = 0.60^{+1.65}_{-0.05}$ and $g_2 = 0.51^{+1.81}_{-0.25}$, for case A and B respectively.
Although these estimates are larger than most recent extractions (see for instance Ref.~\cite{Bacchetta:2022awv}) they are definitely closer than the values obtained in the preliminary fit. Thus, we can conclude that the description of the non-perturbative effects associated to thrust is essential for a well-constrained extraction of the TMD FF, at least in this energy range. We recall that we have included these effects following the simplified strategy encoded in Eq.~\eqref{eq:xs_thrustNP}. However,  more sophisticated methods might lead to more precise estimates of the TMD parameters.
As far as the description of the experimental measurements of the cross section is concerned, Figs.~\ref{fig:step2_A} and~\ref{fig:step2_B} show how, despite a significant difference in the overall $\chi^2$, both parametrizations A and B offer similarly good descriptions of the cross sections measured by BELLE. Notice that the parameter values determined by fit B offer a more direct physical interpretation compared to the values extracted from fit A. For example, the thrust shift $T_0$ is slightly above one bin in fit A, while it remains within the width of one bin in case B. For this reason, whenever suitable, we only show the results of fit B.  
Finally, in Fig.~\ref{fig:Tdep} we show the thrust dependent distributions of the \sia cross section, as obtained from our final fit (case B), compared to the BELLE experimental measurements.
Case A is not shown as it produces analogous curves. It is interesting to notice that the optimal description obtained at small $P_T$ values starts worsening as we get closer to the phase space boundaries, especially at large transverse momentum and low $z$ (lower panel, blue line).
We stress the importance of these results, as this is the first time that the thrust dependence of the \sia~ cross section, $\epm \to hX$ data is described phenomenologically, although by adopting a simple recipe to include the non-perturbative effects associated with thrust. 

\subsection{Unpolarized TMD FF}

\begin{figure}
\centering
\includegraphics[width=15cm]{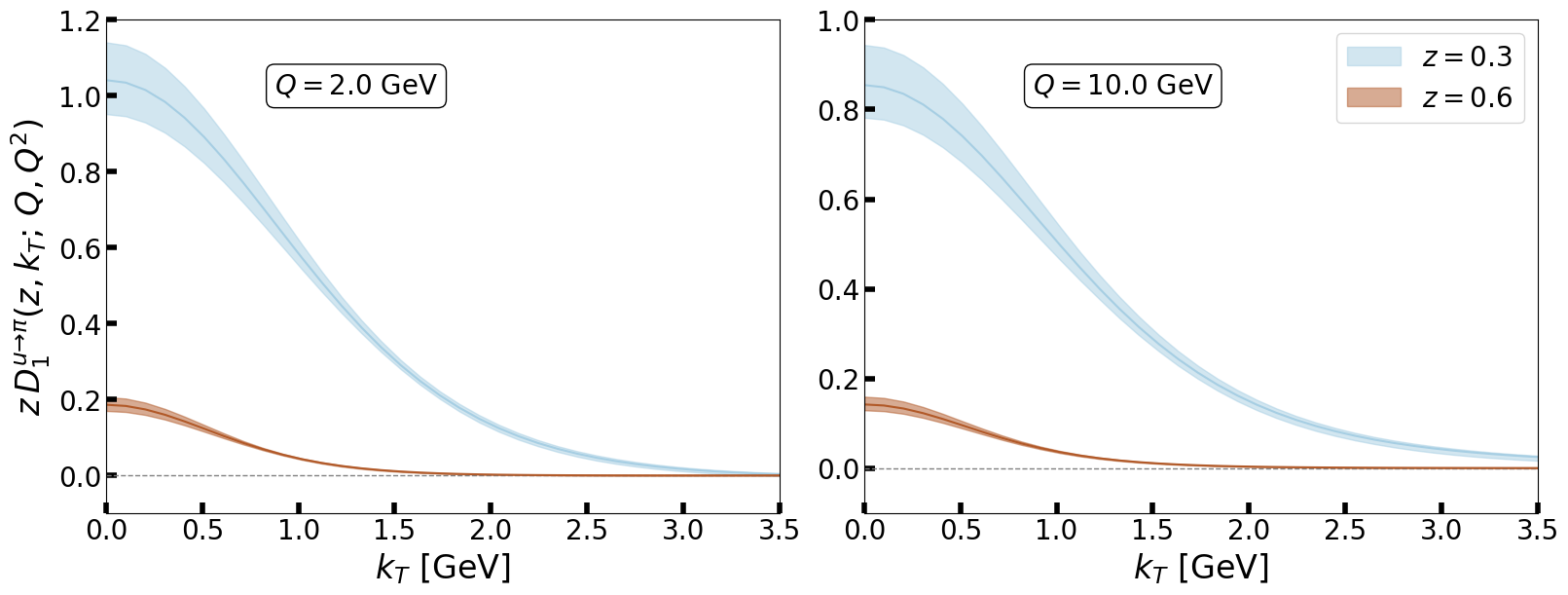}
\caption{Extraction of the unpolarized TMD FF as obtained using model B, for different values of $z$ and $Q$. The case of a fragmenting up/down quark into a charged pion is shown.  Error bands represent the statistical uncertainty of the fit at 2$\sigma$ confidence level. The rapidity scale has been set to $\zeta=Q^2$ as in standard TMD factorization.} 
\label{fig:tmds}
\end{figure}

The unpolarized TMD FF obtained from the fit corresponding to model B is shown in Fig.~\ref{fig:tmds} (the extraction obtained using model A leads to analogous results). 
In order to facilitate the comparison with extractions obtained in standard TMD factorization, we have set the rapidity scale to $\zeta=Q^2$. Remarkably, the behavior at low-$k_T$ is significantly different from that obtained in Ref.~\cite{Bacchetta:2022awv}. This should not be surprising, as the definition of TMD FF that we adopt for the \sia \, process is different from that used in standard TMD factorization~\cite{Boglione:2020cwn}. Indeed, such difference is expected to be particularly evident at low transverse momentum, where the effects of the inclusion of contributions associated with soft radiation into the standard definition become significant. In the future, this kind of comparison will be crucial for exploring the long-distance behavior of the the soft factor contributing to SIDIS and $\epm$ TMD cross sections.

\begin{figure}
\centering
\includegraphics[width=8cm]{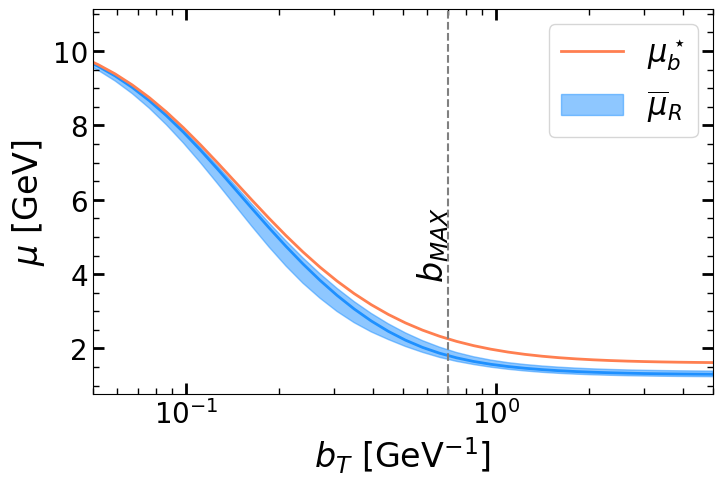}
\caption{Comparison of the scales $\mu_R = {Q e^{\overline{y}_1}}/{u_E}$ and $\mu_b^\star$. The behavior of $\mu_R$ depends on $g_K$, here in our case B. Even at large distances, the scale $\mu_R$ remains large enough to allow a treatment in terms of perturbation theory.} 
\label{fig:scales_y1}
\end{figure}

The extraction of the TMD FF implies the determination of the function $g_K$. This allows to verify \emph{a posteriori} whether the assumption on the positivity of the rapidity cut-off is satisfied by our selection on BELLE data (Table~\ref{tab:data_sel}). Let's first consider the condition of Eq.~\eqref{eq:y1SOL_hp1} and, in particular, the asymptotic limit reached by $u_E^{\text{\tiny MIN}}$ at large $b_T$. With our choices for $b_{\text{\tiny MAX}}$ and log-accuracy, the two constants appearing in its asymptotic behaviour, Eq.~\eqref{eq:uEmin_asy}, are $A_1 = 1$ and $A_2 \sim 0.85$. 
Thus, using the values of $g_0$ obtained using model A, Eqs.~\eqref{eq:gKA},~\eqref{eq:fNPA}, and model B, Eqs.~\eqref{eq:gKB},~\eqref{eq:fNPB}, we obtain ${u_E^{\text{\tiny MIN}} \to {7.953}^{+0.620}_{-0.979}}$ and ${u_E^{\text{\tiny MIN}} \to {8.175}^{+0.428}_{-0.499}}$, respectively.
Such estimates are still given in Laplace space. A very crude estimate of the corresponding quantities in thrust space can be obtained by replacing $u_E$ with $1/\tau$. Hence the condition $u_E > u_E^{\text{\tiny MIN}}$ becomes $T \gtrsim 1 - {1}/{u_E^{\text{\tiny MIN}}}$. This implies that the size of the errors associated with the positivity of the rapidity cut-off is under control up to very small values of $P_T/z$ if ${T \gtrsim {0.874}^{+0.009}_{-0.018}}$ and ${T \gtrsim {0.878}^{+0.006}_{-0.008}}$, respectively. It is important to stress that such estimates should only be regarded as an indication of the values of thrust that can be considered large enough to extend the kinematics of Region 2 to very low values of transverse momentum, compatibly with the increasing dominance of the Region 1. Analogously, they do not imply that the factorization theorem of Region 2 abruptly ceases to be valid if $T$ is lower than the limit associated with $u_E^{\text{\tiny MIN}}$, nor should they be trusted for determining the value of  $P_T$ at which the size of the errors starts getting too large. In fact, we have found a satisfactory agreement with experimental data starting from the lowest data points in $P_T$ not only for $T \gtrsim 0.875$, but also in the lowest thrust bin, as showed in Section~\ref{subsec:step0} and in Fig.~\ref{fig:Tdep}.
The other two conditions encoded in Eq.~\eqref{eq:y1SOL_hp2} and Eq.~\eqref{eq:y1SOL_hp3} can instead be easily verified, with the latter in particular being automatically satisfied for any $g_0 < \infty$. The hypothesis on the size of the scale $\mu_R = {Q e^{\overline{y}_1}}/{u_E}$, that should always be large enough to avoid the Landau pole, is checked by comparing it to the scale $\mu_b^\star$. As shown in Fig.~\ref{fig:scales_y1}, they follow a similar behavior and, most importantly, they saturate to similar values at large $b_T$. More specifically, ${\overline{\mu}_R \to 1.330^{+0.187}_{-0.096}\text{ GeV}}$ for model A, while ${\overline{\mu}_R \to 1.294^{+0.084}_{-0.064} \text{ GeV}}$ for model B. Since these are scales where we can reasonably apply perturbation theory, the condition of Eq.~\eqref{eq:y1SOL_hp2} can be considered to be satisfied.

\bigskip

\subsection{Comparison with previous work \label{sub:approx_formalism}}

In this section, we examine the formalism proposed in Ref.~\cite{Boglione:2020auc} and tested in Ref.~\cite{Boglione:2022nzq} in the light of the rigorous scheme devised in this paper. The comparison with the results for Region 2 obtained in the framework of SCET~\cite{Makris:2020ltr} has already been addressed in Section~\ref{sec:match}.

\bigskip

The cross section presented in Ref.~\cite{Boglione:2020auc} had been obtained by introducing a further artificial ``topology" cut-off $\lambda$ in the formalism, forcing the cross section to describe the $2$-jet limit of the $\epm \to hX$ process in the limit $\lambda \to 0$. In this sense, the parameter $\lambda$, which acted as an upper limit for integrals over the transverse momentum, was strictly related to  thrust. This introduces an explicit relation between the rapidity of the collinear radiation and the thrust, which ultimately results in a condition that fixes the rapidity cut-off to $y_1 = -\log{(\sqrt{\tau})}$. Notice that this coincides with the minimum allowed value for the rapidity of a particle belonging to the same jet of the detected hadron, according to a simple kinematic argument~\cite{Makris:2020ltr,Boglione:2020auc}.
Because of this strong approximation, the resulting factorization theorem cannot be considered a rigorous treatment of the thrust distribution of the TMD \sia ~cross section; nevertheless it has the undeniable advantage of strongly simplifying the phenomenological analysis of $\epm \to HX$ cross sections, without affecting the genuine spirit of the physics of Region 2. In fact, it restores the naive expectation of a cross section expressed as a convolution of a ``hard" partonic cross section, completely independent of the transverse momentum of the detected hadron, with a TMD FF that encodes all the TMD effects. On the other hand, the TMD FF is defined differently from that employed in SIDIS, as in \sia ~it is unaffected by soft radiation. Moreover, although altered, the relation with the rapidity cut-off and the thrust is made explicit, playing a central role in phenomenological studies.

\bigskip

This result can be recovered starting from the rigorous factorization theorem devised in Ref.~\cite{Boglione:2021wov} and expressed by the cross section in Eq.~\eqref{eq:FactT_R2} by applying two main approximations. First, the whole $b_T$-dependence associated with terms other than the TMD FF in Eq.~\eqref{eq:FactT_R2} has to be eliminated by freezing $b_T = {c_1}/{\sqrt{\tau}\,Q}$, as if such contributions were integrated out up to ${P_T}/{z_h} \leq \sqrt{\tau}\,Q$. This inevitably harms the relation between the rapidity cut-off and the kinematic variables. Next, the TMD FF is re-equipped explicitly with a rapidity cut-off\footnote{Recall that $\zeta=Q^2$ in the TMD FF of Eq.~\eqref{eq:FactT_R2}.}, although this cannot be related rigorously to thrust anymore and, consequently, it is fixed to $y_1^{\text{approx.}} = -\log{(\sqrt{\tau})}$, according to the naive kinematic argument mentioned above. 
The formula presented in Ref.~\cite{Boglione:2020auc} corresponds to the double-log (DL) approximation for the thrust dependence of the resulting cross section, and, neglecting multiplying constants as in Eqs.~\eqref{eq:FactT_R2_schematic},~\eqref{eq:FactT_M_schematic}, in $b_T$-space reads:
\begin{align}
    \label{eq:ft_approx_bT}
    &d\widetilde{\sigma}_{R_2}^{\text{approx.}} 
    \sim 
    -  a_S \, C_F \,
    \frac{3+8\log{\tau}}{\tau}\,
    e^{
    -a_S \,3 C_F \,
    \log^2{\tau}
    }\,
    \widetilde{D}_{h/j}\left(z_h,a_S,L_b,y_1^{\text{approx.}};\,b_T\right).
\end{align}
This is the cross section tested in the phenomenological analysis of Ref.~\cite{Boglione:2022nzq}. Notice that, since in Eq.~\eqref{eq:ft_approx_bT} the actual relation between the thrust and the rapidity cut-off is not properly taken into account, this result undermines any serious attempt of a rigorous thrust resummation. For this reason, any phenomenological analysis based on Eq.~\eqref{eq:ft_approx_bT} cannot be applied to large values of $T$. For instance, in Ref.~\cite{Boglione:2022nzq} the selected range of thrust values is $0.75 \leq T \leq 0.875$, leaving out most of the core of the $2$-jet region of $\epm$ annihilation.

\bigskip

\section{Conclusions \label{sec:concl}}

The highly non-trivial kinematic structure of \sia\ processes makes its study extremely complicated, especially for the formulation of factorization theorems properly describing their  underlying physics. It is nevertheless one of the most relevant processes carrying information about the transverse motion of partons which is not covered by standard TMD factorization. For this reason, it is inevitably acquiring a central role for TMD physics, triggering interest in the theoretical aspects of its factorization properties as well as in the phenomenological extraction of TMD Fragmentation Functions. 
Any reliable phenomenological application must stem from a well defined factorization theorem, which in turn must be devised in its appropriate kinematic domain. Previous works on this subject~\cite{Makris:2020ltr,Boglione:2021wov} highlighted a very rich underlying kinematic structure.
However, the two main different approaches oddly showed some tension in the results associated with the central region, while agreeing at the boundary of the phase space. In Section~\ref{sec:id_R2}, we have defined this central region (Region 2) following in the footsteps of Ref.~\cite{Boglione:2021wov} and we have showed that this is indeed the \emph{only} possible definition allowing for a truly independent kinematic region in the ``bulk" of the phase space. We have devised the corresponding factorization theorem, which has to be regarded as a new and updated version of the analogous result previously devised in Ref.~\cite{Boglione:2021wov}.  However this is not yet the final result for the factorized cross section of Region 2, as it leaves out the dependence on the rapidity cut-off introduced to regularize the rapidity divergences brought about by TMD contributions. This issue has been considered as a signal of the presence of some mechanism relating the thrust dependence to the regularization of the rapidity divergences. This is extremely significant for the role of rapidity regulators in \sia, which have been promoted from mere computational tools to function of measured quantities.
A relation between thrust and rapidity cut-off seems indeed be suggested by simple kinematics considerations, but a formal proof of such intertwining was lacking before this study. 
We have thoroughly studied this correlation in Section~\ref{sec:T_rap} devising, for the first time, a sound proof leading to the explicit relation in Eq.~\eqref{eq:y1SOL}, that links the rapidity cut-off to thrust and transverse momentum. 
This relation follows as a necessary condition for the factorization theorem of Region 2: if the central and most populated region of the phase space is an independent kinematic region, then the rapidity cut-off and the thrust are not independent regulators, but they are related through a very precise equation. Remarkably, this is the equation that locates the minimum of the factorized cross section of Region 2 when regarded as a function of the rapidity cut-off and hence it corresponds to imposing the vanishing of its CS-evolution. Not surprisingly, this is indeed an  expected  requirement for any physical observable.
The final result follows by considering the factorization theorem at this very specific value of the rapidity cut-off, which depends on the behaviour of the CS kernel. It is important to underline that access to the CS kernel is provided by its relation to the rapidity scale of the process, rather than to the energy scale, as it usually happens in other TMD processes. 

In this way, we can reach farther beyond Ref.~\cite{Boglione:2021wov}, overcoming the difficulties related to the thrust resummation. Resummation which could not be performed without a deep understanding of the intimate connection between thrust and rapidity regulators.
In Section~\ref{sec:ft_R2} we showed how to obtain the factorization theorem for Region 2 at NNLL in thrust and NLL in transverse momentum. In this regard Eq.~\eqref{eq:FactT_R2} is our main result and it will be the starting point for future phenomenological studies. 

Furthermore, in Section~\ref{sec:match} we have clarified the connection between our result and the factorization theorem devised in the SCET formalism~\cite{Makris:2020ltr}. We have highlighted how the two theorems rely on different kinematic regions and therefore they lead to different factorization theorems. Moreover, we showed that the SCET result coincides with the factorization theorem holding in the matching region between Region 2 and Region 3. Despite it does not describe the very core of the phase space of \sia, the fact that factorization can be successfully applied in a matching region represents an extraordinary feature by itself. Usually, matching regions do not host a proper factorization and the matching proceeds provided some prescription had   previously been defined. In this case, the factorization theorem of Region 2 can be smoothly mapped into the factorization theorem of Region 3 through the factorization theorem holding in $M_{2,3}$, between the two regions. 

In Section~\ref{sec:pheno} we have applied the formalism developed so far to describe the thrust distribution measured by the BELLE collaboration~\cite{Seidl:2019jei}. In particular, we have considered the case where the detected hadron is a charged pion. 
The agreement with the experimental data is incredibly good. Remarkably,
we are able to describe the behavior of the \sia ~cross section with respect to thrust as well as its $z$ and $P_T$ dependencies  in the central region. This has been particularly troublesome in the past years and it is indeed one of the main results of this work. We have also extracted the \sia ~TMD fragmentation function for pions. This function is not the same used in SIDIS processes, as it is defined as a purely collinear object. However, as pointed out in Refs.~\cite{Boglione:2020auc,Boglione:2021wov,Boglione:2022nzq}, a well defined relation exists which links this peculiar TMD FF to those defined in SIDIS, and it is our plan to perform a global analysis of SIDIS and \sia ~in the near future.  In this perspective we might consider to apply the newly proposed hadron structure oriented scheme presented in Refs.~\cite{Gonzalez-Hernandez:2022ifv,Gonzalez-Hernandez:2023iso}.

In conclusion, this paper sets the proper theoretical formalism to treat the central kinematic region of \sia. All the issues regarding how rapidity divergences are handled and how they intertwine with the thrust dependence have been solved and the final result can now be computed at the desired logarithmic accuracy in thrust. This result is then proposed as a cornerstone for future phenomenological applications, that will allow to investigate the properties of TMD fragmentation functions to a new and unprecedented level of precision.

\bigskip

\section*{Acknowledgements}
We thank T.C. Rogers and J.O. Gonzalez-Hernandez for useful discussions.
This work is supported in part by the US Department of Energy (DOE) Contract No.~DE-AC05-06OR23177, under which Jefferson Science Associates, LLC operates Jefferson Lab. We acknowledge partial support by the European Union’s Horizon 2020 
research and innovation programme, under grant agreement No 824093.

\newpage

%%%%%%%%%%%%%%%%%%%%%%%%%%%%%%%%%%%%%%
%%%%%%%%%%%%%%%%%%%%%%%%%%%%%%%%%%%%%%

\appendix

\section{Notation \label{app:notation}}

In this Section we summarize the compact notation used through this work.
\begin{align}
\label{eq:notation_logs}
    &L_S = \log{\frac{\mu}{\mu_S}} = \log{\left( u_E \, \frac{\mu}{Q}\right)};
    \qquad
    L_{R,L} = \log{\frac{\mu}{\mu_{R,L}}} = L_S \mp y_{1,2};\notag \\
    &L_J = 2\,\log{\frac{\mu}{\mu_J}} = \log{\left( u_E \, \frac{\mu^2}{Q^2}\right)};
    \qquad
    \mathcal{L}_b = \log{\frac{\mu}{\mu_b}} = \log{\left( \frac{b_T \, \mu}{c_1} \right)},
\end{align}
with $u_E = u e^{-\gamma_E}$ and $c_1 = 2 e^{-\gamma_E}$, being $\gamma_E \sim 0.577\dots$ the Euler-Mascheroni constant. 
Often we encounter the logarithm $\mathcal{L}_b$ evaluated at the scale $\mu=Q$. In that case, we label it as $L_b$. We have also used the label $L_u$ for $\log{u_E}$.
When computing quantities at a certain log-accuracy, we are required to express the coupling constant $a_S$ at scale $\mu_1$ in terms of another energy scale $\mu_2$. This is achieved through:
\begin{align}
    \label{eq:running_aS}
    &a_{S,1} = 
    \frac{a_{S,2}}{1-2 \, \beta_0 \, a_{S,2} \,
    L} 
    \left( 
    1 - \frac{\beta_1}{\beta_0}\,
    \frac{a_{S,2}}{1-2 \, \beta_0 \, a_{S,2} \,
    L}\,
    \log{\left( 1-2 \, \beta_0 \, a_{S,2} \,
    L
    \right)}
    +\dots
    \right)
\end{align}
where $a_{S,i} = a_S(\mu_i)$ and $L = \log{(\mu_2/\mu_1)}$. Moreover:
\begin{subequations}
\label{eq:beta}
\begin{flalign}
    &\beta_0 = \frac{11}{3}C_A - \frac{2}{3}n_f;
    \label{eq:beta0}&&\\
    &\beta_1 = \frac{34}{3}C_A^2 - \frac{10}{3}C_A\,n_f - 2 C_F\,n_f,&&
    \label{eq:beta1}
\end{flalign}
\end{subequations}
are the lowest order coefficients of the QCD beta function.
Finally, we denote the quantities in Fourier conjugate space with a $\sim$ symbol, while we use the $\wedge$ symbol for quantities defined in the Laplace conjugate space. All perturbative results have been obtained within the $\overline{\text{MS}}$ regularization scheme.

\bigskip

\section{TMD Fragmentation Functions \label{app:tmdff}}

In this Section, we review some of the theoretical aspects of TMD Fragmentation Functions.
The TMD FF appearing in the factorization theorem of Region 2, Eq.~\eqref{eq:FactT_R2_prelim1}, represents the contribution of collinear radiation, properly subtracted to avoid double countings with the soft sector. 
In terms of field operators, this is given by:
\begin{align}
    \label{eq:fact_def}
    &\widetilde{D}_{h/j}\left(z,  b_T, y_1\right) = \cfrac{\widetilde{D}_{h/j}^{\text{uns.}}\left(z,  b_T\right)}{\widetilde{\mathbb{S}}\left( b_T, y_1\right)},
\end{align}
where the numerator is the squared matrix element accounting for the transition from the (unpolarized) fragmenting quark to the final state with the detected hadron, often simply denoted as ``unsubtracted TMD FF":
\begin{align}
    \label{eq:uns_TMDFF}
   & \widetilde{D}_{h/j}^{\text{uns.}}\left(z,  b_T\right) = \frac{\text{Tr}_C}{N_C}\frac{\text{Tr}_D}{4} \sum_X
\frac{1}{z} \int \frac{d x^-}{2\pi} e^{i k^+ x^-} \notag \\
& \quad \left\langle 0 \right| \gamma^+ \, W_{(-))}\left( {x}/{2} \to \infty\right) \psi_j \left(  {x}/{2} \right) 
\left| P;\,X \right\rangle
\left\langle  P;\,X  \right| \overline{\psi}_j \left(  -{x}/{2} \right)  \, W^{\dagger}_{(-))}\left( -{x}/{2} \to \infty\right) 
\left| 0 \right\rangle.
\end{align}
where the gauge links $W_{(-)}$ along the minus $(-)$ direction are defined as in Eq.~\eqref{eq:WL_def}. 
The definition above does not coincide with the usual TMD FFs appearing in standard TMD factorization. In fact, the commonly used TMDs also include soft physics effects~\cite{Collins:2011zzd,Aybat:2011zv}. For this reason, usual TMDs appear naturally whenever there are significant non-perturbative effects associated with the soft sector, as in Drell-Yan scattering, SIDIS and DIA processes, but also in the Region 1 of \sia, where soft radiation is TMD-relevant. In all these cases, the definition of Eq.~\eqref{eq:fact_def} can be replaced by:
\begin{align}
    \label{eq:sqrt_def}
    &\widetilde{D}_{h/j}^{\text{usual}}\left(z,  b_T, y_1\right) = 
\lim_{y_{A,B} \to \pm \infty}
\widetilde{D}_{h/j}^{\text{uns.}}\left(z,  b_T\right) \, 
\sqrt{
\cfrac{\widetilde{\mathbb{S}}\left( b_T, y_A - y_1 \right)}{\widetilde{\mathbb{S}}\left( b_T, y_A - y_B\right) \, \widetilde{\mathbb{S}}\left( b_T, y_1 - y_B\right)}},
\end{align}
where the square root not only accounts for the subtraction of soft-collinear radiation, but it also reabsorbs part of the soft factor of the process. In practice, the two definition can be matched by introducing a further non-perturbative function~\cite{Boglione:2020cwn}, the soft model $M_S$, describing the long-distance behavior of the soft factor $\widetilde{\mathbb{S}}\left( b_T, y_1 - y_2 \right)$. Then, Eq.~\eqref{eq:fact_def} and Eq.~\eqref{eq:sqrt_def} are simply related as $\widetilde{D}_{h/j}^{\text{usual}} = \widetilde{D}_{h/j} \, \sqrt{M_S}$. Thanks to this relation, global phenomenological analyses now become feasible, while preserving the universality of TMDs.

In the following, we will review the properties of TMD FFs defined by Eq.~\eqref{eq:fact_def}.
We refer to Ref.~\cite{Collins:2017oxh} and references therein for higher order-results of perturbative terms. Notice that, with respect to this reference, our Collins-Soper (CS) kernel (and hence also its anomalous dimension $\gamma_K$) is multiplied by a factor of $2$ because of the different TMD definition we adopt. 
The CS kernel controls the evolution of TMDs with respect to their rapidity cut-off. It possesses an additive anomalous dimension $\gamma_K$, which lowest order coefficients are given by:  
 \begin{subequations}
 \label{eq:gammaK}
 \begin{flalign}
        &\gamma_K^{[1]} = 16 C_F; \label{eq:gammaK_1}&&\\
        &\gamma_K^{[2]} = 
        C_A C_F \left( \frac{1072}{9} - \frac{16 \pi^2}{3}  \right) - C_F \, n_f \, \frac{160}{9}.
        \label{eq:gammaK_2}&&
 \end{flalign}
 \end{subequations}
Hence, the CS kernel can be written as:
    \begin{flalign}
        \label{eq:Ktilde_evosol}
        &\widetilde{K}\left(a_S(\mu),L_b;\,b_T\right) = 
        \widetilde{K}\left(a_S(\mu_b),0;\,b_T\right) - 
        \int_{\mu_b}^{\mu} \frac{d \mu'}{\mu'}\,\gamma_K\left(a_S(\mu')\right),&&
    \end{flalign}
where the $b_T$-dependence in $\widetilde{K}$ not encoded into the logs accounts for the  non-perturbative content of the Collins-Soper kernel. This is usually separated from its perturbative part by using the $b^\star$ prescription~\cite{Collins:2011zzd} and recast into the non-perturbative $g_K$ function~\cite{Collins:2011zzd,Aybat:2011zv}:
    \begin{flalign}
        \label{eq:gK_def}
        &\widetilde{K}\left(a_S(\mu_b),0;\,b_T\right) = 
        \widetilde{K}_{\star}\left(a_S(\mu_b^\star)\right) - \int_{\mu_b^\star}^{\mu_b} \frac{d \mu'}{\mu'}\,\gamma_K\left(a_S(\mu')\right)- g_K(b_T),&&
    \end{flalign}
    where $\widetilde{K}_{\star}\left(a_S(\mu_b^\star)\right) =  \widetilde{K}\left(a_S(\mu_b^\star),0\right)
    \sim \widetilde{K}\left(a_S(\mu_b^\star),0;\,b_T^\star\right)$, as at low distances the $b_T$-dependence outside the logs is strongly suppressed. 
Notably, other approaches~\cite{Vladimirov:2020umg} have been recently developed to investigate the large-distance behavior of the Collins-Soper kernel. 
In this work we have used the common choice:
\begin{align}
    \label{eq:bTstar_def}
    &b_T^\star = \frac{b_T}{\sqrt{1+\frac{b_T^2}{b^2_{\text{\tiny MAX}}}}}.
\end{align}
At perturbative level the CS-kernel is:
\begin{subequations}
 \label{eq:Ktilde_pert}
 \begin{flalign}
        \label{eq:Ktilde_1loop}
        &\widetilde{K} = -a_S(\mu)\, 16 C_F \,
       \mathcal{L}_b + \mathcal{O}\left( a_S^2\right),&&\\
        &\widetilde{K}_{\star}^{[1]} = 0. &&
 \end{flalign}
 \end{subequations}
The RG-evolution of TMDs is determined by the TMD anomalous dimension~\cite{Collins:2011zzd,Aybat:2011zv} $\gamma_D$. Expressing the rapidity cut-off in terms of the variable $\zeta = 2 (k^+)^2 e^{-2y_1}$, where $k^+ \sim Q/{\sqrt{2}}$ is the plus component of the overall collinear momentum,  the TMD anomalous dimension can be expressed as:
    \begin{flalign}
        \label{eq:tmd_anomD_def}
        &\gamma_D\left(a_S(\mu),\,\log{\left(\frac{\sqrt{\zeta}}{\mu}\right)}\right) = 
        \gamma_d\left(a_S(\mu)\right) - 
        \frac{1}{2} \gamma_K\left(a_S(\mu)\right)
        \, \log{\left(\frac{\sqrt{\zeta}}{\mu}\right)},&&
    \end{flalign}
    with:
    \begin{flalign}
        \label{eq:gammad_1}
        &\gamma_d^{[1]} = 6 C_F.&&
    \end{flalign}
    The general solution to the evolution equations of TMDs can be written as:
    \begin{flalign}
    \label{eq:tmd_evosol}
    &\widetilde{D}_{h/j}(z,\, a_S(\mu), \mathcal{L}_b, \log{\frac{\sqrt{\zeta}}{\mu}}; \,b_T) = 
    \widetilde{D}_{h/j}(z,\, a_S(\mu_b), 0, 0; \,b_T) \times 
    \notag &&\\
    & \times \,
    \text{exp} \left\{
    \frac{1}{2} \, \widetilde{K}   \left(a_S(\mu_b),0;\,b_T\right) \, \log{\frac{\sqrt{\zeta}}{\mu_b}} 
    + 
    \int_{\mu_b}^{\mu} \frac{d \mu'}{\mu'} \, \gamma_D\left(a_S(\mu'),\,\log{\left(\frac{\sqrt{\zeta}}{\mu'}\right)}\right)
    \right\},&&
    \end{flalign} 
    where, as did for $\widetilde{K}$, the $b_T$-dependence not encoded into the logs accounts for the large-distance behavior of the TMD. In particular, the fingerprint of the TMD is described by the model function $M_D(z,b_T;\,j,h)$, defined as:
    \begin{flalign}
    \label{eq:model_def}
    &\frac{\widetilde{D}_{h/j}(z,\, a_S(\mu), \mathcal{L}_b, \log{\frac{\sqrt{\zeta}}{\mu}}; \,b_T)}{\widetilde{D}_{h/j}(z,\, a_S(\mu), \mathcal{L}_{b^\star}, \log{\frac{\sqrt{\zeta}}{\mu}}; \,b_T^\star)} = 
     M_D(z,b_T;\,j,h) \,
     \text{exp} \left\{
     -\frac{1}{2}\,g_K(b_T)\,\log{\frac{\sqrt{\zeta}}{\sqrt{\zeta_0}}}
     \right\},&&
    \end{flalign} 
    for any $\mu$, $\zeta$. It is common to relate the rapidity reference scale to the mass of the detected hadron, as\footnote{Also~\cite{Collins:2011zzd}, $\sqrt{\zeta_0} = M_h/z_h$, with $z_h$ being the fractional energy of the detected hadron.} $\sqrt{\zeta_0} = M_h^2$. Therefore, Eq.~\eqref{eq:tmd_evosol} can be recast as:
    \begin{flalign}
    \label{eq:tmd_evosol_star}
    &\widetilde{D}_{h/j}(z,\, a_S(\mu), \mathcal{L}_b, \log{\frac{\sqrt{\zeta}}{\mu}}; \,b_T) = 
    \widetilde{D}_{h/j,\;\star}\left(z,\, a_S(\mu_b^\star)\right) \times 
    \notag &&\\
    & \times \,
    \text{exp} \left\{
    \frac{1}{2} \, \widetilde{K}_{\star} \left(a_S(\mu_b^\star)\right) \, \log{\frac{\sqrt{\zeta}}{\mu_b^\star}} 
    + 
    \int_{\mu_b^\star}^{\mu} \frac{d \mu'}{\mu'} \, \gamma_D\left(a_S(\mu'),\,\log{\left(\frac{\sqrt{\zeta}}{\mu'}\right)}\right)
    \right\}\times 
    \notag &&\\
    & \times \,
    M_D(z,b_T;\,j,h) \,
     \text{exp} \left\{
     -\frac{1}{2}\,g_K(b_T)\,\log{\frac{\sqrt{\zeta}}{M_h}}
     \right\}
    .&&
    \end{flalign} 
    The small-$b_T$ behavior of the (unpolarized) TMD FF can be written as an OPE in terms of integrated (unpolarized) FFs, indicated by lowercase letters. In particular, in the first line of Eq.~\eqref{eq:tmd_evosol_star} we have:
    \begin{flalign}
    \label{eq:wilson_coeffs}
    &\widetilde{D}_{h/j,\;\star}(z,\, a_S(\mu_b^\star)) = \sum_k 
    \int_z^1 \, \frac{d \rho}{\rho} \,
    \left[\rho^2 \, \widetilde{\mathcal{C}}_{k/j}(\rho, a_S(\mu_b^\star))\right] \,
    d_{h/k} ({z}/{\rho}, \mu_b^\star).&&
    \end{flalign} 
    where $\widetilde{D}_{h/j,\;\star}(z,\, a_S(\mu_b^\star)) = \widetilde{D}_{h/j}(z,\, a_S(\mu_b^\star), 0, 0; \,b_T^\star) \sim \widetilde{D}_{h/j}(z,\, a_S(\mu_b^\star), 0, 0)$, as the $b_T$ dependence not encoded into the logs is suppressed at low distances.

\bigskip

The procedure described above entails the standard implementation of the CSS approach.
This is usually equipped with a $b_{\text{\tiny MIN}} = c_1/\mu$ prescription, in order to satisfy the integral constraint of the TMDs, meaning to recover its collinear counterpart when it is integrated over transverse momentum.
As long as $b_{\text{\tiny MIN}} \ll b_{\text{\tiny MAX}}$, this constraint is fulfilled. This modification takes place at low-$b_T$, in the OPE of Eq.~\eqref{eq:wilson_coeffs}, in general by substituting $b_T$ with $\sqrt{b_T^2+b_{\text{\tiny MIN}}^2}$. This pushes the expected matching with the fixed order from $b_T = c_1/Q$ to $b_T = 0$. Thus, this kind of renormalization of the $UV$ divergences of the collinear functions is necessarily associated with a modification of the TMDs at large transverse momentum, starting already where $k_T \approx Q$.
These issues, together with the unavoidable left out dependence on $b_{\text{\tiny MAX}}$ in practical applications, have recently lead to a critique of the standard approach~\cite{Gonzalez-Hernandez:2022ifv,Aslan:2022zkz,Gonzalez-Hernandez:2023iso}, suggesting a novel strategy that automatically satisfies all the necessary constraints associated with the definition of the TMDs, without introducing any arbitrary scale such as $b_{\text{\tiny MAX}}$.

\bigskip

Although the introduction of $b_{\text{\tiny MAX}}$ prevents the logarithms to become large, we still approximate the TMD FF of Eq.~\eqref{eq:tmd_evosol_star} by rearrangin the perturbative expansion in powers of $L_b^\star$ instead of $a_S$, where $L_{b^\star}$ is $L_b$ with the replacement $b_T \to b_T^\star$. This follows the track of most recent applications~\cite{Bacchetta:2019sam,Bacchetta:2022awv,Scimemi:2019cmh,Moos:2023yfa,Barry:2023qqh}. 
At NLL-accuracy, the OPE is required to be evaluated up to NLO. The relevant Wilson coefficients (at scale $\mu = \mu_b^\star$) are:
    \begin{subequations}
    \label{eq:wilson_pert}
    \begin{flalign}
    &z^2 \, \widetilde{\mathcal{C}}_{q/q}(z) =
    \delta(1-z) + a_S \, 2 C_F \, \left(
    1-z + 2 \, \frac{1+z^2}{1-z} \,
    \log{z}
    \right) + \mathcal{O}\left( a_S^2 \right);
    \label{eq:Wqq_1loop}&&\\
    &z^2 \, \widetilde{\mathcal{C}}_{g/q} = 
    a_S \, 2 C_F  \, \left[
    z + 2 \, \frac{1+(1-z)^2}{z} \, \log{z}
    \right]  + \mathcal{O}\left( a_S^2 \right).
    \label{eq:Wgq_1loop}&&
    \end{flalign}
    \end{subequations}
    The exponent in the second line of Eq.~\eqref{eq:tmd_evosol_star} encodes the perturbative contributions to the evolution from $\mu_b^\star$ to $\mu$ and from $(\mu_b^\star)^2$ to $\zeta$. At NLL-accuracy its expression is given by~\cite{Boglione:2020cwn}:
    \begin{align}
        \label{eq:sud_NLL}
        &\left(\substack{\mbox{exp. 2nd line}\\\mbox{Eq.~\eqref{eq:tmd_evosol_star}}}\right) = 
        L_{b^\star} \, g_1^{\text{tmd}}(x_b^\star) + 
        g_2^{\text{tmd}}(x_b^\star) + \mathcal{O}\left( \frac{1}{L_{b^\star}}\right) + 
        \notag \\
        &\hspace{1.5cm}+
        \frac{1}{2}\log{\left( \frac{\sqrt{\zeta}}{\mu} \right)}\,
        \Big[
        g_1^{\text{CS-kernel}}(x_b^\star) + 
        \frac{1}{L_{b^\star}} \, g_2^{\text{CS-kernel}}(x_b^\star)
        + \mathcal{O}\left( \frac{1}{(L_{b^\star})^2}\right)
        \Big],
    \end{align}
    where $x_b^\star = a_S(\mu)\,L_{b^\star}$. The functions $g_i^{\text{tmd}}$, $i = 1,2$ (see Ref.~\cite{Koike:2006fn}) and $g_i^{\text{CS-kernel}}$, $i = 1,2$ are:
    \begin{subequations}
    \label{eq:gi_sud}
    \begin{flalign}
    &g_1^{\text{tmd}}(\lambda) = 
    \frac{\gamma_K^{[1]}}{4\beta_0} \left( 
    1+\frac{\log{\left(1 - \lambda\right)}}{\lambda}
    \right);
    \label{eq:g1tmd}&& \\
    &g_2^{\text{tmd}}(\lambda) = 
    \frac{\gamma_K^{[1]}}{8\beta_0^2}\,
    \frac{\beta_1}{\beta_0}\,
    \frac{\lambda}{1 - \lambda}\,
    \left( 
    1+\frac{\log{\left(1-\lambda\right)}}{\lambda}+\frac{1}{2}\,
    \frac{1 - \lambda}
    {\lambda}\,
    \log^2{\left(1 - \lambda\right)}
    \right) - 
    \notag &&\\
    &\quad-\frac{\gamma_K^{[2]}}{8\beta_0^2}\,
    \left( 
    \frac{\lambda}{1-\lambda}+\log{\left(1-\lambda\right)}
    \right) - 
    \frac{\gamma_d^{[1]}}{2\beta_0}\,\log{\left(1-\lambda\right)};
    \label{eq:g2tmd}&& \\
    &g_1^{\text{CS-kernel}}(\lambda) = 
    \frac{\gamma_K^{[1]}}{2\beta_0}\log{\left(1-\lambda\right)};
    \label{eq:g1csker}&& \\
    &g_2^{\text{CS-kernel}}(\lambda) = 
     \frac{\gamma_K^{[1]}}{4\beta_0^2}\,
     \frac{\beta_1}{\beta_0}\,
     \frac{\lambda^2}{1-\lambda}\,
     \left(1 + \frac{\log{\left(1-\lambda\right)}}{\lambda} \right) -
     \frac{\gamma_K^{[2]}}{4\beta_0^2}\,
     \frac{\lambda^2}{1-\lambda}.
     \label{eq:g2csker}&&
    \end{flalign}
    \end{subequations}
    where $\lambda = 2\,\beta_0\,x_b^\star$. Notice that if $\zeta = \mu^2$, e.g. $y_1 = 0$, then the logarithm in front of the contributions associated to the functions $g_i^{\text{CS-kernel}}$ vanishes and Eq.~\eqref{eq:sud_NLL} reduces to its first line.

\bigskip

\section{Generalized Soft Thrust Function\label{app:gst}}

The generalized soft thrust function introduced in Ref.~\cite{Boglione:2021wov} describes the contribution of the soft radiation when the dependence on the thrust $\tau \ll 1$ is taken into account. In the soft approximation, the interactions of the soft gluons with the collinear particles in the direction $n$ are encoded into a Wilson line:
\begin{align}
    \label{eq:WL_def}
    W_n(x) = \mathcal{P} \text{exp} \left\{
    -i g_0 \int_0^\infty d t \, n \cdot A(x + n t)
    \right\},
\end{align}
where the symbol $\mathcal{P}$ denotes the usual path-ordering. In contrast with the usual soft thrust function, the Wilson lines appearing in the definition are tilted off the light-cone according to the prescription of the factorization procedure presented in Ref.~\cite{Collins:2011zzd}:
\begin{align}
    \label{eq:Sgen_def}
    \mathscr{S}(\tau,\,y_1,y_2) = 
    \frac{\text{Tr}_C}{N_C} \, \sum_X \, 
    \delta\left( \tau - \frac{w \cdot P_X^R + \overline{w} \cdot P_X^L}{Q} \right) \,
    \left|
    \left \langle
    0 \left|
    W_{n_1}(0) \, W^\dagger_{n_2}(0)
    \right| X
    \right \rangle
    \right|^2,
\end{align}
where the vectors $w$ and $\overline{w}$ denote the plus and the minus light-cone directions and $n_1 = w - e^{-2y_1} \overline{w}$, $n_2 = \overline{w} - e^{2y_2}w$ are their tilted counterparts. Moreover, $P_X^{R,L}$ represents the total momentum flowing into the right/left hemisphere defined by the thrust axis.
The operation of tilting the Wilson lines off the light-cone acts as a regulator for rapidity divergences, in addition to the natural regularization induced by the thrust dependence. 
Despite being introduced artificially, such redundancy is manifest when soft-collinear radiation has an active role in generating TMD effects, in at least one of the two hemispheres. 
The central kinematic region of the thrust distribution of $\epm \to h\,X$ offers an example where this exotic configuration is realized~\cite{Boglione:2020cwn,Boglione:2020auc,Boglione:2021wov}. 
On the other hand, in most of the thrust distributions for which data are available at present time, the soft-collinear radiation is on the same footage of the soft contributions and hence integrated over transverse momentum. In this case, the implementation of the subtraction mechanism that avoids the double counting due to the overlapping between soft and collinear contributions, also remove all the rapidity cut-off dependence. 
More specifically, this overlapping is represented by the soft-collinear thrust functions introduced in Ref.~\cite{Boglione:2021wov} and defined as:
\begin{align}
    \label{eq:Ygen_def}
    &\mathscr{Y}_R(\tau,\,y_1) = \lim_{y_2\to-\infty} \mathscr{S}(\tau,\,y_1,y_2);
    &\mathscr{Y}_L(\tau,\,y_2) = \lim_{y_1\to+\infty} \mathscr{S}(\tau,\,y_1,y_2),
\end{align}
from which follows $\mathscr{Y}_R(\tau,\,-y_2) = \mathscr{Y}_L(\tau,\,y_2)$. In the previous equations, the limit operation is intended to be applied to the squared amplitudes, \emph{before} the integration over the phase space of the final state.
Once freed by double counting, the subtracted generalized soft thrust function equals the standard soft thrust function:
\begin{align}
    \label{eq:Sthr_def}
    &S(\tau) = 
    \frac{\mathscr{S}(\tau,\,y_1,y_2)}
    {\mathscr{Y}_R(\tau,\,y_1) \, \mathscr{Y}_L(\tau,\,y_2)} = \\
    &\quad=
    \frac{\text{Tr}_C}{N_C} \, \sum_X \, 
    \delta\left( \tau - \frac{w \cdot P_X^R + \overline{w} \cdot P_X^L}{Q} \right) \,
    \left|
    \left \langle
    0 \left|
    W_{w}(0) \, W^\dagger_{\overline{w}}(0)
    \right| X
    \right \rangle
    \right|^2. 
    \notag
\end{align}
This equation has been checked at 1-loop accuracy in Ref.~\cite{Boglione:2021wov}. Eq.~\eqref{eq:Sthr_def} is crucial to prove standard results of the factorization of thrust distributions in the Collins factorization formalism. In this regard, the standard approaches~\cite{Catani:1992ua,Schwartz:2007ib} to the thrust distribution of $\epm \to X$ never consider rapidity cut-offs in the definition of any of the terms appearing in the factorized cross section, as in the final result \emph{all} the rapidity divergences are naturally regulated by the thrust dependence. As long as this holds true, the various approaches converge to the same factorization theorem and expressing the soft thrust function in terms of rapidity-regulated quantities as in Eq.~\eqref{eq:Sthr_def} may seem an unnecessary complication. In Section~\ref{sec:T_rap} we have provided a concrete example where only part of the rapidity divergences are naturally regulated by the thrust, and the generalized soft thrust function appears explicitly in the final factorization theorem. 

\subsection{Evolution Equations\label{subsec:gst_evo}}

Given the crucial role of the generalized soft thrust function in recent factorization theorem~\cite{Boglione:2021wov}, in this section we will present the evolution equations of $\mathscr{S}$, with respect to both $\mu$ and the rapidity cut-offs.
As common with thrust functions, the natural ground to study evolution equations is the Laplace conjugate space of thrust. Hence, we will consider the Laplace transform of $\mathscr{S}$, defined as:
\begin{align}
    \label{eq:Sgen_lap}
    \widehat{\mathscr{S}}(u,\,y_1,y_2) = 
    \int_0^\infty d\tau 
    \,e^{-u \, \tau}\,\mathscr{S}(\tau,\,y_1-y_2).
\end{align}
Later on, it will be clear how the $u$-space is actually the proper place where perform factorization. In particular, the rapidity cut-offs have to be considered in their proper limits in the Laplace conjugate space. 

The generalized soft thrust function has many similarities with the $2$-h soft factor appearing in TMD factorization~\cite{Collins:2011zzd,Aybat:2011zv,Boglione:2020cwn}. 
In fact, even if there are three arbitrary scales, $\mu$ and $\zeta_{1,2} = Q^2 e^{\mp 2 \, y_{1,2}}$, the whole evolution is controlled solely by the kernels of the evolution with respect to the rapidity cut-offs. We will refer to such quantity as the G-kernels, defined as:
\begin{subequations}
\label{eq:Sgen_evo}
\begin{align}
    &\widehat{G}_R(u,\,y_1) = 
    2 \lim_{y_2\to-\infty} \frac{\partial \log{\widehat{\mathscr{S}}(u,\,y_1,y_2)}}
    {\partial y_1} ;
    \label{eq:Sgen_evo_R} \\
    &\widehat{G}_L(u,\,y_2) = 
    -2 \lim_{y_1\to+\infty} \frac{\partial \log{\widehat{\mathscr{S}}(u,\,y_1,y_2)}}
    {\partial y_2} ,
    \label{eq:Sgen_evo_L} 
\end{align}
\end{subequations}
from which follows that $\widehat{G}_R(u,y_1) = \widehat{G}_L(u,-y_1)$.
The limit operation has to be intended as in Eqs.~\eqref{eq:Ygen_def} and it is justified by the same argument that holds for the Collins-Soper kernel~\cite{Collins:2011zzd}. In fact, the derivative with respect to one of the rapidity cut-offs reduces the momentum regions to be either hard or collinear to the opposite direction, and hence the other Wilson line can be taken along the light-cone without encountering any rapidity divergence. 
However, in contrast with the Collins-Soper kernel, the G-kernels are not rapidity cut-off independent, as the thrust dependence involves the rapidity of each of the gluons crossing the final state cut. For this reason there is a right and a left G-kernel, depending on which rapidity cut-off the derivative acts on. 
The G-kernels have an additive anomalous dimension $\gamma_K$, the same of the Collins-Soper kernel but with reversed sign:
\begin{align}
    \label{eq:G_evo}
    &\frac{\partial \, \widehat{G}_{R,L}}{\partial \log{\mu}} = -\gamma_K.
\end{align}
Notice that this implies that the sum of a G-kernel with the Collins-Soper kernel is RG-invariant.
Eqs.~\eqref{eq:G_pert} and~\eqref{eq:G_evo} allow to recast the G-kernels as:
\begin{align}
    \label{eq:G_evosol}
    \widehat{G}_{R,L}\left( 
    a_S(\mu),\,\log \frac{\mu}{\mu_{R,L}} 
    \right)= 
    \widehat{g}(a_S(\mu_{R,L})) - 
    \int_{\mu_{R,L}}^\mu \frac{d \mu'}{\mu'}
    \gamma_K (a_S(\mu')),
\end{align}
where $\mu_{R,L} = \mu_S e^{\pm y_{1,2}}$. 
The lowest order reads:
\begin{subequations}
    \label{eq:Gkernel_pert}
    \begin{flalign}
        \label{eq:GR_1loop}
        &\widehat{G}_{R,L} = a_S(\mu)\, 16 C_F \,
        L_{R,L} + \mathcal{O}\left( a_S^2\right),&&\\
        &\widehat{g}^{[1]} = 0.&&
    \end{flalign}
\end{subequations}
More generally,  the perturbative expansion of the G-kernel is organized as follows:
\begin{align}
    \label{eq:G_pert}
    \widehat{G}_{R,L} = \sum_{n=1}^\infty a_S^n(\mu)
    \sum_{m=0}^n \, \widehat{G}^{[n],\,m} L_{R,L}^m,
\end{align}
where $a_S = {\alpha_S}/{4\pi}$. Since G-kernels have the same evolution of the CS-kernel, the perturbative coefficients in front of the log powers coincide, i.e. $\widehat{G}^{[n],\,m} \equiv \widetilde{K}^{[n],\,m}$ for $m \geq 1$. Moreover, by comparing the one-loop results of Eq.~\eqref{eq:Gkernel_pert} and Eq.~\eqref{eq:Ktilde_pert}, we see that $\widehat{G}^{[1],\,0} \equiv \widetilde{K}^{[1],\,0} = 0$. This suggests that G-kernel are indeed the same of the CS-kernel, but evaluated on a different variable. At perturbative level, one is obtained from the other with the replacement $b_T \leftrightarrow {c_1 u_E}/ {Q e^{-y_1}}$. On the other hand, the non-perturbative contributions to the CS-kernel are still  being studied and far to be fully determined. This prevent us to apply the same replacement beyond the perturbative regime. Such issues have been widely discussed in Section~\ref{subsec:cond_y1} when we solved the condition Eq.~\eqref{eq:CS_cond} for the rapidity cut-off in Region 2.

\bigskip

Having explored the properties of the G-kernels, we are now ready to write the solution of the evolution equations~\eqref{eq:Sgen_evo} for the generalized soft thrust function. Using $\mu = \mu_S$ and $y_{1,2} = 0$ for the reference values of the evolution, we have:
\begin{align}
    \label{eq:Sgen_evosol_1}
    &\widehat{\mathscr{S}}\left(\mu,\,u,\,y_1,y_2 \right) = 
    \widehat{\mathscr{S}}\left(\mu_S,\,u,\,0 \right) \times
    \notag \\
    &\quad
    \text{exp}\Big\{
    -\frac{1}{2}\int_0^{y_1} dx \, 
    \widehat{G}_R\left( 
    a_S(\mu),\,L_S-x \right) 
    + 
    \frac{1}{2}\int_0^{y_2} dx \, 
    \widehat{G}_L\left( 
    a_S(\mu),\,L_S+x \right) 
    \Big\}.
\end{align}
It is important to stress that the solution above of the evolution equations is valid only provided that the difference $y_1-y_2$ is large, as this is the condition underlying Eqs.~\eqref{eq:Sgen_evo}.
The exponent appearing in Eq.~\eqref{eq:Sgen_evosol_1} can be further manipulated by exploiting the RG-evolution of the G-kernels of Eq.~\eqref{eq:G_evo}:
\begin{subequations}
\label{eq:expG_recasting}
\begin{align}
    &\int_0^{y_1} dx \, 
    \widehat{G}_R\left( 
    a_S(\mu),\,L_S-x \right)  = 
    \notag \\
    &\quad=
    \int_{\mu_S}^{\mu_R} \frac{d \mu'}{\mu'}
    \left[
    \widehat{g}\left( a_S(\mu') \right) - 
    \gamma_K\left( a_S(\mu') \right) \, 
    \log{\left( \frac{\mu'}{\mu_S} \right)}
    \right] - 
    \log{\left( \frac{\mu_R}{\mu_S} \right)}\,
    \int_{\mu_R}^{\mu} \frac{d \mu'}{\mu'}
    \gamma_K\left( a_S(\mu') \right)
   \notag \\
&\quad=
\Phi(\mu_S, \mu_R) + \widetilde{K}\left(a_S(\mu),L_b;\,b_T\right)\,\log{\left( \frac{\mu_R}{\mu_S} \right)} ;
    \label{eq:expG_recasting_R}\\
    &\int_0^{y_2} dx \, 
    \widehat{G}_L\left( 
    a_S(\mu),\,L_S+x \right)  = 
    -\int_0^{-y_2} dx \, 
    \widehat{G}_R\left( 
    a_S(\mu),\,L_S-x \right),
    \label{eq:expG_recasting_L}
\end{align}
\end{subequations}
where the function $\Phi$ has been defined in Eq.~\eqref{eq:Phi_def}.
Therefore, the generalized soft thrust function can be written as:
\begin{align}
    \label{eq:Sgen_evosol_2}
    &\widehat{\mathscr{S}}\left(\mu,\,u,\,y_1,y_2 \right) = 
    \widehat{\mathscr{S}}\left(\mu_S,\,u,\,0 \right) \times
    \notag \\
    &\quad
    \text{exp}\Big\{
    -\frac{1}{2}
    \Big( \int_{\mu_S}^{\mu_R} \frac{d \mu'}{\mu'}
    \left[
    \widehat{g} -
    \gamma_K \, 
    \log{\left( \frac{\mu'}{\mu_S} \right)}
    \right] - 
    \log{\left( \frac{\mu_R}{\mu_S} \right)}\,
    \int_{\mu_R}^{\mu} \frac{d \mu'}{\mu'}
    \gamma_K 
    +
    \left( \mu_R \leftrightarrow \mu_L\right)
    \Big)
    \Big\}.
\end{align}
It is straightforward to check that this result is consistent with Eqs.~\eqref{eq:Sgen_evo}. Perturbatively, the finite part of the generalized soft-thrust function is given by:
\begin{flalign}
        \label{eq:gensoft_1loop}
        &\widehat{\mathscr{S}}\left( a_S(\mu),L_S,y_1,y_2\right) = 
        1 - 
        a_S \, 4 \, C_F \left( 
        L_L^2 + L_R^2 - 2 L_S^2 - \frac{\pi^2}{3}
        \right)
        + \mathcal{O}\left(a_S^2\right).&& 
\end{flalign}
where we highlighted the dependence on $L_S$ and $L_{L,R}$. 

\subsection{Soft-collinear thrust functions \label{subsec:scthr-fun}}

Now that the evolution of the generalized soft thrust function is known, the evolution of the soft-collinear thrust functions introduced in Eqs.~\eqref{eq:Ygen_def} can be deduced thanks to Eq.~\eqref{eq:Sthr_def}. 
In this case, though, the sole kernel ruling the evolution with respect to the rapidity cut-off is not enough to specify the RG-evolution. For this reason, the soft-collinear thrust functions possess two evolution equations, analogously to TMDs~\cite{Collins:2011zzd,Aybat:2011zv}.
The CS-evolution follows straighforwardly from Eqs.~\eqref{eq:Ygen_def}:
\begin{subequations}
\label{eq:Ygen_CSevo}
\begin{align}
    &\frac{\partial \log{\widehat{\mathscr{Y}}_R(u,y_1)}}{\partial{y_1}} 
    = 
    \frac{1}{2} \widehat{G}_R(a_S(\mu),\,L_S-y_1);
    \label{eq:Ygen_CSevo_R} \\
    &\frac{\partial \log{\widehat{\mathscr{Y}}_L(u,y_2)}}{\partial{y_2}} 
    = 
    -\frac{1}{2} \widehat{G}_L(a_S(\mu),\,L_S+y_2).
    \label{eq:Ygen_CSevo_L} 
\end{align}
\end{subequations}
On the other hand, RG-evolution is given by:
\begin{subequations}
\label{eq:Ygen_RGevo}
\begin{align}
    &\frac{\partial \log{\widehat{\mathscr{Y}}_R(u,y_1)}}
    {\partial{\log{\mu}}} 
    = 
    \gamma_R(a_S(\mu),\,L_S-y_1);
    \label{eq:Ygen_RGevo_R} \\
    &\frac{\partial \log{\widehat{\mathscr{Y}}_L(u,y_2)}}
    {\partial{\log{\mu}}} 
    = 
    \gamma_L(a_S(\mu),\,L_S+y_2).
    \label{eq:Ygen_RGevo_L} 
\end{align}
\end{subequations}
Combining the equations above with Eq.~\eqref{eq:G_evo}, we find that:
\begin{subequations}
\label{eq:Ygen_ad}
\begin{align}
    &\frac{\partial \gamma_R(a_S(\mu),\,L_S-y_1)}
    {\partial{y_1}} 
    = 
    -\frac{1}{2}\gamma_K(a_S(\mu));
    \label{eq:Ygen_ad_R} \\
    &\frac{\partial \gamma_L(a_S(\mu),\,L_S+y_2)}
    {\partial{y_2}} 
    = 
    \frac{1}{2}\gamma_K(a_S(\mu));
    \label{eq:Ygen_ad_L} 
\end{align}
\end{subequations}
The structure of the anomalous dimensions $\gamma_{R,L}$ is then fixed by exploiting Eq.~\eqref{eq:Sthr_def}. This relates $\gamma_{R,L}$ with the anomalous dimension $\gamma_S$ of the usual soft thrust function:
\begin{align}
    \label{eq:Ygen_ad_full}
    &\gamma_{R,L}(a_S(\mu),L_{R,L}) = 
    -\frac{1}{2}\gamma_S(a_S(\mu),L_S) \mp y_{1,2} \frac{1}{2}\gamma_K(a_S(\mu)) =
    \notag \\
    &\quad=
    -\frac{1}{2} \Big(
    \gamma_s(a_S(\mu)) 
    - \gamma_K(a_S(\mu)) \,L_{R,L}
    \Big),
\end{align}
where in the last line we used explicitly the structure of $\gamma_S$. Finally, the solution to the evolution equations of the soft-collinear thrust functions is:
\begin{align}
    \label{eq:Ygen_evosol_R}
    &\widehat{\mathscr{Y}}_{R,L}\left(\mu,\,u,\,y_{1,2}\right) = 
    \widehat{\mathscr{Y}}_{R,L}\left(\mu_S,\,u,\,0 \right) \,
    \text{exp}\Big\{
    -\frac{1}{2} \int_{\mu_S}^\mu \frac{d \mu'}{\mu'}
    \gamma_S(a_S(\mu'),\,L_S)
    \Big\}
    \times
    \notag \\
    &\quad
    \text{exp}\Big\{
    -\frac{1}{2}
    \Big( \int_{\mu_S}^{\mu_{R,L}} \frac{d \mu'}{\mu'}
    \left[
    \widehat{g} - 
    \gamma_K \, 
    \log{\left( \frac{\mu'}{\mu_S} \right)}
    \right] - 
    \log{\left( \frac{\mu_{R,L}}{\mu_S} \right)}\,
    \int_{\mu_{R,L}}^{\mu} \frac{d \mu'}{\mu'}
    \gamma_K 
    \Big)
    \Big\} .
\end{align}
Perturbatively, the finite part of the (right) soft-thrust function is given by:
\begin{flalign}
        \label{eq:YgenR_1loop}
        &\widehat{\mathscr{Y}}_R\left( a_S(\mu),L_S,y_1\right) = 
        a_S \, C_F \, \left( 4 L_R^2 -\frac{\pi^2}{6} \right)
        + \mathcal{O}\left(a_S^2\right),&& 
\end{flalign}
where we highlighted the dependence on $L_{R}$. The left case $\widehat{\mathscr{Y}}_L$ follows straightforwardly.
The results above are consistent with Eq.~\eqref{eq:Sthr_def} provided that the following relation holds at the reference scales of the evolution:
\begin{align}
    \label{eq:Sthr_ref}
    &\widehat{S}(\mu_S,\,u) = 
    \frac{\widehat{\mathscr{S}}(\mu_S,\,u,\,0)}
    {\widehat{\mathscr{Y}}_R(\mu_S,\,u,\,0) \, \widehat{\mathscr{Y}}_L(\mu_S,\,u,\,0)}.
\end{align}
The perturbative expansion of such quantity depends on $u$ only through the powers of $a_S(\mu_S)$, as the choice $\mu=\mu_S$ makes all the logarithmic terms to vanish. As a consequence, at very large values of $|u|$ e.g. (roughly speaking) at very low values of $\tau$, the perturbative expansion breaks down and non-perturbative contributions must be considered. In the commonly studied thrust distribution of $\epm$ annihilation, such non-perturbative effects become manifest in the peak region, where they compensate for a shift in the distribution causing a discrepancy between the experimental and perturbative results~\cite{Dokshitzer:1997ew,Davison:2009wzs}.  
In the past, different methods have been proposed to properly take into account such non-perturbative contributions~\cite{Korchemsky:2000kp,Hoang:2007vb,Ligeti:2008ac,Abbate:2010xh}.  

\bigskip

\section{ Soft-collinear thrust factors \label{app:scthr-fact}}

In this section, we review the properties of the soft-collinear thrust factors. These functions were first introduced in SCET-based factorization approaches~\cite{Procura:2009vm,Makris:2020ltr} and called ``Thrust-TMD collinear-soft function". More recently, they have been also presented in Ref.~\cite{Boglione:2021wov} as the large-$b_T$ asymptotic behavior of the TMD relevant soft-collinear radiation contribution to the cross section of Region 2. In this paper, we refer to them simply as  soft-collinear thrust factors.
Analogously to TMDs, these functions are fully determined through a pair of evolution equations. The CS-evolution is equal and opposite to TMDs, making the combination $\mathcal{C}_R \, D_{h/j}$ naturally CS-invariant. Moreover, the RG-evolution is ruled by the anomalous dimension $\gamma_C$, which coincides with the anomalous dimension $\gamma_{R,L}$ of the soft-collinear thrust functions defined in Eq.~\eqref{eq:Ygen_ad_full}. Such analogies are a consequence of the nature of soft-collinear thrust factors, which can be considered as ``bridges" among the collinear and the soft sector in thrust dependent observables. After introducing a $b^\star$ prescription to separate large and small $b_T$ behavior and setting the reference scales for the evolution as $\mu_0 = \mu_b^\star$ and $y_1^{(0)} = L_u - L_b^\star \equiv \overline{y}_1^\star$ in order to kill all the log dependence, the right soft-thrust factor can be written as:
\begin{align}
\label{eq:CR_evosol}
&\mathcal{C}_R\left( a_S(\mu), \mathcal{L}_b, L_R; u, b_T\right) =
\mathcal{C}_{R \,\star}\left( a_S(\mu_b^\star)\right) \times 
\notag &&\\
& \times \,
\text{exp} \left\{
\frac{1}{2} \widetilde{K}_{\star} \left(a_S(\mu_b^\star)\right) \, 
\left(y_1 - \overline{y}_1^\star\right)
%\log{\frac{\mu_R}{\mu_R^\star}} 
+
\int_{\mu_b^\star}^\mu \frac{d \mu'}{\mu'} \, \gamma_R\left( 
a_S(\mu'), \log{\frac{\mu'}{\mu_R}}
\right)
\right\}\times 
\notag &&\\
& \times \,
\Delta_{\mathcal{C}}(b_T,u) \, \text{exp} \left\{
-\frac{1}{2}g_K(b_T) \, y_1
\right\}.
\end{align}
The left case $\mathcal{C}_L$ follows straightforwardly. For simplicity, we have omitted the $\sim$ and the $\wedge$ symbols associated to the Fourier and Laplace conjugate space, respectively.
The last line accounts for the non-perturbative contributions at large $b_T$ isolated by the $b^\star$ prescription. Notice how the model $\Delta_{\mathcal{C}}$ also shows an explicit dependence on $u$. This is a feature inherited by the generalized FJF, as follows  from Eq.~\eqref{eq:coll_mix}.
Perturbatively, the finite part of the (right) soft-thrust factor is given by:
\begin{flalign}
        \label{eq:CR_1loop}
        &\mathcal{C}_R\left(  a_S(\mu), \mathcal{L}_b, L_R\right) = 
        1 + a_S \, C_F \, \left( 
-\frac{\pi^2}{6} + 8 L_R  \mathcal{L}_b - 4 \mathcal{L}_b^2
 \right)
        + \mathcal{O}\left(a_S^2\right),&& 
\end{flalign}
where we have  highlighted the dependence on $L_{R}$ and $\mathcal{L}_b$.

\bigskip

\section{Hard factor and Thrust dependent functions \label{app:thr-fun}}

We refer to Ref.~\cite{Makris:2020ltr} and references therein for the values of the coefficients of the perturbative expansions of the following thrust-dependent functions. Notice that our definition for the anomalous dimension $\gamma_k$ is obtained multiplying by a factor of $4$ the cusp anomalous dimension used in this reference. In the following list, we collect the main results up to one-loop (pole parts not included):
\begin{itemize}    
    \item Hard function. 
    \begin{flalign}
        \label{eq:hard_1loop}
        &|H|^2\hspace{-.1cm}\left( \hspace{-.1cm}a_S(\mu),\log{\frac{\mu}{Q}}\right)\hspace{-.1cm} = 
        1 + a_S C_F \left( \hspace{-.1cm}-16 + \frac{7\pi^2}{3} - 12 \log{\frac{\mu}{Q}} -8 \log^2{\frac{\mu}{Q}}\right) 
        + \mathcal{O}\left(a_S^2\right).&& 
    \end{flalign}
    with anomalous dimension:
    \begin{flalign}
        \label{eq:hard_anomD_def}
        &\gamma_H\left(a_S(\mu),\log{\frac{\mu}{Q}}\right) = 
        \gamma_h\left(a_S(\mu)\right) - \gamma_K \left(a_S(\mu)\right)\,\log{\frac{\mu}{Q}},
        &&\\
        &\gamma_h^{[1]} = -12 C_F.&&
    \end{flalign}
    \item Jet Thrust function.
    \begin{flalign}
        \label{eq:jet_1loop}
        &\widehat{J}\left( a_S(\mu),L_J\right) = 
        1 + 
        a_S C_F \left(
7 + 3 L_J + 2 L_J^2 - \frac{2 \pi^2}{3}
\right)
        + \mathcal{O}\left(a_S^2\right).&& 
    \end{flalign}
with anomalous dimension:
    \begin{flalign}
        \label{eq:jet_anomD_def}
        &\gamma_J\left(a_S(\mu),L_J\right) = 
        \gamma_j\left(a_S(\mu)\right) + \frac{1}{2} \gamma_K \left(a_S(\mu)\right)\,L_J,
         &&\\
        &\gamma_j^{[1]} = 6 C_F&&
    \end{flalign}
    \item Soft Thrust function.
    \begin{flalign}
        \label{eq:soft_1loop}
        &\widehat{S}\left( a_S(\mu),L_S\right) = 
        1 - 
        a_S C_F \left( 8 L_S^2 + \pi^2\right)
        + \mathcal{O}\left(a_S^2\right).&& 
    \end{flalign}
with anomalous dimension:
    \begin{flalign}
        \label{eq:soft_anomD_def}
        &\gamma_S\left(a_S(\mu),L_S\right) = 
        \gamma_s\left(a_S(\mu)\right) - \gamma_K \left(a_S(\mu)\right)\,L_S,
        &&\\
        &\gamma_s^{[1]} = 0.&&
    \end{flalign}
\end{itemize}
The constant terms of the anomalous dimensions are not independent, as further constraints come from the RG-invariance of the factorized cross sections in which the corresponding functions contribute. In particular:
\begin{align}
&\text{RG-invariance of  fIA$^{\text{\tiny thr}}$}
\Rightarrow \gamma_h + 2 \gamma_j + \gamma_s = 0 ,
\label{eq:ad_constr_1}\\
&\text{RG-invariance of  DIA}
\Rightarrow \gamma_h + 2 \gamma_d = 0 . 
\label{eq:ad_constr_2}
\end{align}
Notice how the second of the above equations is strictly related to the particular prescription used to regularize the rapidity divergences. More specifically, Eq.~\eqref{eq:ad_constr_2} is associated with the Collins regularization through the tilting of the soft Wilson lines off their light-cone directions. Since Ref.~\cite{Makris:2020ltr} uses a different regularization scheme, the constant terms of the TMD anomalous dimension will be different beyond the lowest order.

%\clearpage
%
%
%%%%%%%%%%%%%%%%%%%%%%%%%%
\bibliographystyle{JHEP}
\bibliography{sample}

\providecommand{\href}[2]{#2}\begingroup\raggedright\begin{thebibliography}{10}

\bibitem{Collins:1984kg}
J.C.~Collins, D.E.~Soper and G.F.~Sterman, \emph{{Transverse Momentum
  Distribution in Drell-Yan Pair and W and Z Boson Production}},
  \href{https://doi.org/10.1016/0550-3213(85)90479-1}{\emph{Nucl.Phys.}
  {\bfseries B250} (1985) 199}.

\bibitem{Collins:1989bt}
J.C.~Collins, \emph{{Sudakov form-factors}},
  \href{https://doi.org/10.1142/9789814503266_0006}{\emph{Adv. Ser. Direct.
  High Energy Phys.} {\bfseries 5} (1989) 573}
  [\href{https://arxiv.org/abs/hep-ph/0312336}{{\ttfamily hep-ph/0312336}}].

\bibitem{Aybat:2011zv}
S.M.~Aybat and T.C.~Rogers, \emph{{TMD Parton Distribution and Fragmentation
  Functions with QCD Evolution}},
  \href{https://doi.org/10.1103/PhysRevD.83.114042}{\emph{Phys.Rev.} {\bfseries
  D83} (2011) 114042} [\href{https://arxiv.org/abs/1101.5057}{{\ttfamily
  1101.5057}}].

\bibitem{Seidl:2019jei}
{\scshape Belle} collaboration, \emph{{Transverse momentum dependent production
  cross sections of charged pions, kaons and protons produced in inclusive
  $e^+e^-$ annihilation at $\sqrt{s}=$ 10.58 GeV}},
  \href{https://doi.org/10.1103/PhysRevD.99.112006}{\emph{Phys. Rev.}
  {\bfseries D99} (2019) 112006}
  [\href{https://arxiv.org/abs/1902.01552}{{\ttfamily 1902.01552}}].

\bibitem{Kang:2020yqw}
Z.-B.~Kang, D.Y.~Shao and F.~Zhao, \emph{{QCD resummation on single hadron
  transverse momentum distribution with the thrust axis}},
  \href{https://doi.org/10.1007/JHEP12(2020)127}{\emph{JHEP} {\bfseries 12}
  (2020) 127} [\href{https://arxiv.org/abs/2007.14425}{{\ttfamily
  2007.14425}}].

\bibitem{Makris:2020ltr}
Y.~Makris, F.~Ringer and W.J.~Waalewijn, \emph{{Joint thrust and TMD
  resummation in electron-positron and electron-proton collisions}},
  \href{https://doi.org/10.1007/JHEP02(2021)070}{\emph{JHEP} {\bfseries 02}
  (2021) 070} [\href{https://arxiv.org/abs/2009.11871}{{\ttfamily
  2009.11871}}].

\bibitem{Boglione:2020cwn}
M.~Boglione and A.~Simonelli, \emph{{Universality-breaking effects in $e^+e^-$
  hadronic production processes}},
  \href{https://doi.org/10.1140/epjc/s10052-020-08821-y}{\emph{Eur. Phys. J. C}
  {\bfseries 81} (2021) 96} [\href{https://arxiv.org/abs/2007.13674}{{\ttfamily
  2007.13674}}].

\bibitem{Boglione:2020auc}
M.~Boglione and A.~Simonelli, \emph{{Factorization of $e^+e^- \to H \, X$ cross
  section, differential in $z_h$, $P_T$ and thrust, in the $2$-jet limit}},
  \href{https://doi.org/10.1007/JHEP02(2021)076}{\emph{JHEP} {\bfseries 02}
  (2021) 076} [\href{https://arxiv.org/abs/2011.07366}{{\ttfamily
  2011.07366}}].

\bibitem{Boglione:2021wov}
M.~Boglione and A.~Simonelli, \emph{{Kinematic regions in the $e^+e^- \to h \,
  X$ factorized cross section in a $2$-jet topology with thrust}},
  \href{https://doi.org/10.1007/JHEP02(2022)013}{\emph{JHEP} {\bfseries 02}
  (2022) } [\href{https://arxiv.org/abs/2109.11497}{{\ttfamily 2109.11497}}].

\bibitem{Boglione:2022nzq}
M.~Boglione, J.O.~Gonzalez-Hernandez and A.~Simonelli, \emph{{Transverse
  momentum dependent fragmentation functions from recent BELLE data}},
  \href{https://doi.org/10.1103/PhysRevD.106.074024}{\emph{Phys. Rev. D}
  {\bfseries 106} (2022) 074024}
  [\href{https://arxiv.org/abs/2206.08876}{{\ttfamily 2206.08876}}].

\bibitem{Collins:2011zzd}
J.~Collins, \emph{{Foundations of perturbative QCD}}, Cambridge University
  Press (2011).

\bibitem{Chiu:2011qc}
J.-y.~Chiu, A.~Jain, D.~Neill and I.Z.~Rothstein, \emph{{The Rapidity
  Renormalization Group}},
  \href{https://doi.org/10.1103/PhysRevLett.108.151601}{\emph{Phys. Rev. Lett.}
  {\bfseries 108} (2012) 151601}
  [\href{https://arxiv.org/abs/1104.0881}{{\ttfamily 1104.0881}}].

\bibitem{Chiu:2012ir}
J.-Y.~Chiu, A.~Jain, D.~Neill and I.Z.~Rothstein, \emph{{A Formalism for the
  Systematic Treatment of Rapidity Logarithms in Quantum Field Theory}},
  \href{https://doi.org/10.1007/JHEP05(2012)084}{\emph{JHEP} {\bfseries 05}
  (2012) 084} [\href{https://arxiv.org/abs/1202.0814}{{\ttfamily 1202.0814}}].

\bibitem{Echevarria:2011epo}
M.G.~Echevarria, A.~Idilbi and I.~Scimemi, \emph{{Factorization Theorem For
  Drell-Yan At Low $q_T$ And Transverse Momentum Distributions
  On-The-Light-Cone}},
  \href{https://doi.org/10.1007/JHEP07(2012)002}{\emph{JHEP} {\bfseries 07}
  (2012) 002} [\href{https://arxiv.org/abs/1111.4996}{{\ttfamily 1111.4996}}].

\bibitem{Li:2016axz}
Y.~Li, D.~Neill and H.X.~Zhu, \emph{{An exponential regulator for rapidity
  divergences}},
  \href{https://doi.org/10.1016/j.nuclphysb.2020.115193}{\emph{Nucl. Phys. B}
  {\bfseries 960} (2020) 115193}
  [\href{https://arxiv.org/abs/1604.00392}{{\ttfamily 1604.00392}}].

\bibitem{Li:2016ctv}
Y.~Li and H.X.~Zhu, \emph{{Bootstrapping Rapidity Anomalous Dimensions for
  Transverse-Momentum Resummation}},
  \href{https://doi.org/10.1103/PhysRevLett.118.022004}{\emph{Phys. Rev. Lett.}
  {\bfseries 118} (2017) 022004}
  [\href{https://arxiv.org/abs/1604.01404}{{\ttfamily 1604.01404}}].

\bibitem{Echevarria:2016scs}
M.G.~Echevarria, I.~Scimemi and A.~Vladimirov, \emph{{Unpolarized Transverse
  Momentum Dependent Parton Distribution and Fragmentation Functions at
  next-to-next-to-leading order}},
  \href{https://doi.org/10.1007/JHEP09(2016)004}{\emph{JHEP} {\bfseries 09}
  (2016) 004} [\href{https://arxiv.org/abs/1604.07869}{{\ttfamily
  1604.07869}}].

\bibitem{Vladimirov:2020umg}
A.A.~Vladimirov, \emph{{Self-contained definition of the Collins-Soper
  kernel}}, \href{https://doi.org/10.1103/PhysRevLett.125.192002}{\emph{Phys.
  Rev. Lett.} {\bfseries 125} (2020) 192002}
  [\href{https://arxiv.org/abs/2003.02288}{{\ttfamily 2003.02288}}].

\bibitem{Collins:2017oxh}
J.~Collins and T.C.~Rogers, \emph{{Connecting Different TMD Factorization
  Formalisms in QCD}},
  \href{https://doi.org/10.1103/PhysRevD.96.054011}{\emph{Phys. Rev. D}
  {\bfseries 96} (2017) 054011}
  [\href{https://arxiv.org/abs/1705.07167}{{\ttfamily 1705.07167}}].

\bibitem{Moch:2018wjh}
S.~Moch, B.~Ruijl, T.~Ueda, J.A.M.~Vermaseren and A.~Vogt, \emph{{On quartic
  colour factors in splitting functions and the gluon cusp anomalous
  dimension}},
  \href{https://doi.org/10.1016/j.physletb.2018.06.017}{\emph{Phys. Lett. B}
  {\bfseries 782} (2018) 627}
  [\href{https://arxiv.org/abs/1805.09638}{{\ttfamily 1805.09638}}].

\bibitem{Moch:2017uml}
S.~Moch, B.~Ruijl, T.~Ueda, J.A.M.~Vermaseren and A.~Vogt, \emph{{Four-Loop
  Non-Singlet Splitting Functions in the Planar Limit and Beyond}},
  \href{https://doi.org/10.1007/JHEP10(2017)041}{\emph{JHEP} {\bfseries 10}
  (2017) 041} [\href{https://arxiv.org/abs/1707.08315}{{\ttfamily
  1707.08315}}].

\bibitem{Henn:2019swt}
J.M.~Henn, G.P.~Korchemsky and B.~Mistlberger, \emph{{The full four-loop cusp
  anomalous dimension in $\mathcal{N}=4$ super Yang-Mills and QCD}},
  \href{https://doi.org/10.1007/JHEP04(2020)018}{\emph{JHEP} {\bfseries 04}
  (2020) 018} [\href{https://arxiv.org/abs/1911.10174}{{\ttfamily
  1911.10174}}].

\bibitem{Konychev:2005iy}
A.V.~Konychev and P.M.~Nadolsky, \emph{{Universality of the
  Collins-Soper-Sterman nonperturbative function in gauge boson production}},
  \href{https://doi.org/10.1016/j.physletb.2005.12.063}{\emph{Phys.Lett.}
  {\bfseries B633} (2006) 710}
  [\href{https://arxiv.org/abs/hep-ph/0506225}{{\ttfamily hep-ph/0506225}}].

\bibitem{Scimemi:2019cmh}
I.~Scimemi and A.~Vladimirov, \emph{{Non-perturbative structure of
  semi-inclusive deep-inelastic and Drell-Yan scattering at small transverse
  momentum}}, \href{https://doi.org/10.1007/JHEP06(2020)137}{\emph{JHEP}
  {\bfseries 06} (2020) 137}
  [\href{https://arxiv.org/abs/1912.06532}{{\ttfamily 1912.06532}}].

\bibitem{Bacchetta:2019sam}
A.~Bacchetta, V.~Bertone, C.~Bissolotti, G.~Bozzi, F.~Delcarro, F.~Piacenza
  et~al., \emph{{Transverse-momentum-dependent parton distributions up to
  N$^{3}$LL from Drell-Yan data}},
  \href{https://doi.org/10.1007/JHEP07(2020)117}{\emph{JHEP} {\bfseries 07}
  (2020) 117} [\href{https://arxiv.org/abs/1912.07550}{{\ttfamily
  1912.07550}}].

\bibitem{Cerutti:2022lmb}
{\scshape MAP Collaboration} collaboration, \emph{{Extraction of pion
  transverse momentum distributions from Drell-Yan data}},
  \href{https://doi.org/10.1103/PhysRevD.107.014014}{\emph{Phys. Rev. D}
  {\bfseries 107} (2023) 014014}
  [\href{https://arxiv.org/abs/2210.01733}{{\ttfamily 2210.01733}}].

\bibitem{Bacchetta:2022awv}
{\scshape MAP Collaboration} collaboration, \emph{{Unpolarized transverse
  momentum distributions from a global fit of Drell-Yan and semi-inclusive
  deep-inelastic scattering data}},
  \href{https://doi.org/10.1007/JHEP10(2022)127}{\emph{JHEP} {\bfseries 10}
  (2022) 127} [\href{https://arxiv.org/abs/2206.07598}{{\ttfamily
  2206.07598}}].

\bibitem{Bury:2022czx}
M.~Bury, F.~Hautmann, S.~Leal-Gomez, I.~Scimemi, A.~Vladimirov and P.~Zurita,
  \emph{{PDF bias and flavor dependence in TMD distributions}},
  \href{https://doi.org/10.1007/JHEP10(2022)118}{\emph{JHEP} {\bfseries 10}
  (2022) 118} [\href{https://arxiv.org/abs/2201.07114}{{\ttfamily
  2201.07114}}].

\bibitem{Barry:2023qqh}
P.C.~Barry, L.~Gamberg, W.~Melnitchouk, E.~Moffat, D.~Pitonyak, A.~Prokudin
  et~al., \emph{{Tomography of pions and protons via transverse momentum
  dependent distributions}},
  \href{https://arxiv.org/abs/2302.01192}{{\ttfamily 2302.01192}}.

\bibitem{DAlesio:2020wjq}
U.~D'Alesio, F.~Murgia and M.~Zaccheddu, \emph{{First extraction of the
  $\Lambda$ polarizing fragmentation function from Belle $e^+e^-$ data}},
  \href{https://doi.org/10.1103/PhysRevD.102.054001}{\emph{Phys. Rev. D}
  {\bfseries 102} (2020) 054001}
  [\href{https://arxiv.org/abs/2003.01128}{{\ttfamily 2003.01128}}].

\bibitem{DAlesio:2022brl}
U.~D'Alesio, L.~Gamberg, F.~Murgia and M.~Zaccheddu, \emph{{Transverse
  $\Lambda$ polarization in e$^{+}$e$^{-}$ processes within a TMD factorization
  approach and the polarizing fragmentation function}},
  \href{https://doi.org/10.1007/JHEP12(2022)074}{\emph{JHEP} {\bfseries 12}
  (2022) 074} [\href{https://arxiv.org/abs/2209.11670}{{\ttfamily
  2209.11670}}].

\bibitem{Ebert:2018gzl}
M.A.~Ebert, I.W.~Stewart and Y.~Zhao, \emph{{Determining the Nonperturbative
  Collins-Soper Kernel From Lattice QCD}},
  \href{https://doi.org/10.1103/PhysRevD.99.034505}{\emph{Phys. Rev. D}
  {\bfseries 99} (2019) 034505}
  [\href{https://arxiv.org/abs/1811.00026}{{\ttfamily 1811.00026}}].

\bibitem{Ebert:2019tvc}
M.A.~Ebert, I.W.~Stewart and Y.~Zhao, \emph{{Renormalization and Matching for
  the Collins-Soper Kernel from Lattice QCD}},
  \href{https://doi.org/10.1007/JHEP03(2020)099}{\emph{JHEP} {\bfseries 03}
  (2020) 099} [\href{https://arxiv.org/abs/1910.08569}{{\ttfamily
  1910.08569}}].

\bibitem{Schlemmer:2021aij}
M.~Schlemmer, A.~Vladimirov, C.~Zimmermann, M.~Engelhardt and A.~Sch\"afer,
  \emph{{Determination of the Collins-Soper Kernel from Lattice QCD}},
  \href{https://doi.org/10.1007/JHEP08(2021)004}{\emph{JHEP} {\bfseries 08}
  (2021) 004} [\href{https://arxiv.org/abs/2103.16991}{{\ttfamily
  2103.16991}}].

\bibitem{LatticeParton:2020uhz}
{\scshape Lattice Parton} collaboration, \emph{{Lattice-QCD Calculations of TMD
  Soft Function Through Large-Momentum Effective Theory}},
  \href{https://doi.org/10.22323/1.396.0477}{\emph{Phys. Rev. Lett.} {\bfseries
  125} (2020) 192001} [\href{https://arxiv.org/abs/2005.14572}{{\ttfamily
  2005.14572}}].

\bibitem{Shanahan:2021tst}
P.~Shanahan, M.~Wagman and Y.~Zhao, \emph{{Lattice QCD calculation of the
  Collins-Soper kernel from quasi-TMDPDFs}},
  \href{https://doi.org/10.1103/PhysRevD.104.114502}{\emph{Phys. Rev. D}
  {\bfseries 104} (2021) 114502}
  [\href{https://arxiv.org/abs/2107.11930}{{\ttfamily 2107.11930}}].

\bibitem{Collins:2014jpa}
J.~Collins and T.~Rogers, \emph{{Understanding the large-distance behavior of
  transverse-momentum-dependent parton densities and the Collins-Soper
  evolution kernel}},
  \href{https://doi.org/10.1103/PhysRevD.91.074020}{\emph{Phys. Rev. D}
  {\bfseries 91} (2015) 074020}
  [\href{https://arxiv.org/abs/1412.3820}{{\ttfamily 1412.3820}}].

\bibitem{Catani:1991kz}
S.~Catani, G.~Turnock, B.R.~Webber and L.~Trentadue, \emph{{Thrust distribution
  in $e^+ e^-$ annihilation}},
  \href{https://doi.org/10.1016/0370-2693(91)90494-B}{\emph{Phys. Lett.}
  {\bfseries B263} (1991) 491}.

\bibitem{Catani:1992ua}
S.~Catani, L.~Trentadue, G.~Turnock and B.~Webber, \emph{{Resummation of large
  logarithms in e+ e- event shape distributions}},
  \href{https://doi.org/10.1016/0550-3213(93)90271-P}{\emph{Nucl. Phys. B}
  {\bfseries 407} (1993) 3}.

\bibitem{Dokshitzer:1997ew}
Y.L.~Dokshitzer and B.R.~Webber, \emph{{Power corrections to event shape
  distributions}},
  \href{https://doi.org/10.1016/S0370-2693(97)00573-X}{\emph{Phys. Lett. B}
  {\bfseries 404} (1997) 321}
  [\href{https://arxiv.org/abs/hep-ph/9704298}{{\ttfamily hep-ph/9704298}}].

\bibitem{Schwartz:2007ib}
M.D.~Schwartz, \emph{{Resummation and NLO matching of event shapes with
  effective field theory}},
  \href{https://doi.org/10.1103/PhysRevD.77.014026}{\emph{Phys. Rev. D}
  {\bfseries 77} (2008) 014026}
  [\href{https://arxiv.org/abs/0709.2709}{{\ttfamily 0709.2709}}].

\bibitem{Becher:2008cf}
T.~Becher and M.D.~Schwartz, \emph{{A precise determination of $\alpha_s$ from
  LEP thrust data using effective field theory}},
  \href{https://doi.org/10.1088/1126-6708/2008/07/034}{\emph{JHEP} {\bfseries
  07} (2008) 034} [\href{https://arxiv.org/abs/0803.0342}{{\ttfamily
  0803.0342}}].

\bibitem{Monni:2011gb}
P.F.~Monni, T.~Gehrmann and G.~Luisoni, \emph{{Two-Loop Soft Corrections and
  Resummation of the Thrust Distribution in the Dijet Region}},
  \href{https://doi.org/10.1007/JHEP08(2011)010}{\emph{JHEP} {\bfseries 08}
  (2011) 010} [\href{https://arxiv.org/abs/1105.4560}{{\ttfamily 1105.4560}}].

\bibitem{Korchemsky:1999kt}
G.P.~Korchemsky and G.F.~Sterman, \emph{{Power corrections to event shapes and
  factorization}},
  \href{https://doi.org/10.1016/S0550-3213(99)00308-9}{\emph{Nucl. Phys. B}
  {\bfseries 555} (1999) 335}
  [\href{https://arxiv.org/abs/hep-ph/9902341}{{\ttfamily hep-ph/9902341}}].

\bibitem{Korchemsky:2000kp}
G.P.~Korchemsky and S.~Tafat, \emph{{On power corrections to the event shape
  distributions in QCD}},
  \href{https://doi.org/10.1088/1126-6708/2000/10/010}{\emph{JHEP} {\bfseries
  10} (2000) 010} [\href{https://arxiv.org/abs/hep-ph/0007005}{{\ttfamily
  hep-ph/0007005}}].

\bibitem{Gardi:2001ny}
E.~Gardi and J.~Rathsman, \emph{{Renormalon resummation and exponentiation of
  soft and collinear gluon radiation in the thrust distribution}},
  \href{https://doi.org/10.1016/S0550-3213(01)00284-X}{\emph{Nucl. Phys. B}
  {\bfseries 609} (2001) 123}
  [\href{https://arxiv.org/abs/hep-ph/0103217}{{\ttfamily hep-ph/0103217}}].

\bibitem{Gardi:2002bg}
E.~Gardi and J.~Rathsman, \emph{{The Thrust and heavy jet mass distributions in
  the two jet region}},
  \href{https://doi.org/10.1016/S0550-3213(02)00502-3}{\emph{Nucl. Phys. B}
  {\bfseries 638} (2002) 243}
  [\href{https://arxiv.org/abs/hep-ph/0201019}{{\ttfamily hep-ph/0201019}}].

\bibitem{Hoang:2007vb}
A.H.~Hoang and I.W.~Stewart, \emph{{Designing gapped soft functions for jet
  production}},
  \href{https://doi.org/10.1016/j.physletb.2008.01.040}{\emph{Phys. Lett. B}
  {\bfseries 660} (2008) 483}
  [\href{https://arxiv.org/abs/0709.3519}{{\ttfamily 0709.3519}}].

\bibitem{Ligeti:2008ac}
Z.~Ligeti, I.W.~Stewart and F.J.~Tackmann, \emph{{Treating the b quark
  distribution function with reliable uncertainties}},
  \href{https://doi.org/10.1103/PhysRevD.78.114014}{\emph{Phys. Rev. D}
  {\bfseries 78} (2008) 114014}
  [\href{https://arxiv.org/abs/0807.1926}{{\ttfamily 0807.1926}}].

\bibitem{Abbate:2010xh}
R.~Abbate, M.~Fickinger, A.H.~Hoang, V.~Mateu and I.W.~Stewart, \emph{{Thrust
  at $N^{3}LL$ with Power Corrections and a Precision Global Fit for
  $\alpha_{s}(mZ)$}},
  \href{https://doi.org/10.1103/PhysRevD.83.074021}{\emph{Phys. Rev. D}
  {\bfseries 83} (2011) 074021}
  [\href{https://arxiv.org/abs/1006.3080}{{\ttfamily 1006.3080}}].

\bibitem{Davison:2009wzs}
R.A.~Davison and B.R.~Webber, \emph{{Non-Perturbative Contribution to the
  Thrust Distribution in e+ e- Annihilation}},
  \href{https://doi.org/10.1140/epjc/s10052-008-0836-7}{\emph{Eur. Phys. J. C}
  {\bfseries 59} (2009) 13} [\href{https://arxiv.org/abs/0809.3326}{{\ttfamily
  0809.3326}}].

\bibitem{AbdulKhalek:2021gbh}
R.~Abdul~Khalek et~al., \emph{{Science Requirements and Detector Concepts for
  the Electron-Ion Collider}: {EIC Yellow Report}},
  \href{https://doi.org/10.1016/j.nuclphysa.2022.122447}{\emph{Nucl. Phys. A}
  {\bfseries 1026} (2022) 122447}
  [\href{https://arxiv.org/abs/2103.05419}{{\ttfamily 2103.05419}}].

\bibitem{Modarres:2021ffg}
M.~Modarres and R.~Taghavi, \emph{{Applying different angular ordering
  constraints and kt-factorization approaches to the single inclusive hadron
  production in the e+e- annihilation processes}},
  \href{https://doi.org/10.1103/PhysRevD.104.114004}{\emph{Phys. Rev. D}
  {\bfseries 104} (2021) 114004}
  [\href{https://arxiv.org/abs/2111.06190}{{\ttfamily 2111.06190}}].

\bibitem{Kimber:2001sc}
M.A.~Kimber, A.D.~Martin and M.G.~Ryskin, \emph{{Unintegrated parton
  distributions}},
  \href{https://doi.org/10.1103/PhysRevD.63.114027}{\emph{Phys. Rev. D}
  {\bfseries 63} (2001) 114027}
  [\href{https://arxiv.org/abs/hep-ph/0101348}{{\ttfamily hep-ph/0101348}}].

\bibitem{Martin:2009ii}
A.D.~Martin, M.G.~Ryskin and G.~Watt, \emph{{NLO prescription for unintegrated
  parton distributions}},
  \href{https://doi.org/10.1140/epjc/s10052-010-1242-5}{\emph{Eur. Phys. J. C}
  {\bfseries 66} (2010) 163} [\href{https://arxiv.org/abs/0909.5529}{{\ttfamily
  0909.5529}}].

\bibitem{Bertone:2017tyb}
{\scshape NNPDF} collaboration, \emph{{A determination of the fragmentation
  functions of pions, kaons, and protons with faithful uncertainties}},
  \href{https://doi.org/10.1140/epjc/s10052-017-5088-y}{\emph{Eur. Phys. J. C}
  {\bfseries 77} (2017) 516}
  [\href{https://arxiv.org/abs/1706.07049}{{\ttfamily 1706.07049}}].

\bibitem{Boglione:2017jlh}
M.~Boglione, J.~Gonzalez-Hernandez and R.~Taghavi, \emph{{Transverse parton
  momenta in single inclusive hadron production in ${e^ + }{e^ - }$
  annihilation processes}},
  \href{https://doi.org/10.1016/j.physletb.2017.06.034}{\emph{Phys. Lett. B}
  {\bfseries 772} (2017) 78}
  [\href{https://arxiv.org/abs/1704.08882}{{\ttfamily 1704.08882}}].

\bibitem{Bacchetta:2008af}
A.~Bacchetta, F.~Conti and M.~Radici, \emph{{Transverse-momentum distributions
  in a diquark spectator model}},
  \href{https://doi.org/10.1103/PhysRevD.78.074010}{\emph{Phys. Rev. D}
  {\bfseries 78} (2008) 074010}
  [\href{https://arxiv.org/abs/0807.0323}{{\ttfamily 0807.0323}}].

\bibitem{Bacchetta:2007wc}
A.~Bacchetta, L.P.~Gamberg, G.R.~Goldstein and A.~Mukherjee, \emph{{Collins
  fragmentation function for pions and kaons in a spectator model}},
  \href{https://doi.org/10.1016/j.physletb.2007.09.076}{\emph{Phys. Lett. B}
  {\bfseries 659} (2008) 234}
  [\href{https://arxiv.org/abs/0707.3372}{{\ttfamily 0707.3372}}].

\bibitem{Moos:2023yfa}
V.~Moos, I.~Scimemi, A.~Vladimirov and P.~Zurita, \emph{{Extraction of
  unpolarized transverse momentum distributions from fit of Drell-Yan data at
  N$^4$LL}},  \href{https://arxiv.org/abs/2305.07473}{{\ttfamily 2305.07473}}.

\bibitem{Gonzalez-Hernandez:2022ifv}
J.O.~Gonzalez-Hernandez, T.C.~Rogers and N.~Sato, \emph{{Combining
  nonperturbative transverse momentum dependence with TMD evolution}},
  \href{https://doi.org/10.1103/PhysRevD.106.034002}{\emph{Phys. Rev. D}
  {\bfseries 106} (2022) 034002}
  [\href{https://arxiv.org/abs/2205.05750}{{\ttfamily 2205.05750}}].

\bibitem{Aslan:2022zkz}
F.~Aslan, L.~Gamberg, J.O.~Gonzalez-Hernandez, T.~Rainaldi and T.C.~Rogers,
  \emph{{Basics of factorization in a scalar Yukawa field theory}},
  \href{https://doi.org/10.1103/PhysRevD.107.074031}{\emph{Phys. Rev. D}
  {\bfseries 107} (2023) 074031}
  [\href{https://arxiv.org/abs/2212.00757}{{\ttfamily 2212.00757}}].

\bibitem{Gonzalez-Hernandez:2023iso}
J.O.~Gonzalez-Hernandez, T.~Rainaldi and T.C.~Rogers, \emph{{The resolution to
  the problem of consistent large transverse momentum in TMDs}},
  \href{https://arxiv.org/abs/2303.04921}{{\ttfamily 2303.04921}}.

\bibitem{Koike:2006fn}
Y.~Koike, J.~Nagashima and W.~Vogelsang, \emph{{Resummation for polarized
  semi-inclusive deep-inelastic scattering at small transverse momentum}},
  \href{https://doi.org/10.1016/j.nuclphysb.2006.03.009}{\emph{Nucl.Phys.}
  {\bfseries B744} (2006) 59}
  [\href{https://arxiv.org/abs/hep-ph/0602188}{{\ttfamily hep-ph/0602188}}].

\bibitem{Procura:2009vm}
M.~Procura and I.W.~Stewart, \emph{{Quark Fragmentation within an Identified
  Jet}}, \href{https://doi.org/10.1103/PhysRevD.81.074009}{\emph{Phys. Rev. D}
  {\bfseries 81} (2010) 074009}
  [\href{https://arxiv.org/abs/0911.4980}{{\ttfamily 0911.4980}}].

\end{thebibliography}\endgroup
%%%%%%%%%%%%%%%%%%%%%%%%%%
%

\end{document}